\documentclass[nofootinbib]{revtex4-2}
\usepackage[T1]{fontenc}
\usepackage[utf8]{inputenc}
\usepackage{color}
\usepackage{prettyref}
\usepackage{amsmath}
\usepackage{amssymb}
\usepackage{graphicx}
\usepackage{wasysym}
\usepackage{feyn}
\usepackage{microtype}
\usepackage[pdfusetitle,
 bookmarks=false,
 breaklinks=false,pdfborder={0 0 0},pdfborderstyle={},backref=section,colorlinks=true]
 {hyperref}
\hypersetup{
 pdfborderstyle=,linkcolor=blue,citecolor=blue}

\makeatletter

\usepackage{xcolor}
\usepackage{url}
\usepackage{booktabs}
\usepackage{amsfonts}
\usepackage{nicefrac}

\usepackage{feyn}
\usepackage[mathscr]{eucal}

\newrefformat{cap}{\hyperref[#1]{Figure~\ref{#1}}}
\newrefformat{fig}{\hyperref[#1]{Figure~\ref{#1}}}
\newrefformat{tab}{\hyperref[#1]{Table ~\ref{#1}}}
\newrefformat{sec}{\hyperref[#1]{Section~\ref{#1}}}
\newrefformat{sub}{\hyperref[#1]{Section~\ref{#1}}}
\newrefformat{subsec}{\hyperref[#1]{Section~\ref{#1}}}
\newrefformat{cha}{\hyperref[#1]{Chapter~\ref{#1}}}
\newrefformat{app}{\hyperref[#1]{Appendix~\ref{#1}}}

\makeatother

\begin{document}
\global\long\def\bR{\mathbb{R}}%
\global\long\def\bN{\mathbb{N}}%
\global\long\def\llangle{\langle\!\langle}%
\global\long\def\rrangle{\rangle\!\rangle}%
\global\long\def\T{\mathrm{\mathrm{T}}}%
\global\long\def\tf{\tilde{f}}%
\global\long\def\N{\mathcal{N}}%
\global\long\def\cD{\mathcal{D}}%
\global\long\def\cS{\mathcal{S}}%
\global\long\def\cF{\mathcal{F}}%
\global\long\def\bI{\mathbb{I}}%
\global\long\def\tr{\mathrm{tr}}%
\global\long\def\tQ{\tilde{Q}}%
\global\long\def\order{\mathcal{O}}%
\global\long\def\const{\mathrm{const.}}%
\global\long\def\cW{\mathcal{W}}%
\global\long\def\tU{\tilde{U}}%
\global\long\def\cL{\mathcal{L}}%
\global\long\def\tvarphi{\tilde{\varphi}}%
\global\long\def\tF{\tilde{F}}%
\global\long\def\cZ{\mathcal{Z}}%
\global\long\def\tC{\tilde{C}}%

\preprint{APS/123-QED}

\title{Renormalization group for deep neural networks:\\ Universality of learning and scaling laws}

\author{Gorka Peraza Coppola}
\affiliation{
 Institute for Advanced Simulation (IAS-6), \\Computational and Systems Neuroscience, \\Jülich Research Centre, \\Jülich, Germany
}%
\affiliation{
 RWTH Aachen University, \\Aachen, Germany
}%

\author{Moritz Helias}
\altaffiliation{Equal contribution. Correspondence: \texttt{g.peraza.coppola@fz-juelich.de}}
\affiliation{
 Institute for Advanced Simulation (IAS-6), \\Computational and Systems Neuroscience, \\Jülich Research Centre, \\Jülich, Germany
}%
\affiliation{
 Department of Physics, \\RWTH Aachen University, \\Aachen, Germany
}%

\author{Zohar Ringel}
\altaffiliation{Equal contribution. Correspondence: \texttt{g.peraza.coppola@fz-juelich.de}}
\affiliation{%
 The Racah Institute of Physics, \\The Hebrew University of Jerusalem, \\Jerusalem, Israel
}%

\begin{abstract}
Self-similarity, where observables at different length scales exhibit
similar behavior, is ubiquitous in natural systems. Such systems are
typically characterized by power-law correlations and universality,
and are studied using the powerful framework of the renormalization
group (RG). Intriguingly, power laws and weak forms of universality
also pervade real-world datasets and deep learning models, motivating
the application of RG ideas to the analysis of deep learning. In this
work, we develop an RG framework to analyze self-similarity and its
breakdown in learning curves for a class of weakly non-linear (non-lazy)
neural networks trained on power-law distributed data. Features often
neglected in standard treatments---such as spectrum discreteness
and lack of translation invariance---lead to both quantitative and
qualitative departures from conventional perturbative RG. In particular,
we find that the concept of scaling intervals naturally replaces that
of scaling dimensions. Despite these differences, the framework retains
key RG features: it enables the classification of perturbations as
relevant or irrelevant, and reveals a form of universality at large
data limits, governed by a Gaussian Process--like UV fixed point. 
\end{abstract}
\maketitle

\section{Introduction}

Self-similarity is a highly structured property of probabilistic systems
that possess both a notion of scale and a well-defined coarse-graining
procedure under which observables become 
indistinguishable across scales; in particular, correlations decay as power
laws with distance. Many physical systems exhibit near self-similarity---for
example, the Standard Model at high energies, turbulent flows, and
all second-order phase transitions (i.e., critical phenomena). These
behaviors are often accompanied by \textit{universality}, the insensitivity
of observations on one scale to microscopic details or perturbations
introduced at a different scale.

In the realm of machine learning, power-law behavior is similarly
pervasive. Zipf’s law in natural language, heavy-tailed distributions
in vision and audio datasets, and power-law learning curves with respect
to model or dataset scale (known as scaling laws) all point to potential
underlying structure. Moreover, the robustness of these scaling laws
to architectural and algorithmic variations \citep{kaplan2020scaling,hestness2017deeplearningscalingpredictable}
hints at some degree of universality that also aligns with the prevailing
view that scale and compute dominate performance trends \footnote{See Sutton's opinion paper, ``The Bitter Lesson" (2019) }
However, a critical gap remains: unlike in physics, we lack a precise,
let alone a canonical, notion of what defines “scale” in data or models,
and how coarse-graining should be implemented. As a result, the potential
self-similar origin of these behaviors remains obscure, and limits
our ability to import tools from critical phenomena, such as the renormalization group.

The renormalization group (RG) \citep{cardy1996scaling} is one of
the central theoretical tools in physics, particularly for understanding
critical phenomena. At its core, RG describes how probability distributions
evolve under a process of marginalizing high-energy (or small-scale)
degrees of freedom and rescaling the remaining ones. This process
effectively changes the parameters of the theory (e.g., the coefficients
of the action or negative log-probability), often leading to a continuous parameter flow,
and allows for controlled non-perturbative approximations. When the
flow reaches a fixed point, the system exhibits self-similarity and
power-law correlations. Perturbations away from the fixed point may
decay, grow, or remain unchanged under this flow, corresponding to
irrelevant, relevant, or marginal directions, respectively. Via those
notions of relevance, RG is also a central modelling framework, as
the effect of microscopic changes on the coefficients of irrelevant
perturbations to the action can often be ignored.

Motivated by the conceptual parallels between deep learning and field
theory \citep{cohen2021learning,halverson2020neural}, several works
have proposed analogies between RG and neural networks. A brief survey
of these perspectives is provided below. Following some of these works,
we adopt the Neural Network Field Theory viewpoint \citep{halverson2020neural,cohen2021learning},
treating deep neural network (DNN) outputs as fields, and the stochasticity
of training as inducing a distribution over these fields. However,
unlike previous studies that focus primarily on qualitative analogies
or formal RG constructions, we seek a concrete analytical advantage.
To this end, we focus on data exhibiting power-law correlation and confront challenges specific to DNNs---such as
spectral discreteness and lack of translation invariance---that complicate
conventional coarse-graining and rescaling procedures.

Concretely, we propose an RG framework that connects empirically
observed DNN scaling laws to underlying self-similar structures. Starting
from DNNs in the infinite-width (lazy or Gaussian Process) limit,
where the kernel spectrum follows a power law, we introduce weak non-linearities
representing either quartic loss terms or finite-width corrections.
We identify the low eigenmodes of the kernel (the 2-point function,
in physics terms) as the high-energy degrees of freedom and construct
a consistent RG procedure that tracks the impact of removing these
modes and rescaling the remaining ones. This setup enables several
key insights:

\textbf{(1)} We establish a concrete correspondence between neural scaling
laws \cite{kaplan2020scaling}, self-similarity, and RG fixed points.

\textbf{(2)} We show that peculiarities of neural network field theories---namely,
their non-locality and discrete kernel spectra---render the conventional
notion of scaling dimension ill-defined. In its place, we introduce
the concept of a scaling interval.

\textbf{(3)} We develop a dimensional analysis framework that connects
these scaling intervals to power laws in learning curves as a function
of dataset size $P$.

\textbf{(4)} For the class of models we study, we demonstrate
a form of universality at large $P$, analogous to asymptotic freedom,
governed by a UV fixed point corresponding to a non-interacting Gaussian
Process and derive leading order corrections to the GP scaling laws.

\textbf{(5)} Finally, we propose a novel view on hyperparameter transfer as pairs of systems with small and large $P$, respectively, that reside on the same RG trajectory and thus have statistically comparable properties.

\paragraph*{Related works. }

Several recent lines of work have developed formal analogies between
exact RG (ERG) and learning treated as Bayesian Inference \citep{Erbin_2022,erbin2023functional,Cotler_2023,Berman_2023}, formulated notions of coarse-graining on untrained neural networks,
or augmented or interpreted neural network training with regards to
performing RG on data \citep{ mehta2014exact,Koch_Janusz_2018,Beny:2013pmv}.
Only the first line of work considers RG on trained neural networks. The suggested coarse-graining procedures are, however, quite different and do not include a rescaling step.
On the technical level, our choice of marginalization or integration of modes
is that of Ref. \citep{howard2024}, which reduces the resolution
of the DNN by marginalizing the low-lying principal components (PC) of
network outputs and pre-activation over the data \citep{Bradde_2017}.
This approach is also closely connected to compression techniques \citep{paccolat2020geometric, Jaderberg14}.
However, unlike these earlier works, we develop a notion re-scaling
allowing us to define relevant and irrelevant perturbations as well
as proper RG flows.

Another branch of related works studies random feature models with
an underlying power-law data distribution in terms of regression \citep{bahri2021explaining,maloney2022solvablemodelneuralscaling,defilippis2024dimensionfreedeterministicequivalentsscaling}
where the analog of interactions in these works are finite sample
effect and RG techniques, central to this work, are not being used
(see however the work \citep{howard2024}, which, as a aforementioned,
avoids the re-scaling step). Other works focus on dynamical effects in
such linear networks \citep{bordelon2024dynamicalmodelneuralscaling,lin2025scalinglawslinearregression,paquette202543phasescomputeoptimalneural}.
A recent work along these lines \citep{2025JSMTE2025h4002B} 
focuses on two-layer linear networks in the rich learning regime. The authors find
that for fat-tailed distributions of the training target  ($\beta<0$ in our terminology below),
the rich regime alters the time exponent of the learning curve. While we mainly study the power-law exponent of the learning curve as a function of the number of training samples here, it is interesting to note that our UV fixed-point
becomes unstable for $\beta<0$, consistently with their findings. In other settings, some authors find that feature learning effects are decoupled from generalization  \cite{goering2025featurelearningdecoupledgeneralization}. 

\section{Field Theory of Weakly non-Gaussian DNNs}

Consider, for concreteness \footnote{similar derivation applies for CNNs and transformers by trading width
with suitable redundancy hyperparameters}, a deep non-linear fully-connected network with width $N$ at each
layer with a scalar output. For each choice of network parameters
$\theta$ such DNN generates a function on the input space. Thus,
stochasticity in the weights, as induced by training the network using
Langevin dynamics \citep{Welling2011} induces a distribution on functions.
Using properly scaled mean-squared error (MSE) loss $\cL(f,y)=1/2\,\sum_{\alpha=1}^{P}(f_{\alpha}-y_{\alpha})^{2}$
on the $P$ training points and a weight decay term (quadratic potential
$\propto\|\theta\|^{2}$) as the potential for the Langevin training
dynamics \citep{Seung1992}
\begin{align*}
\partial_{t}\theta & =-\nabla_{\theta}\,\Big[\cL+\kappa\,\frac{\|\theta\|^{2}}{2g}\Big]+\xi(t)\\
\langle\xi(t)\xi(s)\rangle & =2\kappa\,\delta(t-s)\,,
\end{align*}
the stationary distribution of network parameters is of Boltzmann
form $\theta\sim\exp\big(-\kappa^{-1}\cL-\|\theta\|^{2}/2g\big)$
(see \prettyref{sec:Langevin-training} for the example of a linear
network). Furthermore taking $N\rightarrow\infty$, the distribution
of $f$ becomes Gaussian at any point during the Langevin dynamics.
This is a considerable simplification, dubbed lazy learning \citep{lee2017deep,Jacot2018,chizat2018lazy},
which quantitatively describes learning of networks at large width.
This description here serves us the non-interacting theory around
which we will perform the renormalization group (RG) treatment.

More specifically, our starting point for analysis is such a field
theory description of wide neural networks governing the equilibrium
distribution of outputs of the trained network ($f(x)$) which is
furthermore averaged over all choices of datasets of size $P$ that
are drawn from the dame data distribution $p_{\text{data}}(x)$. For
large ridge ($\kappa$, corresponding to Langevin noise temperature)
leads to \citep{cohen2021learning,naveh2021predicting} \footnote{Decimation only RG capturing finite sample effects was carried in
Ref. \citep{howard2024}} 
\begin{align}
Z[j] & =\int df_{1}\ldots df_{\Lambda}\;e^{-\frac{1}{2}\int d\mu_{x}d\mu_{y}f(x)K^{-1}(x,y)f(y)-\frac{P}{2\kappa}\int d\mu_{x}(f(x)-y(x))^{2}+\int d\mu_{x}j(x)f(x)}\label{Eq:Introducing_Z_EK}
\end{align}
where $d\mu_{x}=p_{\text{data}}(x)\,d^{d_{\text{in}}}x$ is the integration
measure and $K(x,y)$ is the neural network Gaussian process (NNGP)
kernel \citep{neal1996priors,lee2017deep} which acts on a function
$g$ as $\int d\mu_{y}K(x,y)g(y)$ (its inverse above is defined w.r.t.
to this action). We may expand $f(x)=\sum_{k=1}^{\Lambda}\phi_{k}(x)f_{k}$,
into $\phi_{k}(x)$, the eigenfunctions of the kernel with respect
to the data measure $p_{\mathrm{data}}(x)$, where $f_{k}=\int d\mu_{x}\phi_{k}(x)f(x)$,
and the cutoff $\Lambda$ regulates the rank of the kernel operator
and will be taken to infinity at the end of the computation. Quantities
of interest, such as the  DNN predictor
averaged over said ensemble of datasets, are obtained using functional
derivatives ($\delta_{j(x)}j(y)=\delta(x-y)/p_{\text{data}}(x)$)
via $\bar{f}(x_{*})=\delta_{j(x_{*})}\log(Z[j])|_{j=0}$. Here we
also note that constant multiplicative factors in front of the partition
function ($Z$) are inconsequential and hence, from now on, when comparing
partition functions we ignore such factors.

Next, it is computationally convenient to separate the modes of the
field $f(x)$ by rewriting $Z[j]$ in the spectral representation
(analogous to $k-$space in standard RG) 
\begin{align}
Z[j] & =\int df_{1}\ldots df_{\Lambda}\;e^{S(f)+\sum_{k}j_{k}f_{k}}\,,\\
\cS(f) & :=-\frac{1}{2}\sum_{k=1}^{\Lambda}\frac{f_{k}^{2}}{\lambda_{k}}-\frac{P}{2\kappa}\sum_{k=1}^{\Lambda}(f_{k}-y_{k})^{2}\,\label{eq:action_main}
\end{align}
where we defined the action $S$ which is, conveniently, diagonal
in the eigenmodes numbered by the $k$-index. Following this one can
derive the equivalent kernel (EK) estimator \citep{Silverman1984}
(cf. \prettyref{sec:Equivalent-kernel}) and for the dataset averaged
GPR predictor for an input $x_{*}$ given by 
\begin{align}
\bar{f}(x_{*}) & =\delta_{j(x_{*})}\log(Z[j])|_{j=0}=\sum_{k=1}^{\Lambda}\phi_{k}(x)\partial_{j_{k}}\log(Z[j])|_{j=0}\label{Eq:EKPred}\\
 & =\sum_{k=1}^{\Lambda}\frac{\lambda_{k}}{\lambda_{k}+\frac{\kappa}{P}}y_{k}\phi_{k}(x)\,,\nonumber 
\end{align}
which predicts that target modes with $\lambda_{k}P\ll\kappa$ are
copied from the target to the predictor, whereas modes with $\lambda_{k}P\gg\kappa$
are effectively projected out.

\paragraph*{Power-law spectrum.}

Next, we note that for many real-world datasets
and kernels the eigenmode spectrum follows a power law \citep{bahri2021explaining}
\begin{align}
\lambda_{k}\propto & k^{-1-\alpha}\,.\label{eq:power_law_lambda_main}
\end{align}
We show an example of such a power law in \prettyref{fig:Power-law-falloff-CIFAR}
in the appendix \prettyref{sec:Overlaps}. Furthermore, it is common
to assume a power law for the target function as well $y_{k}^{2}\propto k^{-1-\beta}$ \cite{bahri2021explaining,paquette202543phasescomputeoptimalneural}.
Using the above formula for $\lambda_{k}$ in our action \prettyref{eq:action_main},
the first term appears as $\sum_{k=1}^{\Lambda}k^{1+\alpha}f_{k}^{2}$,
which for $\alpha=1$ and, pretending $k$ is momentum, appears as
a standard kinetic term. Notwithstanding, the discrete summation over
$k$ which is related to the effectively finite support of $p_{\text{data}}(x)$,
both suggest an analogy with finite-size systems\footnote{In periodic or finite systems, the allowed $k$ values are discrete}.

Within this analogy, the second term appears as a mass term $P/\kappa$,
which, in a machine learning context (see also Eq. \prettyref{Eq:EKPred}),
sets the scale under which modes are learnable. The learnable modes
are massive while unlearnable modes with $\lambda_{k}^{-1}\gg P/\kappa$
are governed by the first term and are hence termed massless or critical.
Our RG will focus on integrating out those latter critical modes until
the mass scale $k_{P}$ set by $\lambda_{k_{P}}^{-1}=P/\kappa$ or
equivalently 
\begin{align}
k_{P} & =\left(P/\kappa\right)^{\frac{1}{1+\alpha}}\label{eq:k_P}
\end{align}
Notably, $k_{P}$ is set both by $P$ and the type of data and network
architecture, which determine $\alpha$ as well as $\kappa$.

\paragraph*{Adding interactions.}

Various real-world effects induce non-Gaussian terms in this action,
which correspond to interactions in field-theory language. These include
finite $N$ corrections \citep{yaida2020non,naveh2021predicting},
as demonstrated in \prettyref{sec:Quartic-theory-from-finite-size},
changes to the loss (e.g. cross entropy loss instead of MSE loss),
 different scaling of weight decay terms at infinite $N$ \citep{yang2019scaling},
and finite learning rates. Taking a physics-like RG perspective, it
is plausible and in fact quite common that very different microscopics
underlying these corrections all lead to the same leading relevant
term in the action, only with different coefficients. In physics,
relevancy is easily determined using the scaling dimensions of the
fields. One aim of the current work is to establish an analogous tool at hand for neuronal networks that would
allow us to predict relevance or irrelevance of specific terms of
the neuronal network action.

We here proceed concretely by adding a $\int d\mu_{x}(f(x)-y(x))^{4}$
interaction, which from a microscopic perspective can either be considered
as a finite-width correction (cf. \prettyref{sec:Quartic-theory-from-finite-size})
or a quartic addition to the MSE loss function. As shown in \prettyref{sec:Additional-quartic-loss}
under reasonable assumptions on the eigenmodes $\phi_{k}(x)$ (which
we empirically verify in \prettyref{sec:Overlaps}) one has 
\begin{align}
\cS_{\mathrm{int}} & =\frac{P\,U}{3}\,\int d\mu_{x}(f(x)-y(x))^{4}=\frac{P\,U}{3}\,\sum_{k,k',q,q'=1}^{\Lambda}(f_{k}-y_{k})(f_{k'}-y_{k'})(f_{q}-y_{q})(f_{q'}-y_{q'})\int d\mu_{x}\phi_{k}(x)\phi_{k'}(x)\phi_{q}(x)\phi_{q'}(x)\nonumber \\
 & \simeq P\,U\,\left[\sum_{k=1}^{\Lambda}(f_{k}-y_{k})^{2}\right]^{2}=P\,U\,\left[\int d\mu_{x}(f(x)-y(x))^{2}\right]^{2}\,.\label{eq:quartic_interaction_main}
\end{align}
The form of this interaction terms reveals a subtlety in neural network
field theories, which is the lack of locality and translation invariance.
The former is manifested by the last term on the r.h.s. and the latter
by the lack of $k$ conservation in the interaction. These two properties
will turn out to be crucial in shaping the analysis of relevancy of
terms in the action.

\section{The continuum theory and tree-level RG}

Next, we set up the basic form of the renormalization group that at
first neglects all fluctuation effects. We follow the common nomenclature
and call it ``tree-level RG'', referring to the fact that only tree-like
diagrams with no loops are taken into account. We here pay specific
attention to the two peculiarities of the interaction term, the non-locality
and the lack of translation invariance. Our first modest goal is to
integrate out a thin shell $k\in[\Lambda/\ell,\Lambda]$ ($\ell\apprge1$)
of modes $f_{k}$ in the free ($U=0$) theory, then rescale $k$ such
that the cutoff goes from its new $\Lambda'$ value back to $\Lambda$.
This rescaling prescription is important in standard RG as it determines
the scaling dimension of perturbations to the model, which implies
their relevancy.

To address the discreteness of $k$ we take cues from physics, where
this discreteness can be understood by working in a finite-size system.
As long as all the terms in the action are smooth on the discretization
scale of $k$ (i.e. $1$), physical locality means that this discreteness
can be traded from a continuum, with suitable changes reminiscent
of the transition from Fourier series to Fourier transforms. We follow
this route here by defining the following auxiliary (or infinite)
theory with continuous $k$. The process of construction is described
in detail in \prettyref{sec:Continuum-limit}. It proceeds by first
defining a family of systems which, between any pair of discrete modes,
possesses $n$ additional discrete modes. Second, we define all quantities
such that they become intensive in $n$, so that in the third step
the limit $n\to\infty$ can be taken which by construction then approximates
the original, discrete system. The result of this procedure yields
an intuitive form for the partition function and action 
\begin{align}
\tilde{Z} & =\int Df\,e^{\tilde{S}[f]}\\
\tilde{S} & =-\frac{1}{2}\int_{1}^{\Lambda}dkf(k)\left[\lambda_{k}^{-1}+\frac{P}{\kappa}\right]f(k)+\frac{P}{2\kappa}\left[y(k)^{2}-2f(k)y(k)\right]\,,\label{eq:cont_action}
\end{align}
which leads to the following expectation values, which we refer to
as the two-point and one-point correlation functions, 
\begin{align}
\langle f(k)f(l)\rangle & =\delta(k-l)\,[\lambda_{k}^{-1}+P/\kappa]^{-1}+\langle f(k)\rangle\langle f(l)\rangle\label{Eq:FreeTheoryContDef}\\
\langle f(k)\rangle & =\frac{\lambda_{k}}{\lambda_{k}+\kappa/P}\;y(k)\nonumber 
\end{align}
where these correlations, via Wick's theorem, determine all higher-order
correlations.

\paragraph*{Correspondence between the original and continuum observables.}

As shown in \prettyref{sec:Continuum-limit}, the above continuum
theory is set up such that computing an observable in the discrete theory
and taking a continuum approximation for the resulting summations
over $k$'s ($\sum_{k}\rightarrow\int dk$) would be the same as first
replacing the observable by a continuum observable (defined below),
then taking its expectation value under the continuum theory.

The simplest observables to take a continuum version of are ones where
each eigenvalue index (``momentum'') is accompanied by a distinct
summation, e.g. $\sum_{k_{1},\ldots,k_{4}}V(k_{1},\ldots,k_{4})f_{k_{1}}f_{k_{2}}f_{k_{3}}f_{k_{4}}$
and where $V(k_{1},\ldots,k_{4})$ is a smooth function. Here, the
continuum limit amounts to replacing summations by integrals $\int dk_{1}\ldots dk_{4}\;V(k_{1},\ldots,k_{4})\;f(k_{1})f(k_{2})f(k_{3})f(k_{4})$
as one would expect also naively.

Observables containing more $f$'s than summations, and hence products
of two or more $f(k)$ with the same argument, have a more subtle
continuum limit. Indeed taking a naive continuum limit as done above,
averaging $f(k)^{2}$ would result in a term $\delta(k-k)$, which
is of course infinite. This is the same situation in physics where
these terms are removed by normal-ordering; in the current setting,
however, normal ordering becomes cumbersome due to the lack of locality
and translation invariance. Instead, we show in \prettyref{sec:continuum-limit-Non-Gaussian-terms}
(i.p. in \prettyref{eq:Kronecker_replacement}) that such partly diagonal
terms need to be treated in a point-splitting-like procedure to obtain
an intensive limit $n\to\infty$. Here a quantity such as $\sum_{k}f(k)^{2}$
is taken, in the continuum limit to 
\begin{align}
\sum_{k}f(k)^{2}\rightarrow & \int dk\,dq\,f(k)f(q)\tilde{\delta}(k-q)\,,\label{eq:point_splitting_delta_diag}
\end{align}
where $\tilde{\delta}(k-q)$ obeys \textbf{(i)} $\int dk\,\tilde{\delta}(k-q)\,f(k)=f(q)+O(f'(q))$.
\textbf{(ii)} $\int dq\,\tilde{\delta}(k-q)\,\tilde{\delta}(q-k')=\tilde{\delta}(k-k')$
and \textbf{(iii)} $\tilde{\delta}(0)=1$. In the appendix we show
that the here appearing modified $\tilde{\delta}$-functions may for
example be implemented by smearing out the unit mass on an interval
of unit interval width; a concrete way to interpret $\tilde{\delta}$
is to define the limit 
\begin{align}
\int_1^\Lambda \int_1^\Lambda\,dk\,dk^{\prime}\,\tilde{\delta}(k - k^{\prime}) \ldots & \stackrel{n\to\infty}{:=}\int_n^{n\, \Lambda} \int_n^{n\,\Lambda}dk\,dk^{\prime}\,\delta_{\lfloor k/n\rfloor\lfloor k^{\prime}/n\rfloor}\,,\label{eq:point_splitting_main}
\end{align}
where $\delta_{ij}$ is the usual Kronecker $\delta$ and $\lfloor x\rfloor$
denotes the next lower integer of $x$. In the limit $n\to\infty$ the
dependence on $k/n$ and $k'/n$ of the right hand side effectively behaves as a dependence
on $k-k'$. As advertized in the beginning
of this subsection, following this procedure taking the continuum
version of an observable and averaging it under the continuum theory,
reproduces the continuum limit of the original discrete observable
under the original/discrete theory.

\paragraph*{Tree level RG in the continuum theory.}

Next we define the RG flow of the action \prettyref{eq:cont_action}.
Being non-interacting and already diagonal in the $k$ index, the
first stage of RG, which is the integration over the $f_{k}$'s with
$k\in[\Lambda/\ell,\Lambda]$ ($\ell\apprge1$) in the partition function
--- is trivial here, and amounts to simply yields a constant. As
a result, the modes are removed from the action. The emphasis here
is thus on the second part of the RG procedure which is the rescaling
step. Here define the scaled momentum $k'=\ell k$ in terms of which
the cut-off is again at $\Lambda$ and similarly define 
\begin{align}
\ell^{\gamma_{f}}f'(k') & =f(k)
\end{align}
where $\gamma_{f}$ is the, soon to be determined, native (or mean-field)
scaling dimension of $f$.

We proceed by rewriting the action in terms $k'$ and $f'(k')$. Concretely,
we first rewrite the action in terms of $k'$, yielding (up to constants)
\begin{align}
\tilde{S} & =-\frac{\ell^{-1}}{2}\int_{\ell}^{\Lambda}dk'\,f(k'/\ell)\left[\lambda_{k'/\ell}^{-1}+\frac{P}{\kappa}\right]f(k'/\ell)-\frac{P}{\kappa}f(k'/\ell)y(k'/\ell)
\end{align}
next we use $\lambda(k'/\ell)=\lambda(k')\,\ell^{1+\alpha}$, $y(k'/\ell)=y(k')\ell^{[1+\beta]/2}$
following from their power-law dependencies \prettyref{eq:power_law_lambda_main}
and $f(k'/\ell)=f(k)=\ell^{\gamma_{f}}f(k')$ to obtain 
\begin{align}
\tilde{S} & =-\frac{\ell^{-1}}{2}\int_{\ell}^{\Lambda}dk'\,l^{2\gamma_{f}}\,f'(k')\,\left[\lambda_{k'}^{-1}\ell^{-1-\alpha}+\frac{P}{\kappa}\right]\,f'(k')-\frac{P}{\kappa}\ell^{-\gamma_{f}}f'(k')y(k')\ell^{[1+\beta]/2}\,.
\end{align}
Seeking to conserve the scaling of the first ``kinetic-like'' term,
we choose $-1+2\gamma_{f}-1-\alpha=0$, implying $\gamma_{f}=1+\alpha/2$.
We further define $r(\ell):=$$P/\kappa\,\ell^{1+\alpha}$, which
can be viewed an $\ell^{1+\alpha}$ rescaling of the $P/\kappa$ mass-like
term. Turning to the final term, rewriting it using $r(\ell)$, and
gathering all the power of $\ell$, one obtains $\ell^{-1-\gamma_{f}+[1+\beta]/2+1+\alpha}=\ell^{[\alpha+\beta]/2-1/2}$.
Note that for $\beta=\alpha+1$, it does not acquire any scaling beyond
that of $r(\ell)$. In the main text, we focus on this case for simplicity.
For general $\alpha,\beta$, one requires introducing more ``dimension-full''
quantities such as a temperature scale, which we show in the appendix
\prettyref{sec:Rescaling-maintaining-target}. Interestingly, the
condition $\beta=\alpha+1$ is also the threshold value for $\beta$
separating the in-RKHS case ($\sum_{k=1}^{\Lambda}\lambda_{k}^{-1}y_{k}^{2}<\infty$
as $\Lambda\rightarrow\infty$), where the target function $y$ may
be expanded in terms of eigenmodes of the kernel, and the out-of-RKHS
scenario ($\sum_{k=1}^{\Lambda}\lambda_{k}^{-1}y_{k}^{2}\rightarrow\infty$),
where this is not possible.

Under the above choices of $\gamma_{f}$ and $\beta$, the tree-level
RG amounts to rescaling $\kappa$ by $\ell^{-\gamma_{M}},\gamma_{M}=1+\alpha$.
This has the effect of raising the $k$ at which modes are half-learnable
($k_{P}$), which is a direct consequence of our rescaling of $k$.

\paragraph*{Rescaling of the interaction $U$.}

Similarly, rescaling can now be applied to any perturbation to the
action, in particular the quartic perturbation \prettyref{eq:quartic_interaction_main}.
Specifically, adding a power of $\gamma_{f}$ for each $f(k)$ and
$y(k)$ and a power of $-1$ for each $k$ integral we obtain 
\begin{align*}
P\,U\left[\int dk\,dq\,\tilde{\delta}(k-q)\,\big[f(k)-y(k)\big]\,\big[f(q)-y(q)\big]\right]^{2}\\
\rightarrow P\,U\ell^{-4+4\gamma_{f}}\left[\int dk'\,dq'\,\tilde{\delta}(k'/\ell-q'/\ell)\,\big[f'(k')-y'(k')\big]\,\big[f'(q')-y'(q')\big]\right]^{2}\,.
\end{align*}
Had we worked with regular Dirac delta distributions, $\delta(k'/\ell-q'/\ell)=\ell\,\delta(k'-q')$
and we would have obtained a $\ell^{-2+4\gamma_{f}}=\ell^{2(1+\alpha)}=\ell^{2\beta}$
scaling factor
\begin{align}
\gamma_{U} & =2\beta\,.\label{eq:native_scaling_U}
\end{align}
The exponent being positive indicates that the interaction term grows
under the RG flow. We refer to this scaling dimension as the native
scaling dimension. As we next show, this replacement is correct under
certain momentum integrals but wrong for others, which contain a free
momentum loop. Using the concrete implementation of $\tilde{\delta}$
given by \prettyref{eq:point_splitting_main}, the replacement $\delta(k'/\ell-q'/\ell)=\ell\,\delta(k'-q')$
corresponds to trading a change of the integration boundaries into
a multiplicative factor $\ell$
\begin{align}
\int_{\ell\lfloor k^{\prime}/\ell\rfloor}^{\ell\lfloor k^{\prime}/\ell\rfloor+\ell}\ldots\to & \ell\,\int_{\lfloor k^{\prime}\rfloor}^{\lfloor k^{\prime}\rfloor+1}\,\ldots\,,\label{eq:replacement_boundaries}
\end{align}
which is only approximately correct if the integral is smooth, in
particular, if the integrand does not contain any Dirac distribution
in the integration variable, for example from a connected propagator.
Discerning those cases and judging their effect on the actual scaling
of the perturbation is the topic of the next section.

\section{Relevance and irrelevance}

In standard RG, the scaling dimension is the strongest source of renormalization
whereas decimation effects lead to smaller $O(U)$ corrections to
the flow. Perturbations which grow under rescaling (have positive
dimension $\gamma$), such as the above $U$ or $r=P/\kappa$, are
thus deemed Infrared (IR) relevant, meaning that their effect appears
larger when focusing on low energy (high kernel) modes. Next, we establish
a similar notion for our neural network field theory. The main complication
here is that, due to the aforementioned lack of locality and momentum
conservation, which we solve by the point-splitting $\tilde{\delta}$-delta
functions appearing in the replacement of diagonal fields \prettyref{eq:point_splitting_delta_diag}---
The interaction vertex has a scale-full momentum dependence and does
not, strictly speaking, maintain its functional form after rescaling.
One potential route of treating this is to simply keep track of this
scale and work with explicit $\ell$-dependent $\tilde{\delta}$ functions.
However, we argue below that an alternative route is possible, which
helps us understand the relevancy of scale-full perturbations such
as $U$, via the notion of scaling intervals introduced below.

To simplify the discussion it is advantageous to express the action
instead of in the fields $f$ in terms of the discrepancy $\Delta=f-y$,
so that the Gaussian part of the action takes the form \prettyref{eq:s_0_as_Delta}
and the interaction term is given by \prettyref{eq:interaction_for_perturb}

\begin{align}
S(\Delta,y)= & -\frac{s(\ell)}{2}\int_{1}^{\Lambda}\,\Big\{ r(\ell)\,\Delta(k)^{2}+\frac{(\Delta(k)+y(k))^{2}}{\lambda(k)}\Big\}\,dk+S_{\mathrm{int}}(\Delta)\,,\label{eq:RG_action_main_Delta}\\
S_{\mathrm{int}}(\Delta):= & -U(\ell)\,P\int_{1}^{\Lambda}dk\,\int_{1}^{\Lambda}dk^{\prime}\,\int_{1}^{\Lambda}\,dq\,\int_{1}^{\Lambda}dq^{\prime}\,\tilde{\delta}(k-k^{\prime})\,\tilde{\delta}(q-q^{\prime})\nonumber \\
 & \times\,\Delta(k)\,\Delta(k^{\prime})\,\Delta(q)\,\Delta(q^{\prime})\,,\nonumber 
\end{align}
where any terms that are independent of $\Delta$ have been dropped.
The parameter $s(\ell)=\ell^{\beta-\alpha-1}$ in front of the Gaussian
part captures the overall change of the amplitude of fluctuations
and is required to obtain a fixed point in the presence of the non-zero
target $y$ in the general case $\beta\neq\alpha+1$, as outlined
in \prettyref{sec:Rescaling-maintaining-target}. The ridge parameter
at the initial point of the RG flow $\ell=1$ is given by $r(1):=P/\kappa$
and the strength of the interaction is controlled by $U(\ell)$. The
connected correlators are correspondingly expressed as \prettyref{eq:mean_prop_rescaled}

\begin{align}
d(k;\ell) & := \frac{1}{r(\ell)\,\lambda(k)+1}\,y(k)\, =: \vcenter{\hbox{\includegraphics{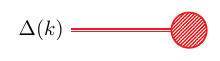}}} \\ \label{eq:mean_prop_Delta_main}
\delta(k-k^\prime)\,c(k;\ell) & := \delta(k-k^\prime)\,s(\ell)^{-1}\,\frac{\lambda(k)}{r(\ell)\,\lambda(k)+1} =: \vcenter{\hbox{\includegraphics{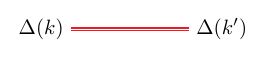}}} \\ \nonumber
\end{align}

To illustrate how the interaction term may contribute term that have
different scaling, we consider two examples that illustrate the two
different scenarios in which the point-splitted $\tilde{\delta}$
may appear when computing observables or perturbative corrections.
Subsequently we will generalize these examples to a generic rule.

As a first example, we show how the interaction vertex may contribute
to the decimation step of the RG flow a term that follows the native
scaling dimension \prettyref{eq:native_scaling_U}. To this end, consider
the diagrammatic contribution from integrating out the modes at the
cutoff $|k|=\Lambda$ to the self-energy, the quadratic part of the
action. Here the two legs $\Delta(q)$ and $\Delta(q^{\prime})$ of
the four-point interaction vertex $\cS_{\mathrm{int}}$ in \prettyref{eq:RG_action_main_Delta}\begin{align}
\cS_\mathrm{int}(\Delta) & =: \vcenter{\hbox{\includegraphics{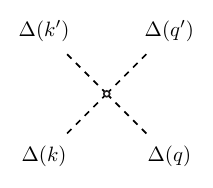}}} \\ \nonumber
\end{align} are contracted, which amounts to computing the second moment $\langle\Delta(q)\,\Delta(q^{\prime})\rangle$.
First consider the contribution due to the unconnected part of this
correlation $\langle\Delta(q)\rangle\,\langle\Delta(q^{\prime})\rangle=d(q,\ell)\,d(q^{\prime},\ell)$,
which is a smooth function in both momenta $q$ and $q^{\prime}$
and which yields\begin{align}
 4\,\,U \int_{|q'|=\Lambda} \int_{|q|=\Lambda} d(q,\ell) d(q',\ell) \; dq \, dq' &= \vcenter{\hbox{\includegraphics{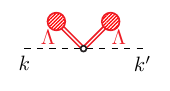}}}\,.
\end{align}

Here the $\tilde{\delta}$ of the interaction has eliminated the second
momentum integral over $l^{\prime}$.

Now consider the corresponding contribution, but with the connected
propagator $c(q,q';\ell)=\langle\Delta(q)\Delta(q')\rangle^{c}\propto\delta(q-q')$,
which hence contributes a diagram of the form

\begin{align}
 4\,U \int_{|q|=\Lambda} c(q,\ell)/\ell \;dq &= \vcenter{\hbox{\includegraphics{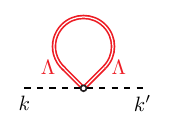}}}\,,
\end{align}where an additional factor $1/\ell$ appears, because this time the
Dirac $\delta$ of the propagator has eliminated one of the integrals;
in consequence the replacement \prettyref{eq:replacement_boundaries}
cannot be made, thus we lack one factor $\ell$ compared to the case
of a smooth integrand. Alternatively, we may regard this as a factor
$\tilde{\delta}(0)$ appearing.

To generalize these results, we first note that in an $n$-th order perturbative
correction, a maximum of $n$ such $\tilde{\delta}(0)$ factors may
appear in a connected diagram, because at each four-point vertex at most two fields may be contracted to one another, leaving the other two free to connect with other elements of the perturbative expansion. Thus, while in standard field theories the $n$-th order correction would scale as $\ell^{n \gamma_U}$, here its scaling would range between $\ell^{n \gamma_U}$, when no $\tilde{\delta}(0)$ factor appears, down to $\ell^{n \gamma_U-n}$, when the maximal number of such factors appear. 

The above implies that our interaction does not have a well-defined scaling dimension, which is no surprise given that it contains the scale-dependent $\tilde{\delta}$-function. Still, we may bound its scaling behavior, by attributing to the interaction a range of scaling dimensions $\gamma \in [\gamma_U,\gamma'_U]$, such that $\ell^{\gamma n}$ can produce all the scaling dimensions seen in $n$'th order perturbation theory. Specifically for our interaction we find $\gamma \in [\gamma_U-1,\gamma_U]$ which we refer to as its ``scaling interval". While short of being a proper RG scaling-dimension, it still allows us to treat $\gamma_U<0$ as a strictly irrelevant interaction (i.e. all its perturbative contributions decay under rescaling)  
and $\gamma_U>1$ as a strictly relevant interaction (i.e. all its perturbative contributions increase under rescaling).   

Generalizing this to other types of interactions, the scaling interval may grow
or shrink compared to $[\gamma_U,\gamma_U-1]$. For instance, interaction such as  $[\sum_{k}k^{a}f_{k}]^{M}$, which do not involve contributions of two or more fields
with identical momentum,  would follow the native scaling dimension; thus the scaling interval contains a single point.
On the other hand, higher order generalizations of the previous interaction such as $V[\sum_{k}k^{a}f_{k}^{2}]^{3}$, will allow a maximal number of two $\tilde{\delta}(0)$ factors per interaction term, by self-contracting four of its legs and leaving two for connectedness. This implies a scaling interval of $[\gamma_V-2,\gamma_V]$ where $\gamma_V$ is the native scaling dimension associated with that interaction.  

Finally, we note that any interaction without explicit negative powers of $k$ will have its lower end of the scaling interval larger
than zero. Hence any such interaction is IR-relevant. 
Indeed, consider an $n$-th order interaction with $m$ integrals which yields
a native scaling dimension $n\gamma_{f}-m$. Taking into account
the implied $n-m$ $\tilde{\delta}$ factors, coming from point-splitting
the required pairings\footnote{Note that triplet pairing such as $\sum_{k}f_{k}^{3}$ have a continuum
limit of $\int dkdqdlf_{k}f_{q}f_{l}\tilde{\delta}(k-q)\tilde{\delta}(q-l)$,
and so forth with higher order pairings}, the lowest contributions can scale as $n\gamma_{f}-m-(n-m-1)=n\gamma_{f}-n+1>1$,
and is hence relevant. Similar reasoning shows that since $\gamma_{f}>1$, each integral
contributes $-1$ to the exponent and, within a perturbation, there
cannot be more integrals than $f(k)$'s -- any perturbation which
does not contain explicit negative powers of $k$ is hence IR relevant.
Obtaining IR-irrelevant observables requires introducing inverse powers
of $\lambda_{k}$, and hence kernel inverses as in the kinetic term.

\paragraph*{UV Irrelevance and large $P$ universality.}

The mass term, $\frac{P}{\kappa}\int dk\,\Delta^{2}(k)$ in \prettyref{eq:RG_action_main_Delta}
acts as an IR regulator for the theory, so that all modes $k<k_{P}$
(cf. \prettyref{eq:k_P}) for which $\lambda^{-1}(k)<P/\kappa$ are
learnable and largely set by $y(k)$, as seen from \prettyref{Eq:FreeTheoryContDef}.
Well above this scale, modes fluctuate and interact
in a critical and therefore scale-free manner.
It is thus natural to apply RG
reasoning for those critical modes and proceed until an $\ell$ such
that the new cutoff $\Lambda/\ell$ agrees to the learnability threshold
$k_{P}$ \prettyref{eq:k_P}, which means that all critical modes
have been removed. This can also be thought of as setting $\ell=\ell_{P}$
given by 
\begin{align}
\ell_{P} & =\Lambda\left(\frac{P}{\kappa}\right)^{\frac{-1}{1+\alpha}}\,.\label{eq:ell_P}
\end{align}
Thus the more data we have (large $P$) the closer $\ell_{P}$ approaches
$1$, and relevant terms grow less until the point where RG stops.

We may also take an opposite perspective (common in high-energy physics)
on relevance and define the notion of UV irrelevance. To this end,
say we observe the model behavior at some value of $P$ at which $\ell_{P}\gg1$
and tune the model parameters to get some behavior which differs,
say by $10\%$ from that of a Gaussian theory by setting $U$ such
that $U\ell_{P}^{\gamma_{U}-1}=O(1/10)$ is small but non-negligible.
As $\ell_{P}$ decreases with $P$, this means that as we increase
$P$, the effect of $U$ measured at the initial point of the flow for that higher $P$ needs to be determined such
that its effect on the learnable modes stays the same; it thus needs to 
shrink, implying a GP-like behavior at large $P$. Thus, we may say
that an IR relevant perturbation is an UV irrelevant one. This can
also be understood as a statement about universality at large $P$
-- two models which differ at small $P$ due, say, to different values
of an IR relevant $U$ would become more similar and GP-like as $P$
grows.

\begin{figure}
\begin{centering}
\includegraphics[scale=0.8]{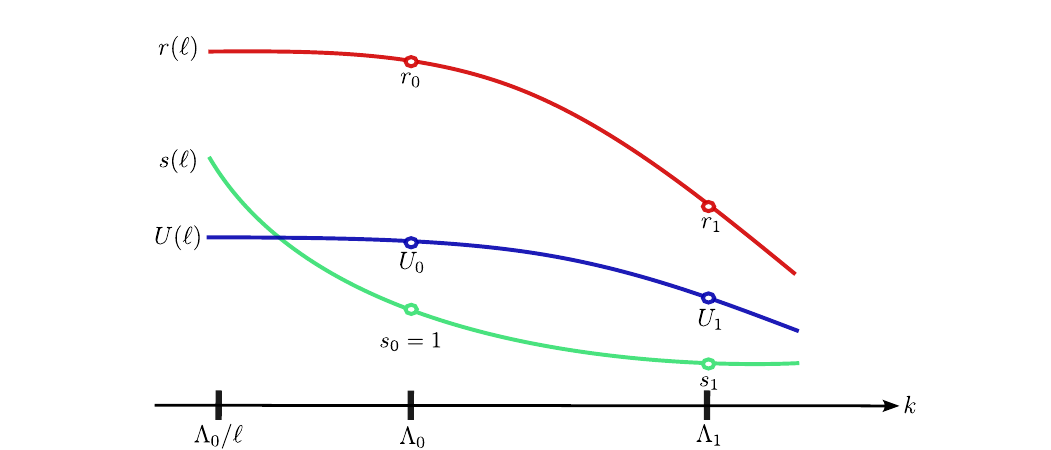}
\par\end{centering}
\caption{\textbf{Initial conditions for the flow equation and the meaning of
the cutoff.} One may choose different cutoffs, denoted as $\Lambda_{0}$
and $\Lambda_{1}$ with corresponding initial conditions $(s_{0},r_{0},U_{0})$
and $(s_{1},r_{1},U_{1})$, respectively. The two corresponding systems
behave the same in the low momentum range for any $k=\Lambda_{0}/\ell$,
if the two systems' parameters lie on the same RG trajectory. This
also allows the definition of the $\Lambda\to\infty$ limit.}\label{fig:Specification-of-initial-conditions}
\end{figure}

An alternative but equivalent view is illustrated in \prettyref{fig:Specification-of-initial-conditions}).
To determine the renormalization group flow, we need to specify the
initial conditions on the parameters $r(\ell)$, $U(\ell)$, and $s(\ell)$
that define the action \prettyref{eq:RG_action_main_Delta}. A special
property of this action is the appearance of the $\tilde{\delta}$-functions
in the interaction term, which possess an explicit scale dependence.
We will subsequently see that this also implies an explicit dependence
of the set of renormalization group equations on the scale $\ell$

The initial condition for $s(\ell=\Lambda_0)=1$ is given by construction,
as in the original action this factor is unity. Choosing the cutoff
$\Lambda_{0}$ we may choose the initial values $r_{0}=P/\kappa$
and $U_{0}$ at this scale, as illustrated in \prettyref{fig:Specification-of-initial-conditions}.
To obtain predictions for a mode $k$ below this cutoff, we
employ the renormalization group to integrate out the modes between
$k$ and $\Lambda_{0}$, so choosing $\ell=\Lambda_{0}/k$. The resulting
parameters $(s(\ell),r(\ell),U(\ell))$ together with the action \prettyref{eq:RG_action_main_Delta}
then describe the statistics of the mode $k$ while implicitly taking
into account the indirect effect of all modes beyond it.

An entirely equivalent choice of initial conditions is the cutoff
$\Lambda_{1}>\Lambda_{0}$ where, in order to observe the same behavior
of mode $k$ as before, we now need to specify the initial values
$(s_{1},r_{1},U_{1})$ such that integrating out the modes $k\in[\Lambda_{0},\Lambda_{1}]$
by choosing $\ell_{1}=\Lambda_{1}/\Lambda_{0}$ one obtains the renormalized
values $r(\ell_{1})=r_{0}$, $U(\ell_{1})=U_{0}$, and $s(\ell_{1})=1$
that agree to the first ones, as illustrated in \prettyref{fig:Specification-of-initial-conditions}.
In brief, we choose the initial conditions such that they lie on
the same RG trajectory. As a result, all modes below $\Lambda_{0}$
are again described by the same effective theory.

As a consequence, one may define the limit $\Lambda_{1}\to\infty$
by considering a family of models and each time setting the corresponding
initial values $(s_{1}, r_{1},U_{1})$ such that one stays on the same
RG trajectory. In high energy physics this is sometimes called the
freedom of choice of the renormalization point \citep{ZinnJustin96}.
This limit, by construction, produces for all modes $k<\Lambda_{0}$
the same predictions as for finite cutoff $\Lambda_{0}$, independent
of how large $\Lambda_{1}$ is chosen. As the set of RG equations
shows that both, $r(\ell)$ and $U(\ell)$, are infrared relevant,
taking the limit $\Lambda_{1}\to\infty$, conversely, both of their
initial values need to converge to zero, thus approaching a free critical
theory; a property called asymptotic freedom.

\section{Full RG treatment and the separatrix}

We now want to employ the renormalization group approach to quantitatively
predict the behavior of a non-Gaussian process. In particular, we
would like to predict the discrepancies between target and network
output and obtain the non-Gaussian corrections to the scaling of the
loss as a function of the number of training points, a characteristics
known as neural scaling law.

\subsection{Full set of renormalization equations}

To describe the renormalization group flow, we study the flow of the
three renormalized parameters $(s(\ell),r(\ell),U(\ell))$ that control
the action \eqref{eq:RG_action_main_Delta}, which we show in \prettyref{sec:Renormalization}
to obey the set of RG equations 

\begin{align}
\ell\,\frac{d\,\big[s(\ell)r(\ell)\big]}{d\ell} & =\beta\,s(\ell)r(\ell)+4\,\Lambda\,P\,U(\ell)\,\big[d(\Lambda,\ell)^{2}+c(\Lambda,\ell)/\ell\big]\,,\label{eq:RG_r_main}\\
\ell\frac{dU(\ell)}{d\ell} & =2\beta\,U(\ell)-2\,\Lambda\,P\,U(\ell)^{2}\,\big[c(\Lambda,\ell)^{2}/\ell+c(\Lambda,\ell)\,d(\Lambda,\ell)^{2}\big]\,,\label{eq:RG_U_main}\\
s(\ell) & =\ell^{\beta-\alpha-1},\label{eq:s_ell_main}
\end{align}
 The respective first, linear terms in the set of RG equations \prettyref{eq:RG_r_main}
and \prettyref{eq:RG_U_main} stem from the rescaling of the fields
which is here chosen such that the target $y$ and the network output
$f$ have identical scaling (described in detail in \prettyref{sec:Rescaling-maintaining-target}).
The decimation step that integrates out an infinitesimally thin shell
of modes just below the cutoff $\Lambda$ contributes to the flow
of $s(\ell)r(\ell)$ in \prettyref{eq:RG_r_main} the terms\begin{align}
 4\,\Lambda\,U(\ell) d(\Lambda,\ell)^{2} &= \vcenter{\hbox{\includegraphics{figures/self_energy_2_renormalized.pdf}}}\,,
\end{align}where a pair of fields $\Delta$ has been contracted and each yields
the mean of the field $d$. The factor $4$ here comes from the combinatorial
factor $2$ of choosing either pair of fields $\Delta(l),\Delta(l^{\prime})$
or $\Delta(k),\Delta(k^{\prime})$ of the interaction term \prettyref{eq:RG_action_main_Delta}
to be contracted and another factor of $2$, because $s(\ell)r(\ell)$
appears with a factor $1/2$ in the action \prettyref{eq:RG_action_main_Delta}.
Likewise, the connected propagator contributes

\begin{align}
 4\,\Lambda\,U(\ell) c(\Lambda,\ell)/\ell &= \vcenter{\hbox{\includegraphics{figures/self_energy_1_renormalized.pdf}}}\,,
\end{align}where a factor $1/\ell$ stems from the discreteness of the scale-dependent
interaction term, as explained above (as described in detail in the
context of \prettyref{eq:U_l}).

Analogously, the RG equation for the interaction $U(\ell)$ \prettyref{eq:RG_U_main}
is composed of a linear rescaling term with the native rescaling term
$2\beta\,U(\ell)$ which requires us to take into account the additional
factor $1/\ell$ in each contraction with a diagonal connected propagator
between two modes that are constrained by a $\tilde{\delta}$. The
decimation step involves two diagrams (as outlined in detail in \prettyref{sec:Decimation-interaction}),

\begin{align}
2U(\ell)^{2}P\,\Lambda\, c(\Lambda,\ell) \, d(\Lambda, \ell)^{2} &= \vcenter{\hbox{\includegraphics{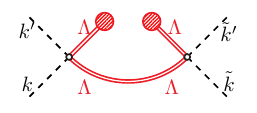}}}\,,
\end{align}and

\begin{align}
2U(\ell)^{2}P\,\Lambda\, c(\Lambda,\ell)^{2}/\ell &= \vcenter{\hbox{\includegraphics{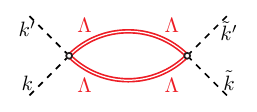}}}\,,
\end{align}where each diagram comes with a combinatorial factor $2\cdot2$, because
at each vertex we may choose either pair $(k,k^{\prime})$ or $(l,l^{\prime})$
to be contracted to the respective other vertex and another factor
$1/2!$, because there are two interaction vertices involved, which
would appear at second order in perturbation theory.

The RG flow also produces a six point vertex, which requires two interaction
vertices, to its magnitude is $\propto U^{2}$. Its flow is neglected
here.

\begin{figure}
\begin{centering}
\includegraphics{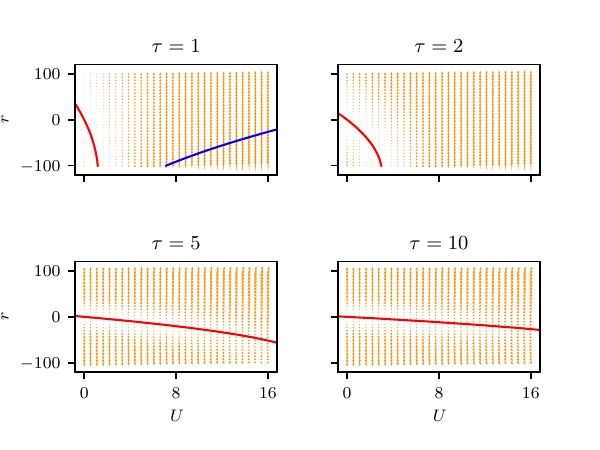}
\par\end{centering}
\caption{\textbf{Renormalization group flow field.} Renormalization group flow field.
given by \eqref{eq:RG_U_main} and \eqref{eq:RG_r_main} at different
decimation scales $\tau:=\ln\,\ell$. The plot represents the normalized
rate of change of the vector field $(U,r)\protect\mapsto(U+\mathsf{d}U,r+\mathsf{d}r)/N(U,r)$,
where $\mathsf{d}U$ and $\mathsf{d}r$ follow the flow equations
\eqref{eq:RG_U_main} and \eqref{eq:RG_r_main}, respectively, and
$N(U,r)$ denotes the vector norm at each point. Parameters are set
to $\alpha=0.2$ with the special choice $\beta=1+\alpha$ which ensures
that $s(\ell)\equiv1$.  The color scale indicates the logarithmic
magnitude of the flow vectors, $\ln\,N(U,r)$. Black dashed curves
show the $r$-nullcline given by \prettyref{eq:nullcline_limit_r}. 
}\label{fig:Phase-diagram-flow}
\end{figure}

In \prettyref{eq:RG_r_main} and \prettyref{eq:RG_U_main} we obtain
a system of inhomogeneous, nonlinear, coupled differential equations
that are reminiscent of the RG equations of a $\phi^{4}$-theory known
from statistical physics. There are, however, important differences.
First, the presence of a non-zero mean of the field shows up by the
flow equations depending on this mean (terms $\propto d$). Second,
the explicit appearance of the flow parameter $\ell$ on the right
hand side renders the flow field a function of the scale; in an ordinary
$\phi^{4}$-theory, the flow field may be studied as a function of
$r$ and $U$ alone, without explicit reference to the scale. As a
result, the flow equations do not exhibit a Wilson--Fisher fixed
point in the conventional sense. Nevertheless, we may identify scale-dependent
regions in which both $r$ as well as $U$ display qualitatively different
behaviors, separated by nontrivial nullclines for each parameter.
An example of such a flow field for the case $\beta=1+\alpha$ for
which $s\equiv1$ is constant is shown in \prettyref{fig:Phase-diagram-flow}.
The flow field can hence be expressed in terms of $r$ and $U$ alone
and depends on the value of $\ell=e^{\tau}$: At small scales ($\tau=1)$,
the decimation term in the flow of $r$ drives the system away from
criticality, towards larger values of $r$. This is because the positive
decimation contribution (second term in \prettyref{eq:RG_r_main})
dominates. For larger scales ($\tau\geq5)$, the contribution to the
decimation part from the connected correlator is reduced by $\ell^{-1}$,
so that for negative $r$ a nullcline emerges in the $r$-direction.
This line depends approximately linearly on $U$ with negative slope
due to the linear dependence of the decimation contribution and corresponds
to the set of points in parameter regime where the rescaling and decimation
contributions in \prettyref{eq:RG_r_main} cancel each other. As $\tau$
increases further, the nullcline keeps on bending up and in the limit
approaches the expression
\begin{align}
0 & \stackrel{!}{=}\beta\,r(\ell)+4\,P\,\Lambda\,U(\ell)\,d(\Lambda,\ell)^{2}\,,\label{eq:nullcline_limit_r}
\end{align}
which is shown in the figure. The points on the nullcline correspond
to the critical point in the usual $\phi^{4}$ theory, which here,
too, requires the fine-tuning of the relevant mass-like parameter
$r$; the parameter is relevant, as a slight detuning from the nullcline
leads to a departure from that line. 

In principle, the evolution equation for $U$ for large $\ell$ possesses
a nullcline as well when
\begin{align}
0 & \stackrel{!}{=}U(\ell)\,\big(2\beta-2U(\ell)\,P\,\Lambda\,c(\Lambda,\ell)\,d(\Lambda,\ell)^{2}\big)\,,\label{eq:nullcline_limit_u}
\end{align}
so either at the trivial choice $U=0$ or at $U(\ell)=2\beta/\big[\,P\,\Lambda\,c(\Lambda,\ell)\,d(\Lambda,\ell)^{2}\big]$.
The latter condition, however, typically leads to values that are
outside the validity of the approximation for small $U$ employed
here. This nullcline in principle limits the growth of $U$ along
the RG flow. The fact that it is lying far above the typical values
of $U$ explains why the RG flow for $U$ is dominated by the rescaling
term $2\beta\,U(\ell)$ alone. 

\subsection{Predictor in non-Gaussian process regression}\label{sec:Predictor-in-non-Gaussian}

We now use the renormalized action to obtain an improved estimate
for the mean predictor in form of the mean discrepancy $\langle\Delta\rangle$.
In the simplest case, this is achieved by computing the stationary
point $\delta S/\delta\Delta|_{\Delta=\Delta_{\ast}}=0$ of the action
to obtain an implicit equation for the mean as the stationary point
$\langle\Delta\rangle\simeq\Delta_{\ast}$. To compute this mean discrepancy
for a mode $k$, we first integrate out all modes above $k$, hence
setting $\ell=\Lambda/k$ and finally using the relation between the
mean discrepancy $D$ of the discrete system \eqref{eq:relation_D_Delta},
the EK's $\langle\Delta\rangle$ and the renormalized system's $\langle\Delta^{\prime}\rangle$
\eqref{eq:rescaling_wavefunction} given by

\begin{align}
\langle D(k)\rangle & =\sqrt{P}\,\langle\Delta(k)\rangle=\sqrt{P}\,z(\ell)\,\langle\Delta^{\prime}(\Lambda)\rangle\,\big|_{\,r(\ell),u(\ell),\ell=\frac{\Lambda}{k}}\simeq\sqrt{P}\,z(\ell)\,\Delta_{\ast}^{\prime}(\Lambda)\,\big|_{\,r(\ell),u(\ell),\ell=\frac{\Lambda}{k}}\label{eq:mean_disc_main}
\end{align}
The mean discrepancy is hence given by the highest mode $\Lambda$
in the renormalized system. We find in \prettyref{sec:Saddle-point-approximation-of-mean}
that the stationary point $\Delta_{\ast}^{\prime}(\Lambda)$ is mainly
determined by the Gaussian terms of the renormalized actions -- the
terms due to the interaction $S_{\mathrm{int}}$ only contribute little,
so that we here approximate $\Delta_{\ast}^{\prime}(\Lambda)\simeq d^{\prime}(\Lambda,\ell)=y(\Lambda)/\big(r(\ell)\,\lambda(\Lambda)+1\big)\,$
by the mean of the Gaussian part.

\begin{figure}
\begin{centering}
\includegraphics{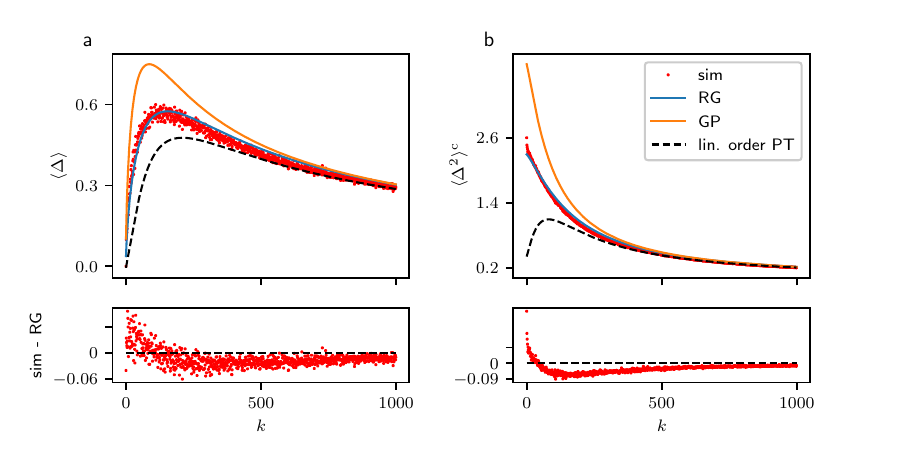}
\par\end{centering}

\caption{\textbf{Predictor in non-Gaussian regression.} \textbf{a} Upper panel:
Numerical result for the mean discrepancy $\langle\Delta\rangle$
(red dots) obtained by Langevin sampling until equilibrium. Gaussian
process prediction that neglects the non-Gaussian terms (orange).
Perturbative result to linear order in $U$ (black dashed). Prediction
of Gaussian discrepancy $\sqrt{P}\,z(\ell)\,d(\Lambda,\ell)|_{\ell=\Lambda/k}$
from the renormalized theory (blue). Lower panel: Difference between
numerical result and prediction from RG. \textbf{b} Corresponding
numerical and theoretical results for the variance $\langle\Delta^{2}\rangle^{c}$,
same color code as in panel a. Other parameters: $\alpha=0.2$, $\beta=0.3$,
$U=0.05$. Numerical results obtained by sampling the learning dynamics
for $T=50\cdot10^{6}$ steps with time resolution $\delta t=10^{-4}$,
measuring each $\Delta T=10$ steps and an initial equilibration time
of $T0=20\cdot10^{3}$ steps.}\label{fig:Mean-discrepancies}
\end{figure}

\prettyref{fig:Mean-discrepancies}\textbf{a} shows a this theoretical
prediction to agree well with the numerically measured discrepancy.
In particular, the prediction is more accurate than neglecting the
non-Gaussian terms alltogether (GP prediction) and is also improved
compared to a first order perturbative treatment of the interaction
term. In summary, knowing the renormalized ridge parameter $r(\ell)$
is sufficient to obtain an accurate estimate for the mean predictor.

Scrutinizing the difference between the numerical result and the RG
prediction in the lower panel of \prettyref{fig:Mean-discrepancies}\textbf{a}
shows that deviations are mainly observed in the intermediate momentum
range of $k$ -- this is to be expected, because for small $k$,
the effective mass term $r(\ell)$ is large, so that fluctuations
are small and hence the saddle point approximation becomes accurate.
Likewise for large $k$, because $U(\ell)$ is UV irrelevant, the
interaction becomes small and hence the mean-predictor is effectively
given by the perturbative result or, for larger $k$, even by the
Gaussian part of the theory. This also explains why all curves converge
in the large $k$ limit. Only in the intermediate momentum regime,
close to the transition between learnable ($r(\ell)>\lambda(\Lambda)$)
and non-learnable ($r(\ell)<\lambda(\Lambda)$) modes, fluctuations
remain and interaction terms are non-negligable, leading to an improvement
by RG and yet to small deviations due to taking a simple saddle-point
approximation that neglects the renormalized interaction term $\propto U(\ell)$.
A quantitative improvement in the intermediate regime could be obtained
by computing corrections (e.g., one-loop or perturbation theory) on
the renormalized theory.

Analogously, we use the renormalization group to obtain a theoretical
prediction for the covariance. We here again use that the discrepancies
in the discrete system $\langle D(k)^{2}\rangle^{c}=P\,\langle\Delta(k)^{2}\rangle^{c}$
and then express the variance of the original system in terms of the
variance in the renormalized system where all modes beyond the mode
$k$ of interest have been integrated out
\begin{align}
\langle D(k)^{2}\rangle^{c}=P\,\langle\Delta(k)^{2}\rangle^{2}= & P\,z^{2}(\ell)\,\langle\Delta^{\prime}(\Lambda)^{2}\rangle^{c}\,\big|_{\,r(\ell),u(\ell),\ell=\frac{\Lambda}{k}}\simeq P\,z^{2}(\ell)\,c^{\prime}(\Lambda,\ell)\,/\delta,\label{eq:var_disc_main}
\end{align}
where in the last step we again used the finite part of the variance
$c^{\prime}(\Lambda,\ell)/\delta=s(\ell)^{-1}\,\lambda(k^{\prime})/(r(\ell)\,\lambda(k^{\prime})+1)$
given by \prettyref{eq:mean_prop_rescaled} of the Gaussian process,
albeit with renormalized ridge parameter $r(\ell)$. The comparison
of this prediction to the numerical result is shown in \prettyref{fig:Mean-discrepancies}\textbf{b
}to agree well, considerably better than the bare Gaussian variance
and also compared to the first order perturbative result.

\subsection{Hyperparameter transfer}\label{sec:Hyperparameter-transfer}

Since the RG allows us to relate systems with different numbers of degrees of freedom, it is natural to ask whether the approach can be used to make predictions for hyperparameter transfer, namely, predicting how to rescale the optimal hyperparameters of one system at model scale $L$ (e.g. width or depth) to the optimal ones at model scale $L^{\prime}$. One strategy of achieving this is to formulate an effective theory at low $P$'s (e.g. using dynamic mean-field theory \cite{yang2022tensorprogramsvtuning,bordelon2023depthwise}), which depends on $L$ and on the other hyperparameters such as weight-decay, learning rate, and ridge-parameter. The central idea is then to scale those other hyperparameters such that the $L$-dependence drops out of this effective theory. Provided that what is optimal for low $P$'s is also optimal for all higher $P$'s, a plausible statement given neural scaling laws \footnote{More specifically, assuming that the power-law is independent/universal across the relevant hyperparameter range.}, such scaling with $L$ allows us to increase model capacity while staying on the optimal learning curve.

From a standard RG viewpoint, the ability to change parameters of the low-energy theory in a way that compensates a microscopic change to the model (e.g. the change of network width and all compensating parameters from $L$ to $L^{\prime}$), implies that increasing $L$ is an IR-irrelevant or marginal perturbation. This then suggests a more general formulation of optimal hyperparameter transfer, namely, to determine how $L$ affects the renormalization group flow of all compensating parameters and thus find their joint effect on the low energy theory. The flow equations may then be used to counter the change of $L$ by adapting the bare values of the compensating parameters to leave the effective low-energy theory invariant. 

We here want to illustrate the concept on a slightly simpler problem:
Suppose we have trained the system with a large number of data points
$P$ and a given optimal set of hyperparameters $(r_{0}=P/\kappa,U_{0})$.
How would one need to change the hyperparameters so that one obtains
a comparable accuracy using a smaller number of training samples $P^{\prime}<P$?
This can be regarded as a question about sample efficiency.

\begin{figure}
\begin{centering}
\includegraphics{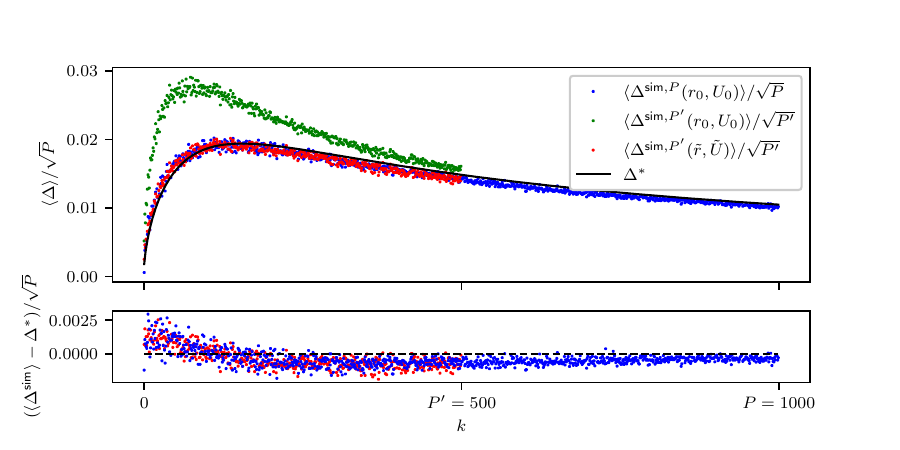}
\par\end{centering}
\caption{\textbf{Hyperparameter transfer.} Upper panel: Numerical result for
the mean discrepancy $\langle\Delta\rangle$ (dots) obtained by Langevin
sampling until equilibrium. Three systems are compared. System 1 was
trained with $P=1000$ samples and ridge parameter $r_{0}=P/\kappa$
(blue). System 2 was trained with $P^{\prime}=500$ samples and ridge
parameter $r_{0}^{\prime}=P^{\prime}/\kappa$ (green); system 1 and
2 both use the same $\kappa\simeq3.98$ and $U_{0}=0.05$. System
3 (red) was trained with $P^{\prime}=500$ samples but with parameters
$\tilde{r}$ and $\tilde{U}$ given by \eqref{eq:hyperparameter_transform}
so that it resides on the same RG trajectory as system 1. Prediction
of the discrepancy $\sqrt{P}\,z(\ell)\,d(\Lambda,\ell)|_{\ell=\Lambda/k}$
from the renormalized theory (black) for system 1. Lower panel: Difference
between numerical results (system 1 and system 3) and prediction from
RG. Other parameters: $\alpha=0.2$, $\beta=0.3$. Numerical results
obtained by sampling the learning dynamics for $T=50\cdot10^{6}$
($T=55\cdot10^{6}$ for the red dots to suppress noise) steps with
time resolution $\delta t=10^{-4}$, measuring each $\Delta T=10$
steps and an initial equilibration time of $T_{0}=20\cdot10^{3}$
steps.}\label{fig:Hyperparameter-transfer.}
\end{figure}

To achieve this goal, we start with system 1 which is trained on $P=1000$
training samples, which achieves a certain accuracy, shown in \prettyref{fig:Hyperparameter-transfer.}
in terms of the mean discrepancies $\langle\Delta^{P}\rangle$. Keeping
all hyperparameters, such as $\kappa$ and $U$ identical as before
but reducing the number of training samples to $P^{\prime}=500$,
one obtains larger discrepancies per square root of $P$ in system
2, as seen in \prettyref{fig:Hyperparameter-transfer.}. We now want
to use the RG to obtain a set of new parameters $(\tilde{\kappa},\tilde{U})$
that define system 3, which is trained on $P^{\prime}=500$ samples
so that its discrepancies agree to those in the large system 1 within
the overlapping range $k\in[1,\ldots,P^{\prime}]$. To this end, we
need to find the effective theory in the larger system which describes
the statistics of its lower $P^{\prime}$ degrees of freedom, hence
we set the RG time to $\ell=P/P^{\prime}$. This yields renormalized
parameters
\begin{align*}
r^{\prime} & =r(\ell)|_{\ell=\frac{P}{P^{\prime}}}\,,\\
U^{\prime} & =U(\ell)|_{\ell=\frac{P}{P^{\prime}}}\,,
\end{align*}
which we obtain by integrating the set of RG equations \prettyref{eq:RG_r_main}
and \prettyref{eq:RG_U_main} with initial condition $r(1)=r_{0}$
and $U(1)=U_{0}$.

The renormalization group transform, however, rescales the momentum
range such that before and after the RG transform the two ranges apparently
agree, $k\in[1,\Lambda]$. To make predictions for the smaller system,
which indeed only has $P^\prime  = \Lambda/\ell$ degrees of freefom, we thus need to undo this rescaling. As shown in detail in \prettyref{sec:RG_parameter_transfer}
the resulting action $\tilde{S}(\Delta;\,\tilde{r},\tilde{U},P^{\prime})$
for the smaller number of degrees of freedom then has the same form
as \prettyref{eq:action_main}, only with the parameters replaced
as
\begin{align}
\Lambda & \to P^{\prime}\,,\label{eq:hyperparameter_transform}\\
s(\ell) & \to1\,,\nonumber \\
r(\ell) & \to\tilde{r}=s(\ell)r(\ell)\,\ell^{-\beta}\,|_{\ell=\frac{P}{P^{\prime}}}=\ell^{-(1+\alpha)}\,r^{\prime}\,|_{\ell=\frac{P}{P^{\prime}}}\,,\nonumber \\
U(\ell) & \to\tilde{U}=\frac{P}{P^{\prime}}\,U\,(\ell)\ell^{-2\beta}\,|_{\ell=\frac{P}{P^{\prime}}}=\frac{P}{P^{\prime}}\ell^{-2\beta}\,U^{\prime}\,|_{\ell=\frac{P}{P^{\prime}}}\,.\nonumber 
\end{align}
These expressions show that the trivial change due to the rescaling
is undone: $\tilde{r}$, for example changes by a factor of $\ell^{-(1+\alpha)}$;
the factor $P/P^{\prime}$ in front of the interaction part, in addition,
arises because of the explicit appearance of $P$ in \prettyref{eq:action_main}.
The system 3 with $P^{\prime}$ degrees of freedom and parameters
$(\kappa^{\prime}=P^{\prime}/\tilde{r}$, $\tilde{U})$ then by construction
lies on the same RG trajectory as system 1: its $P^{\prime}$ modes
behave according to the renormalized theory $\tilde{S}$, which takes
into account the indirect effect of the presence of the $P-P^{\prime}$
upper modes in system 1. The discrepancies for comparable modes in
systems 1 and 3 are thus the same $\langle\Delta(k)\rangle_{S(\Delta;\,r_{0},U_{0},P)}=\langle\Delta(k)\rangle_{\tilde{S}(\Delta;\,\tilde{r},\tilde{U},P^{\prime})}$,
so
\begin{align*}
\langle D_{P}(k)\rangle/\sqrt{P} & =\langle\Delta(k)\rangle_{S(\Delta;\,r_{0},U_{0},P)}=\langle\Delta(k)\rangle_{\tilde{S}(\Delta;\,\tilde{r},\tilde{U},P^{\prime})}=\langle D_{P^{\prime}}(k)\rangle/\sqrt{P^{\prime}}\,.
\end{align*}
Since the contribution of the mean discrepancy to the loss per sample
is given by $1/(2P)\,\sum_{k}\langle D_{P}(k)\rangle^{2}$, this implies
an identical contribution to the loss in the two systems. The validation
of this prediction is shown in \prettyref{fig:Hyperparameter-transfer.}:
The mean discrepancies per square root of samples is preserved by
these rules of hyperparameter transfer \eqref{eq:hyperparameter_transform}.

\subsection{Neural scaling laws}

Next we aim to compute the neural scaling law for the non-Gaussian
process and compare it to the Gaussian process. We want to show that
in the limit of large data $P\to\infty$ the two converge to one another
due to the UV irrelevance of the interaction $U$. The observable
of interest is the mean loss per sample that can readily be expressed
as
\begin{align*}
\langle\cL\rangle/P= & \frac{1}{2P}\int_{1}^{P}\,\big\langle\big(f(k)-y(k)\big)^{2}\big\rangle\,dk\,.
\end{align*}
We decompose the expected loss into the loss due to the mean discrepancy
$\langle\Delta(k)\rangle=\langle f(k)\rangle-y(k)$ and due to the
variance $\langle\Delta^{2}(k)\rangle^{c}=\langle f^{2}(k)\rangle^{c}$
of the predictor which we determine in the same approximations as
before, integrating out all modes beyond the mode $k$ of interest,
so $\ell=\Lambda/k$ to get (cf. \prettyref{eq:L_bias_discrete} and
\prettyref{eq:L_var_discrete})

\begin{align}
\langle\cL\rangle/P\simeq & \frac{1}{2}\,\int_{1}^{P}\,\frac{k^{-(1+\beta)}}{[\tilde{r}(\ell)\,k^{-(1+\alpha)}+1]^{2}}\Big|_{\ell=\Lambda/k}\,dk\label{eq:loss_discrete}\\
+ & \frac{1}{2}\,\int_{1}^{P}\,\frac{k^{-(1+\alpha)}}{\tilde{r}(\ell)\,k^{-(1+\alpha)}+1}\Big|_{\ell=\Lambda/k}\,dk\,,\nonumber 
\end{align}
where the first line is the bias part and the second the variance
contribution and $\tilde{r}(\ell):=r(\ell)\,\ell{}^{-(1+\alpha)}$
is the renormalized ridge parameter, where $r(\ell)$ solves the RG
equation \prettyref{eq:RG_r_main} and the factor $\ell{}^{-(1+\alpha)}$
undoes the rescaling part, as in the case of hyperparmeter transfer
(cf. \prettyref{sec:RG_parameter_transfer}).

\begin{figure}
\centering{}\includegraphics{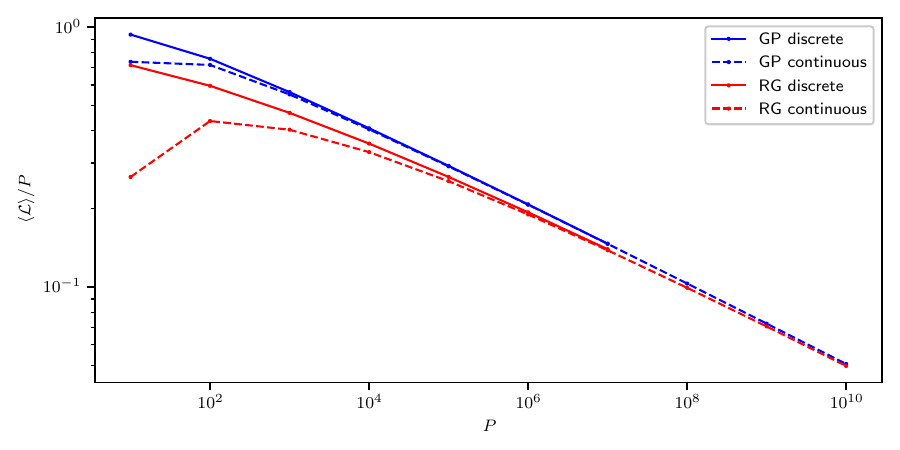}\caption{\textbf{Neural scaling law.} Expected training loss $\langle\protect\cL\rangle/P$
per data sample of a system trained with power-law exponents $\alpha=0.2$
and $\beta=0.3$. Full lines describe the prediction from the discrete
sum \prettyref{eq:loss_discrete} obtained by solving the full flow
equations \prettyref{eq:RG_r_main} and \prettyref{eq:RG_U_main}
numerically for the Gaussian process ($U=0$, full blue curve) and
the non-Gaussian process ($U=0.05$, full red curve). The dashed lines
describe the analytical continuous approximation \prettyref{eq:L_bias_GP_main_continuous}
and \prettyref{eq:L_var_GP_main_continuous} for the Gaussian process
(dashed blue curve) and the continuous approximation \prettyref{eq:L_cont_main_non_GP}
for the non-Gaussian process (dashed red curve). }\label{fig:Neural-scaling-law.}
\end{figure}

For the Gaussian case $U\equiv0$, we have the solution of the RG
equation \prettyref{eq:RG_r_main} $r_{\mathrm{GP}}(\ell)=P/\kappa\,\ell^{(1+\alpha)}$,
so that $\tilde{r}=P/\kappa$ is constant. The predictions of these
theories are shown in \prettyref{fig:Neural-scaling-law.} to produce
different power-laws for the Gaussian and for the non-Gaussian process.
For large $P$, however, the two curves approach the same exponent.

To quantitatively understand this convergence, we extract the scaling
of the mean loss \prettyref{eq:loss_discrete} with $P$ by first
treating the terms due to the finiteness of the discrete system, which
causes the finite summation boundaries, from the $P$-dependence inherent
in the parameter $\tilde{r}$. We exemplify this here for the bias
part; the variance part is treated analogously (see \prettyref{sec:Scaling-of-the-loss}
for details). These operations yield
\begin{align}
\cL_{\mathrm{bias}}/P & \simeq\frac{1}{2}\,\int_{0}^{\infty}\,\frac{k^{-(1+\beta)}}{[\tilde{r}(\ell)\,k^{-(1+\alpha)}+1]^{2}}\Big|_{\ell=\Lambda/k}\,dk-\frac{P^{-\beta}}{2\beta}\,,\label{eq:scaling_L_bias}
\end{align}
where the second term stems from the $-\int_{P}^{\infty}\ldots dk$
and using the fact that for large $k$, the integrand approaches $\to k^{-(1+\beta)}$
in either case. In the Gaussian case, the remaining $P$-dependence
can readily be extracted by substitution of the integration variable
$\tilde{r}_{\mathrm{GP}}\,k^{-(1+\alpha)}=P/\kappa\,k^{-(1+\alpha)}\to\tilde{k}^{-(1+\alpha)}$,
which yields a $P$-independent integral $I_{\mathrm{bias}}$ (defined
in \prettyref{eq:def_I_bias}) and a $P$-dependent power law

\begin{align}
\cL_{\text{bias,GP}}/P & \simeq I_{\mathrm{bias}}\cdot\,\big(\frac{P}{\kappa}\big)^{-\frac{\beta}{1+\alpha}}\,-\frac{P^{-\beta}}{2\beta}\,,\label{eq:L_bias_GP_main_continuous}
\end{align}
where we see that first term from the integral part dominates at large
$P$ due to its slower decay over the second, which stems from the
finiteness of the system. Analogous computations for the variance
contribution to the loss of the GP yield
\begin{align}
\cL_{\text{var,GP}}/P & \simeq I_{\text{var}}\cdot\,\big(\frac{P}{\kappa}\big)^{-\frac{\alpha}{1+\alpha}}-\frac{P^{-\alpha}}{2\alpha}-\frac{\kappa P^{-1}}{2}\,,\label{eq:L_var_GP_main_continuous}
\end{align}
where the last term comes from the replacement of the lower integration
bound $1\to0$. Again, the contribution due to the $P$-dependence
of the parameter $\tilde{r}$ in the integral yields the dominant
scale at large $P$. This analytical result for the Gaussian process
is shown to agree well to the discrete expression \prettyref{eq:loss_discrete}
for $P\apprge10^{3}$ in \prettyref{fig:Neural-scaling-law.}.

For the non-Gaussian process, the contributions due to the boundary
terms can be treated analogously (see \prettyref{sec:Extraction-of-the-P-dependence}
for details). We here only focus on the contribution due the the integral
part in \prettyref{eq:scaling_L_bias}. To extract the resulting corrections
to the scaling law, we make a couple of approximations. First, we
use the fact that the non-Gaussian corrections to the ridge parameter
can be treated as a small perturbation $r(\ell)=r_{\mathrm{GP}}(\ell)+\epsilon(\ell)$,
which allows us to expand the denominator in \prettyref{eq:scaling_L_bias}
into a geometric series, only keeping first order terms in the self-energy
correction $\epsilon(\ell)$. Second, we solve the set of RG equations
\prettyref{eq:RG_r_main} by neglecting the decimation contributions
for $U(\ell)$ and expand the flow equation for $r(\ell)$ up to linear
order in $\epsilon(\ell)$, which yields a closed form solution for
the resulting linear flow equation for $\epsilon(\ell)$ in terms
of an integral \prettyref{eq:non_gauss_r_flow_solution}. Performing
a corresponding substitution in the integral \prettyref{eq:scaling_L_bias}
as in the Gaussian case then allows us to extract the $P$-dependence
as a prefactor and we are left with multiple $P$-independent integrals,
which we denote as $I_{\ldots}$. We thus obtain correction terms
to the Gaussian scaling \prettyref{eq:L_bias_GP_main_continuous}
that are of linear order in the interaction $U$ \prettyref{eq:loss_bias_final}
given by
\begin{align}
\delta\langle\mathcal{L}_{\text{bias}}\rangle/P= & -4U\kappa\,I_{\text{bias}}^{\epsilon,I}\cdot\left(\frac{P}{\kappa}\right)^{-\frac{2\beta}{1+\alpha}}-4U\kappa\,I_{\text{bias}}^{\epsilon,II}\cdot\left(\frac{P}{\kappa}\right)^{-\frac{\beta+\alpha}{1+\alpha}}+F(\Lambda^{-1})\,,\label{eq:delta_L_bias}
\end{align}
where the term $F(\Lambda^{-1})$ that depends on the cutoff vanishes
in the limit $\Lambda\to\infty$ and in the finite system with $\Lambda=P$
can be absorbed as changed prefactors of the other two terms. The
analogous steps for the variance terms yield a correction to the Gaussian
scaling \prettyref{eq:L_var_GP_main_continuous} given by \prettyref{eq:loss_var_final}
\begin{align*}
\delta\langle\mathcal{L}_{\text{var}}\rangle/P & \simeq-4U\kappa\,I_{\text{var}}^{\epsilon,I}\cdot\left(\frac{P}{\kappa}\right)^{-\frac{\beta+\alpha}{1+\alpha}}-4U\kappa\,I_{\text{var}}^{\epsilon,II}\cdot\left(\frac{P}{\kappa}\right)^{-\frac{2\alpha}{1+\alpha}}+G(\Lambda^{-1})\,,
\end{align*}
where again the cutoff dependent term $G$ vanishes for $\Lambda\to\infty$
and can be absorbed in the the other two terms for the finite system
with $\Lambda=P$.

The resulting continuous prediction for the resulting mean loss
\begin{align}
\cL_{\text{non-GP}}^{\text{contiuous}}/P & =\cL_{\text{bias,GP}}/P+\cL_{\text{var,GP}}/P\label{eq:L_cont_main_non_GP}\\
 & +\delta\langle\mathcal{L}_{\text{bias}}\rangle/P+\delta\langle\mathcal{L}_{\text{var}}\rangle/P\,,\nonumber 
\end{align}
is shown compared to the discrete explicit expression \prettyref{eq:loss_discrete}
in \prettyref{fig:Neural-scaling-law.} to agree well for $P\apprge10^{6}$.
Due to the steeper slopes of the correction terms $\delta\langle\mathcal{L}_{\text{bias}}\rangle/P$
and $\delta\langle\mathcal{L}_{\text{var}}\rangle/P$ compared to
the Gaussian scaling of $\cL_{\text{bias,GP}}/P$ and $\cL_{\text{var,GP}}/P$,
the scaling law for the non-Gaussian process converges to the one
of the Gaussian process in the large-$P$-limit. This is what we expected
based on the UV irrelevance of the interaction term $U$, which makes
the non-Gaussian process approach the Gaussian process for high modes.

\section{Effect of perturbations on learning curves -- Heuristic derivation}

Next we provide a heuristic explanation for how the perturbation $U$
affects the learning curve, to demonstrate the power of the scaling
dimensions identified for the different quantities.

\begin{figure}
\begin{centering}
\includegraphics{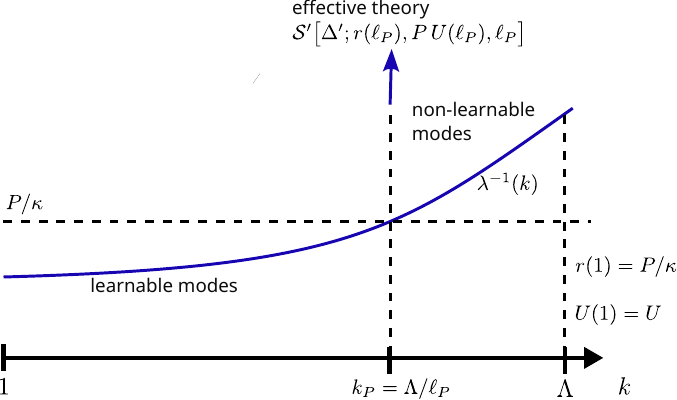}
\par\end{centering}
\caption{\textbf{Extraction of scaling laws from critical action and scaling
dimensions of operators.} The original theory is defined in terms
of its parameters $r(1)$ and $U(1)$ when all modes $k\in[0,\Lambda]$
are present. Modes with $\lambda^{-1}(k)>P/\kappa$ have a high mean
discrepancy $\langle\Delta(k)\rangle\simeq y(k)$ and are hence termed
\textquotedblleft non-learnable\textquotedblright . Modes with $\lambda^{-1}(k)<P/\kappa$
have a small discrepancy $\langle\Delta(k)\rangle\simeq0$ and are
hence called learnable. Since the overall magnitude of $y(k)$ declines,
the main contribution to the loss comes from the modes at the intersection
of these two regimes at $k\simeq k_{P}$. We hence use RG with $\ell_{P}=\Lambda/k_{P}$
to obtain an effective theory $\mathcal{S}^{\prime}\big[\Delta^{\prime};r(\ell_{P}),U(\ell_{P}),\ell_{P}\big]$
to predict the loss for modes as $k_{P}$, which in the renormalized
action appear to lie at the cutoff $k_{P}^{\prime}=\Lambda$, due
to the rescaling $\ell_{P}k_{P}^{\prime}=k_{P}$.}\label{fig:Extraction-of-scaling}
\end{figure}

As a preliminary computation, let us evaluate the learning curves
in the free theory using a scaling argument. To this end we perform
RG to integrate out all critical (non-learnable) modes $k\in[k_{P},\Lambda]$,
where $k_{P}$ is defined by \prettyref{eq:k_P} as the mode for which
the kinetic and the mass term agree, $r(1)=P/\kappa=\lambda^{-1}(k_{P})$,
as illustrated in \prettyref{fig:Extraction-of-scaling}. This implies
that we choose $\ell=\ell_{P}=\Lambda/k_{P}\stackrel{(\ref{eq:ell_P})}{=}\Lambda\left(\frac{P}{\kappa}\right)^{\frac{-1}{1+\alpha}}.$
In the renormalized theory, the mode at the cutoff $\Lambda$ is thus
massive, because by construction its mass agrees to the kinetic term,
$r(\ell_{P})=\lambda^{-1}(\Lambda)$; all non-learnable modes beyond
this scale have thus been integrated out.  Evidently, any explicit
$P$ dependence has been removed from the free part of the renormalized
action $S_{0}^{\prime}\big[\Delta^{\prime};r(\ell_{P}),\ell\big]$
(cf. first line of \prettyref{eq:RG_action_main_Delta}). So no observable
computed from $S_{0}^{\prime}$ contains any $P$ dependence. Also
the parameter $r(\ell_{P})=\lambda^{-1}(\Lambda)$ is independent
of $P$, as it is fixed by the cutoff alone. Obviously $P$ dependence
must enter and it does so by noting that the mean discrepancy of the
original system $\langle\Delta(k)\rangle$ and the renormalized one
$\langle\Delta^{\prime}(\Lambda)\rangle$ relate by \prettyref{eq:mean_disc_main}
$\langle\Delta(k_{P})\rangle=z(\ell_{P})\,\langle\Delta^{\prime}(\Lambda)\rangle\,\big|_{\,S^{\prime}(\ell=\ell_{P})}$,
where the wavefunction renormalization factor $z(\ell)$ appears.
We finally note that the scaling of the loss must be identical to
how the mode $k_{P}$ at the learnability threshold scales, since
by self-similarity in the critical regime, the entire critical region
must scale identically with $P$. We get one additional factor $\ell^{-1}$,
because the loss operator contains an integral $\int_{1}^{\Lambda}\ldots dk$
over all modes, which gets rescaled by the RG, $dk=dk^{\prime}/\ell$
(cf. \prettyref{eq:rescaling_momentum}). We thus obtain the scaling
of the bias part of the loss as
\begin{align}
\cL_{\mathrm{bias}}/P & \sim\langle\Delta(k_{P})\rangle^{2}\sim z(\ell_{P})^{2}\,\ell_{P}^{-1}\stackrel{(\prettyref{eq:z_rescale})}{\sim}P^{-\frac{\beta}{1+\alpha}}\,,\label{eq:L_bias_scaling_heuristic}
\end{align}
which agrees to the dominant term of \prettyref{eq:L_bias_GP_main_continuous}
(the sub-leading term there stems from considering a system with a
finite set of modes $\Lambda=P$).

On a more abstract level, we may summarize the argument very briefly:
The bare loss $1/2\,\sum_{k=1}^{\Lambda}\Delta_{k}^{2}$ in the continuum
limit becomes $1/2\,\int dk\,dq\,\tilde{\delta}(k-q)\,\Delta(k)\Delta(q)$,
which has a native scaling of $z(\ell)^{2}\,\ell^{-1}=\ell^{\beta}$,
so a scaling dimension of $\gamma_{\cL_{\mathrm{bias}}}=\beta$. This
implies that the bias part of the loss declines as $\ell_{P}^{\beta}\propto P^{-\beta/(1+\alpha)}$,
which agrees with the argument above \prettyref{eq:L_bias_scaling_heuristic}
and with the explicit calculation (cf. \prettyref{sec:Scaling-for-the-loss-GP}
i.p. \eqref{eq:def_I_bias} and with \prettyref{eq:L_bias_scaling_heuristic}).
This argument shows the power of the dimensional analysis: Determining
the dimension of the operator alone is sufficient to derive its $P$-dependence.

To extract the scaling of the variance in a similar manner, we need
to take into account that the renormalized variance $\langle\Delta^{2}\rangle^{c}$
receives another factor $\ell^{-\beta+\alpha+1}=s(\ell)^{-1}$ from
\prettyref{eq:s_ell_main}, so $\langle\Delta(k_{P})^{2}\rangle^{c}\sim z(\ell_{P})^{2}\,s(\ell)^{-1}$
and the scaling dimension of the variance part has an additional factor
$\ell^{-1}$ due to the $\tilde{\delta}(0)$ appearing in the definition
of the loss, which together yields the expected scaling with $P$
of the variance contribution to the loss as $\cL_{\mathrm{var}}/P\sim P^{-\alpha/(1+\alpha)}$
(in line with \prettyref{sec:Scaling-for-the-loss-GP} i.p. \prettyref{eq:loss_var_GP}).

Next we extend the argument to extract the leading order contribution
to the scaling law in the interacting theory. We here focus on the
special case $\beta=1+\alpha$ for which $s(\ell)\equiv1$ and for
which $P\,U(\ell_{P})=P\,U(1)\ell_{P}^{2\beta}\sim P^{-1}\searrow0$
for large $P$ to simplify the argument. Following again the RG flow
up to the learnability threshold, $\ell=\ell_{P}$, the renormalized
action $S^{\prime}\big[\Delta^{\prime};r(\ell_{P}),\,P\,U(\ell_{P}),\ell_{P}\big]$
contains the only $P$-dependence in the interaction $P\,U(\ell_{P})$.
There is no implicit $P$-dependence from $r(\ell_{P})=\lambda^{-1}(\Lambda)$,
which is fixed by the cutoff. For sufficiently large $P$ we will
ultimately reach a regime in which we may treat $P\,U(\ell_{P})\sim P^{-1}$
perturbatively; say to first order perturbation theory. 

The effect of the interaction term on the loss depends on the second
moment
\begin{align*}
\langle\Delta^{\prime}(\Lambda)^{2}\rangle_{S^{\prime}\big[\Delta;\,r(\ell_{P}),\,U(\ell_{P}),\,\ell_{P}\big]}= & \cZ^{-1}\,\int\cD\Delta^{\prime}\,\Delta^{\prime}(\Lambda)^{2}\,e^{S^{\prime}\big[\Delta^{\prime};\,r(\ell_{P}),\,P\,U(\ell_{P}),\,\ell_{P}\big]}\\
\stackrel{\text{perturbation theory}}{\simeq} & \cZ_{0}^{-1}\,\Bigg\{\int\cD\Delta^{\prime}\,\Delta^{\prime}(\Lambda)^{2}\,e^{S_{0}^{\prime}\big[\Delta^{\prime};\,r(\ell_{P}),\,\ell_{P}\big]}\,\big(1+P\,U(\ell_{P})\,\Delta^{\prime4}\big)\Bigg\}_{\mathrm{connected}}+\order((PU)^{2})\,.
\end{align*}
The leading order terms $\propto\order([PU]^{0})$ of course reproduce
the bias and variance contributions of the Gaussian case above, $\langle\Delta^{\prime}(\Lambda)^{2}\rangle=\langle\Delta^{\prime}(\Lambda)\rangle^{2}+\langle\Delta^{\prime}(\Lambda)\rangle_{c}^{2}$.
For the perturbative corrections $\delta\langle\Delta^{\prime}(\Lambda)^{2}\rangle_{S^{\prime}\big[\Delta,r(\ell_{P}),\,PU(\ell_{P}),\ell_{P}\big]}\propto P\,U(\ell_{P})$
we only need to take into account connected diagrams (the others are
cancelled by the normalization $\cZ^{-1}$). The diagram appearing
requires two connected propagators to attach each $\Delta^{\prime}(\Lambda)$
of the integrand to one leg of the interaction vertex. The remaining
two legs of the interaction vertex may either be contracted with a
connected correlator, which yields a factor $\propto\ell_{P}^{-1}$
or to a mean, which yields a factor $\propto1$, so together we have

\begin{align*}
\delta\cL\sim\delta\big\{\langle\Delta(k_{P})^{2}\rangle\big\} & =z(\ell_{P})^{2}\ell_{P}^{-1}\,\delta\big\{\langle\Delta^{\prime}(\Lambda)^{2}\rangle\big\}\\
 & =z(\ell_{P})^{2}\ell_{P}^{-1}\,P\,U(\ell_{P})\,\big[c_{1}+c_{2}\,\ell_{P}^{-1}\big]\\
 & \sim c_{1}\,P\ell_{P}^{3\beta}+c_{2}\,P\ell_{P}^{3\beta-1}\\
 & \stackrel{\ell_{P}\sim P^{-\frac{1}{1+\alpha}}}{\sim}c_{1}\,P^{1-\frac{3\beta}{1+\alpha}}+c_{2}\,P^{1-\frac{3\beta-1}{1+\alpha}}\\
 & \stackrel{\beta=1+\alpha}{\sim}c_{1}\,P^{-2}+c_{2}\,P^{-1-\frac{\alpha}{1+\alpha}}\,,
\end{align*}
which agrees to the two terms we have found in the quantitative computation
\eqref{eq:delta_L_bias}.

Again, on an abstract level this result can be understood very simply
in terms of the scaling dimension of the loss operator, which is $z(\ell)^{2}\,\ell^{-1}=\ell^{\beta}$
and the scaling interval defined by the two parts of the interaction
vertex, which are $\ell^{2\beta}$ and $\ell^{2\beta-1}$, respectively;
the product of the operator's scaling dimension and the dimensions
of the two parts of the interaction vertex then yields the two exponents
of the perturbative correction, demonstrating the power of this dimensional
analysis.

\section{Discussion and outlook}

An important insight of modern machine learning and AI is the benefit
of highly overparameterized neuronal networks \citep{kaplan2020scaling}.
Such networks, whose expressivity is much higher than required to
fit the training data, show the remarkable feature of neuronal scaling
laws, where the test loss behaves as a power law in the number of training
samples and in the number of network parameters. Such power laws point
towards scale-free behavior of the training process, which is found
to hold across many neuronal architectures and thus exposes a form
of universality of the learning problem per se. Understanding the
resulting exponents is of practical importance to predict the required
resources (network size, training time, and compute) to reach a desired
accuracy. It is also of theoretical importance, because it may expose
the deeper physical nature of the learning problem. The simplest possible
theory to explain such behavior employs the Gaussian process limit
\citep{bahri2021explaining}, which corresponds to a non-interacting
field theory. Careful analysis of taking the limit of the number of
network parameters and the number of training samples, however, shows
that marked departures from this limit are expected and are also crucial
to explain the power of modern deep learning, for example in terms
of the reached sample efficiency \citep{geiger2020perspective,yang2020feature,yu2013feature}.

From the physics point of view, the renormalization group is the tool
of choice to analyze phenomena of statistical self-similarity that
underlie scaling laws and which are known from high-energy physics
and from statistical physics of critical phenomena. We here specifically
develop the Wilsonian renormalization group for neuronal networks
and apply it to the neural field theory of learning and inference.
The neural field theory is derived from the underlying learning problem
and its structure bears important consequences: The momentum space
of field theories in physics seamlessly maps to the eigenmodes of
principal components of the data that enter our derivation. For the
typical case of power law spectra, the role of the spatial dimension
is here played by the dispersion relation of the spectrum; a connection
that has been noted in the context of biological neuronal networks
previously \citep{Bradde_2017, tiberi2023hiddenconnectivitystructurescontrol}.
As in physics, we also here find that the leading order correction
to the non-interacting limit of infinite networks $N\to\infty$ is
a $\phi^{4}$ interaction vertex. Different to previous work \citep{Demirtas_2024}
however, we find that the interaction vertex does not conserve momentum;
instead, it has an internal structure where pairs of legs are local
in momentum space. This structural difference compared to an ordinary
$\phi^{4}$-theory has marked consequences for the RG analysis. It
requires us to develop a novel concept for relevance and irrelevance
of interaction terms. We find that, in general, interactions of this
form can be thought of as multiple operators of different scaling
dimensions. The difference in scaling results from the locality of
the interaction vertex in conjunction with the locality of the connected
propagator. More formally, we show that this finding corresponds to
the effective Wilsonian action being a functional that not only depends
on a set of renormalized couplings $r(\ell)$, $u(\ell)$, etc., but
that is also explicitly a function of the coarse-graining scale $\ell$.
As a result, also the $\beta$-functions of the renormalization group
equations are inhomogeneous in $\ell$. An important consequence is
that, despite the similarity to an ordinary $\phi^{4}$-theory, there
is no Wilson-Fisher-like fixed point in the usual manner. Rather we
find that the phase space spanned by the renormalized parameters shows
separatrices whose positions depend on the coarse-graining scale.

Despite this slightly more involved picture, qualitative and quantitative
statements can be made. We find that the field theory of neuronal
networks is asymptotically free, which means that at high momenta,
which are non-learnable PC modes of low spectral power, the interaction
terms decline to zero, recovering the Gaussian process behavior of
the $N\to\infty$ limit. This insight allows us to derive the scaling
exponents in the limit of large numbers of training points $P\to\infty$,
as this limit shifts the transition between low-momentum, learnable
PC modes and high-momentum, non-learnable modes up in the spectrum,
into the realm of validity of the free Gaussian theory with its known
critical exponents. We demonstrate that the extended rules of relevance
and irrelevance can be brought to action to predict corrections to
these neural scaling laws. Likewise, the analytical framework to deal
with the Wilsonian renormalization group of non-momentum conserving
and momentum-local interactions that we derive here allows us to derive
approximations of the full $\beta$-function. We obtain quantitative
predictions for neural scaling laws beyond the Gaussian limit that
are in line with numerical simulations obtained in a teacher-student
setting \citep{hinton2015distillingknowledgeneuralnetwork}.
Importantly we find that even though the Gaussian limit is reached
ultimately, corrections decay as power laws and may thus cause significant
deviations at any realistic, finite $P$.

Another potential application of the presented RG approach is hyperparameter transfer. The aim here is to extrapolate from (cheap) numerical experiments on small models, both in terms of capacity and training-set size, the expected behavior of (costly) larger models. The RG setting, in particular the critical nature at large $P$ that leads to power laws, provides a concrete link between small-scale and large-scale behaviors, thus placing the question of optimal hyperparameter-transfer on firm analytical grounds. This link is demonstrated by performing an artificial form of hyperparameter-transfer where hyperparameters are tuned to maintain performance as the dataset size decreases. While this demonstrates the feasibility of quantitatively relating systems of different sizes, maintaining performance with less data relied heavily on working in the regime of strong-regularization, employing the equivalent-kernel approximation.  Studying more practical notions of transfer, for instance, those involving increasing model width, requires a more detailed analysis of generalization and overfitting effects and is left for future work. Similarly, it would be interesting to explore our conjectured RG prescription for optimal transfer, inspired by \cite{yang2022tensorprogramsvtuning,bordelon2023depthwise}: Co-tuning hyperparameters together with the width so as to maintain the relevant and marginal parameters of the renormalized action at $\ell = \ell_P$, for $P$ associated with the smaller model.

The presented theoretical framework presents a stepping-stone towards
a mechanistic understanding of universality of learning. For concreteness
we here employed the equivalent kernel approximation \citep{Silverman1984,Rasmussen2005},
which is valid in the limit of sufficiently strong regularization
of the training process. Concretely, it neglects fluctuations across
drawings of data samples, formally by replacing the quenched average
$-\beta\bar{F}:=\langle\ln\tr e^{-\beta S}\rangle_{\mathrm{data}}$
by the average $\ln\tr e^{-\beta\langle S\rangle_{\mathrm{data}}}$.
Also, we employed the empirically found approximate orthogonality
between higher order products of principal component modes. An exciting
step for further developments of the theory is to include fluctuation
effects beyond these idealizations and study their impact. This typically
requires the use of established replica approaches to compute the
data-averaged free energy $\bar{F}$ \citep{Bordelon2021review},
which needs to be combined with the renormalization group presented
here. A main insight of such an extension may include the separation
of the training and test error, for which the equivalent kernel theory
yields identical predictions.

Another line of future investigations concerns the quantitative comparison
of scaling laws observed in real-world networks with the theoretical
predictions obtained from the $\phi^{4}$-like momentum-local RG theory
derived here. Previous work has shown that the fluctuations of the
Gaussian process kernel indeed capture leading order corrections due
to feature learning beyond the Gaussian limit across many architectures
\citep{naveh2021self,zavatone2021asymptotics, seroussi2023separation,fischer2024criticalfeaturelearningdeep,rubin2025kernelsfeaturesmultiscaleadaptive}.
We therefore expect that qualitative insights regarding the leading
order corrections to neuronal scaling to carry over. We believe that
the presented work provides a stepping stone to a quantitative analysis
of universality of learning with help of RG methods that have uncovered
the true nature of universality in physics within the past decades.

\bibliographystyle{apsrev}
\bibliography{references}

\appendix

\section{Gaussian process regression}

\subsection{Setup}

Consider a linear network 
\begin{align*}
f & =w^{\T}\,x
\end{align*}
with a scalar output $f\in\bR$ and $P$ tuples of training data $\cD=\{(x_{\alpha},y_{\alpha})\}_{1\le\alpha\le P}$.
The data points are combined to form the matrix $\{\bR^{P\times d}\ni X\}_{\alpha i}=x_{\alpha i}$.
Assume a Gaussian prior $w_{i}\stackrel{\text{i.i.d.}}{\sim}\N(0,g_{w}/d)$
then the $f_{\alpha}$ follow a multivariate Gaussian distribution
\begin{align*}
\{f_{\alpha}\} & \sim\N(0,C^{(xx)})\,,\\
C_{\alpha\beta}^{(xx)} & =\frac{g_{w}}{d}\,x_{\alpha}^{\T}x_{\beta}\,.
\end{align*}
In addition assume a readout noise $\xi_{\alpha}\stackrel{\text{i.i.d.}}{\sim}\N(0,\kappa)$,
so that $y_{\alpha}=f_{\alpha}+\xi_{\alpha}$. Then the joint prior
distribution of the network output $f$ and the $y$ is
\begin{align}
p(y,f|X) & =\N(y|f,\kappa)\,\N(f|0,C^{(xx)})\label{eq:joint_y_f}\\
 & \propto e^{\cS(f,y)}\,,\nonumber 
\end{align}
where we defined the action
\begin{align}
\cS(y,f) & =-\frac{1}{2\kappa}\|f-y\|^{2}-\frac{1}{2}f^{\T}[C^{(xx)}]^{-1}f\,.\label{eq:action_joint}
\end{align}
In particular we are interested in the posterior $f$ when conditioning
on the training data $\cD$, which is given by
\begin{align}
p(f|\cD) & =\frac{p(y,f|X)}{\int df\,p(y,f|X)}\,.\label{eq:posterior}
\end{align}
We have the simple relation for the free energy $\cF$
\begin{align}
\cF:= & \ln\,\int df\,e^{\cS(y,f)}\label{eq:free_energy}\\
= & -\frac{1}{2}y^{\T}(C^{(xx)}+\kappa)^{-1}y-\frac{1}{2}\ln\det([C^{(xx)}]^{-1}+\kappa^{-1})\,.\nonumber 
\end{align}

\subsection{Langevin training}\label{sec:Langevin-training}

The action \eqref{eq:action_joint} can alternatively be considered
as resulting from Langevin training of the weights until equilibrium,
following the stochastic differential equation
\begin{align}
\frac{\partial}{\partial t}w_{i} & =-\frac{\partial}{\partial w_{i}}\,\big[\cL(w^{\T}x;y)+\kappa\,\frac{\|w\|^{2}}{2g_{w}/d}\big]+\sqrt{2\kappa}\,\xi(t)\,,\label{eq:Langevin_training}
\end{align}
where $\langle\xi(t)\xi(s)\rangle=\delta(t-s)$ is a unit variance
white noise and
\begin{align}
\cL(f;y) & :=\frac{1}{2}\|f-y\|^{2}\,\label{eq:squared_error_loss}
\end{align}
is the squared loss, because the stationary distribution of the stochastic
differential equation is 
\begin{align}
p_{0}(w) & \propto\exp\big(-\frac{1}{\kappa}\,\big[\cL(\underbrace{w^{\T}x}_{f};y)+\kappa\,\frac{\|w\|^{2}}{2g_{w}/d}\big]\,\big)\,,\label{eq:stationary_distribution_Langevin}
\end{align}
which is hence identical to \eqref{eq:joint_y_f}.

\subsection{Training loss}

In particular we may be interested in averages over the posterior
\eqref{eq:posterior}, such as the training loss
\begin{align}
\langle\cL\rangle & =\frac{1}{2}\langle\|f-y\|^{2}\rangle=\frac{1}{2}(\langle f\rangle-y)^{2}+\frac{1}{2}(f-\langle f\rangle)^{2}\,.\label{eq:training_loss}
\end{align}
Such observables thus only require the free energy \eqref{eq:free_energy}.
We further note that $y$ plays the role of a source field for $f-y$.
The free energy \eqref{eq:free_energy} has two terms, one that depends
on $y$ and another that is independent of $y$. Only the first one
determines correlations of $f-y$ and hence correlations of $f$.

Another important observable is the mean discrepancy
\begin{align}
\langle\Delta_{\alpha}\rangle & :=\langle y_{\alpha}-f_{\alpha}\rangle\label{eq:mean_discrepancy}\\
 & =-\kappa\,\frac{\partial}{\partial y_{\alpha}}\,\cF=:-\kappa\,\cF_{\alpha}^{(1)}\,.\nonumber 
\end{align}
The training loss \prettyref{eq:training_loss}, correspondingly,
may be expressed in terms of derivatives by $y$ alone, namely with
\begin{align*}
\kappa^{2}\cF_{\alpha\beta}^{(2)}:=\kappa^{2}\frac{\partial^{2}}{\partial y_{\alpha}\partial y_{\beta}}\,\cF & =\delta_{\alpha\beta}\,\kappa+\langle(y_{\alpha}-f_{\alpha})(y_{\beta}-f_{\beta})\rangle-\langle y_{\alpha}-f_{\alpha}\rangle\langle y_{\beta}-f_{\beta}\rangle\\
 & =\delta_{\alpha\beta}\,\kappa+\langle\Delta_{\alpha}\Delta_{\beta}\rangle-\langle\Delta_{\alpha}\rangle\langle\Delta_{\beta}\rangle,
\end{align*}
the loss follows as

\begin{align}
\langle\cL\rangle & \equiv\frac{1}{2}\,\tr\,\langle\Delta\Delta^{\T}\rangle\label{eq:loss_as_derivatives}\\
 & =\frac{\kappa^{2}}{2}\,\tr\,\big[\cF^{(2)}+\cF^{(1)}\cF^{(1)\T}-\kappa^{-1}\,\bI\big]\,.\nonumber 
\end{align}

\subsection{Transformation to eigenspace}\label{sec:Transformation-to-eigenspace}

We may transform the action into the eigenspace (principal components)
of $C^{(xx)}$
\begin{align}
C^{(xx)}\,u_{\mu} & =\Lambda_{\mu}\,u_{\mu}\,.\label{eq:PCA_basis}
\end{align}
The set of $U:=\{u_{\mu}\}$ being a complete basis, we write the
target vector $y\in\bR^{P}$ as $y=\sum_{\mu}\,Y_{\mu}\,u_{\mu}$.
In particular we will be interested in power law decays of the eigenvalues
\begin{align}
\Lambda_{\mu} & =\Lambda_{1}\,\mu^{-1-\alpha},\label{eq:power_law_lambda}\\
Y_{\mu}^{2} & =Y_{1}^{2}\,\mu^{-1-\beta}\,,\label{eq:power_law_y}
\end{align}
where $\alpha>0$ and $\beta>0$, because otherwise the variance of
the kernel or of the target would be infinite.

Likewise the vector $f\in\bR^{P}$ is $f=\sum_{\mu}\,F_{\mu}\,u_{\mu}$.
The basis being orthonormal $u_{\mu}^{\T}u_{\nu}=\delta_{\mu\nu}$,
the integration measures for $y$ and $f$ remain unchanged, so we
get the action \eqref{eq:action_joint}
\begin{align}
\cS(F,Y) & =-\frac{1}{2}\sum_{\mu=1}^{P}\frac{\big(F_{\mu}-Y_{\mu}\big)^{2}}{\kappa}+\frac{F_{\mu}^{2}}{\Lambda_{\mu}}\,.\label{eq:joint_action_eigen}
\end{align}
A completion of the square

\begin{align}
\cS(F,Y) & =-\frac{1}{2}\sum_{\mu}\,\Big\{\big(\Lambda_{\mu}^{-1}+\kappa^{-1}\big)\,\big[F_{\mu}-\frac{\kappa^{-1}}{\Lambda_{\mu}^{-1}+\kappa^{-1}}\,Y_{\mu}\big]^{2}\label{eq:action_completed-1}\\
 & -\big[\frac{\kappa^{-1}}{\Lambda_{\mu}^{-1}+\kappa^{-1}}Y_{\mu}(l)\big]^{2}\,\Big\},\nonumber 
\end{align}
allows us to read off the mean and covariance of each mode as

\begin{align}
m_{\mu} & :=\langle F_{\mu}\rangle=\frac{\Lambda_{\mu}}{\Lambda_{\mu}+\kappa}\,Y_{\mu}\,,\label{eq:mean_orig}\\
c_{\mu\nu} & :=\langle F_{\mu}F_{\nu}\rangle-\langle F_{\mu}\rangle\langle F_{\nu}\rangle=\delta_{\mu\nu}\,\big(\Lambda_{\mu}^{-1}+\kappa^{-1}\big)^{-1}\label{eq:cov_orig}\\
 & =\delta_{\mu\nu}\,\frac{\kappa\,\Lambda_{\mu}}{\Lambda_{\mu}+\kappa}=:\delta_{\mu\nu}\,c_{\mu}\,,\nonumber 
\end{align}
where we defined the non-trivial part of $c_{\mu\nu}$ as $c_{\mu}$.
Since $Y$ does not fluctuate, the covariance of $F$ is the same
as the covariance of $\Delta_{\mu}:=Y_{\mu}-F_{\mu}$
\begin{align}
c_{\mu\nu} & =\langle\Delta_{\mu}\Delta_{\nu}\rangle-\langle\Delta_{\mu}\rangle\langle\Delta_{\nu}\rangle\,.\label{eq:cov_Delta}
\end{align}
The mean discrepancy is
\begin{align}
d_{\mu}:=\langle\Delta_{\mu}\rangle & =Y_{\mu}-\langle F_{\mu}\rangle=\frac{\kappa}{\Lambda_{\mu}+\kappa}\,Y_{\mu}\,.\label{eq:mean_Delta}
\end{align}
The expected loss \eqref{eq:training_loss} can be expressed in terms
of $m_{\mu}$ and the non-trivial part $c_{\mu}$ as
\begin{align}
\langle\cL\rangle= & \frac{1}{2}(\langle f\rangle-y)^{2}+\frac{1}{2}(f-\langle f\rangle)^{2}\label{eq:expected_loss_orig}\\
= & \frac{1}{2}\sum_{\mu=1}^{P}(m_{\mu}-Y_{\mu})^{2}+c_{\mu}\,.\nonumber 
\end{align}
The free energy $F$ takes the form
\begin{align}
\cF:= & \ln\,\int df\,e^{\cS(y,f)}\label{eq:free_energy_eigenspace}\\
= & -\frac{1}{2}\sum_{\mu=1}^{P}\frac{Y_{\mu}^{2}}{\Lambda_{\mu}+\kappa}+\ln\big(\Lambda_{\mu}^{-1}+\kappa^{-1}\big)\,,\nonumber 
\end{align}
which again has a first term the controls the correlations of $f-y$
and a second term that is independent of $y$ and hence does not affect
correlation functions.

\subsection{Additional quartic loss term}\label{sec:Additional-quartic-loss}

Now consider an additional quartic part of the loss in \eqref{eq:Langevin_training}
and \eqref{eq:squared_error_loss} of the form
\begin{align}
\cL(f;y):=\frac{1}{2}\|f-y\|^{2}+ & \frac{\kappa\,U}{3}\,\sum_{\alpha=1}^{P}\,\big(y_{\alpha}-f_{\alpha}\big)^{4}\,,\label{eq:quartic_loss}
\end{align}
so in total one obtains the action

\begin{align}
\cS(y,f) & =-\frac{1}{2\kappa}\|f-y\|^{2}-\frac{U}{3}\,\sum_{\alpha=1}^{P}\,\big(y_{\alpha}-f_{\alpha}\big)^{4}-\frac{1}{2}f^{\T}[C^{(xx)}]^{-1}f\,.\label{eq:action_quartic_final}
\end{align}
Expressed in the eigenbasis \eqref{eq:PCA_basis} the quartic term
reads
\begin{align*}
 & \frac{\kappa U}{3}\,\sum_{\alpha=1}^{P}\,\big(\sum_{\mu=1}^{P}\,u_{\mu\alpha}(Y_{\mu}-F_{\mu})\big)^{4}\\
= & \frac{\kappa U}{3}\,\sum_{\mu_{1},\mu_{2},\mu_{3},\mu_{4}=1}^{P}\big(\sum_{\alpha=1}^{P}\,u_{\mu_{1}\alpha}\,u_{\mu_{2}\alpha}\,u_{\mu_{3}\alpha}\,u_{\mu_{4}\alpha}\big)\,(Y_{\mu_{1}}-F_{\mu_{1}})\,(Y_{\mu_{2}}-F_{\mu_{2}})\,(Y_{\mu_{3}}-F_{\mu_{3}})\,(Y_{\mu_{4}}-F_{\mu_{4}})\,.
\end{align*}
We now use that the eigenmodes behave similar as Gaussian variables
which fulfill Wick's theorem (this is shown to hold empirically for
real data sets in \eqref{sec:Overlaps})
\begin{align*}
V_{\mu_{1}\mu_{2}\mu_{3}\mu_{4}}^{\text{approx}} & =\sum_{\alpha=1}^{P}u{}_{\mu_{1}\alpha}u_{\mu_{2}\alpha}u_{\mu_{3}\alpha}u_{\mu_{4}\alpha}\simeq\sum_{\alpha=1}^{P}\langle u{}_{\mu_{1}\alpha}u_{\mu_{2}\alpha}u_{\mu_{3}\alpha}u_{\mu_{4}\alpha}\rangle\\
 & =\frac{3}{P}\,\delta_{\mu_{1}\mu_{2}\mu_{3}\mu_{4}}+\frac{1}{P}\,(1-\delta_{\mu_{1}\mu_{2}\mu_{3}\mu_{4}})\,\big(\delta_{\mu_{1}\mu_{2}}\,\delta_{\mu_{3}\mu_{4}}+\delta_{\mu_{1}\mu_{3}}\delta_{\mu_{2}\mu_{4}}+\delta_{\mu_{1}\mu_{4}}\delta_{\mu_{2}\mu_{3}}\big)\,,
\end{align*}
which reduces the above expression to

\begin{align}
 & \frac{\kappa U}{3}\,\sum_{\alpha=1}^{P}\,\big(\sum_{\mu=1}^{P}\,u_{\mu\alpha}(Y_{\mu}-F_{\mu})\big)^{4}\label{eq:quartic_rewritten}\\
= & \frac{\kappa U}{3}\,\frac{3}{P}\,\sum_{\mu=1}^{P}\,(Y_{\mu}-F_{\mu})^{4}\nonumber \\
 & +\frac{\kappa U}{3}\,\frac{3}{P}\,\sum_{\mu_{1}\neq\mu_{2}=1}^{P}\,(Y_{\mu_{1}}-F_{\mu_{1}})^{2}\,(Y_{\mu_{2}}-F_{\mu_{2}})^{2}\nonumber \\
= & \frac{\kappa U}{P}\,\Big[\sum_{\mu=1}^{P}\,\big(Y_{\mu}-F_{\mu}\big)^{2}\Big]^{2}\,,\nonumber 
\end{align}
which yields the action in eigenspace of $C^{(xx)}$

\begin{align}
\cS(F,Y) & =-\frac{1}{2}\sum_{\mu=1}^{P}\frac{\big(F_{\mu}-Y_{\mu}\big)^{2}}{\kappa}+\frac{F_{\mu}^{2}}{\Lambda_{\mu}}\label{eq:action_eigenspace_F}\\
 & -\frac{U}{P}\,\Big[\sum_{\mu=1}^{P}\,\big(Y_{\mu}-F_{\mu}\big)^{2}\Big]^{2}\,.\nonumber 
\end{align}
Likewise, we may introduce the discrepancies $\Delta_{\mu}:=Y_{\mu}-F_{\mu}$
and rewrite this action as
\begin{align}
\cS(\Delta,Y) & =-\frac{1}{2}\sum_{\mu=1}^{P}\frac{\Delta_{\mu}^{2}}{\kappa}+\frac{(\Delta_{\mu}-Y_{\mu})^{2}}{\Lambda_{\mu}}\label{eq:action_eigenspace}\\
 & -\frac{U}{P}\,\Big[\sum_{\mu=1}^{P}\,\Delta_{\mu}^{2}\Big]^{2}\,.\nonumber 
\end{align}

\subsection{Equivalent kernel}\label{sec:Equivalent-kernel}

Instead of operating on one concrete data set, we are interested in
the behavior on average over data sets. To this end we follow \citep[Sec 7.1]{Rasmussen2005}
and assume that data and labels come from a joint distribution 
\begin{align}
p(y,x)\label{eq:p_x_y_joint}
\end{align}
 of which we denote as $p(x)=\int dy\,p(y,x)$ its marginal for $x$.

To change back from this continuum formulation to the discrete formulation,
we set the probability measure as the empirical measure
\begin{align}
p(x) & =\frac{1}{P}\,\sum_{\alpha=1}^{P}\,\delta(x-x_{\alpha})\,.\label{eq:empirical_measure}
\end{align}
In the continuum assume a pairwise orthogonormal basis with regard
to the integration measure $p(x)\,dx$
\begin{align}
\delta_{\mu\nu} & =\int\phi_{\mu}(x)\,\phi_{\nu}(x)\,p(x)\,dx\,.\label{eq:orthonormality}
\end{align}
These basis functions are eigenfunctions of the kernel $K:\bR^{d}\times\bR^{d}\mapsto\bR$
defined as
\begin{align}
K(x,x^{\prime}) & =\sum_{\mu}\lambda_{\mu}\,\phi_{\mu}(x)\phi_{\mu}(x^{\prime})\,,\label{eq:kernel_decomp_eig}
\end{align}
in the sense

\begin{align}
\int\,K(x,x^{\prime})\,\phi_{\mu}(x^{\prime})\,p(x^{\prime})\,dx^{\prime} & =\lambda_{\mu}\phi_{\mu}(x)\,.\label{eq:kernel_eigvals}
\end{align}

We would like to know how the eigenvalues $\lambda_{\mu}$ relate
to the eigenvalues $\Lambda_{\mu}$ introduced in \eqref{eq:PCA_basis}.
To this end, we insert the empirical measure \eqref{eq:empirical_measure}
into \eqref{eq:kernel_eigvals} to obtain

\begin{align*}
\lambda_{\mu}\phi_{\mu}(x) & =P^{-1}\,\sum_{\beta=1}^{P}K(x,x_{\beta})\,\phi_{\mu}(x_{\beta})\,.
\end{align*}
Evaluated on the set of data points, the right hand side is now of
similar form as \eqref{eq:PCA_basis} if we set
\begin{align*}
K(x_{\alpha},x_{\beta}) & =C_{\alpha\beta}^{(xx)}\,,
\end{align*}
except the additional factor $P^{-1}$. We thus identify the eigenvalues
as
\begin{align}
\Lambda_{\mu} & =P\,\lambda_{\mu}\,.\label{eq:eig_rel_EK}
\end{align}
Likewise, the orthogonality relation \eqref{eq:orthonormality} with
the empirical measure inserted reads
\begin{align*}
\delta_{\mu\nu} & =P^{-1}\,\sum_{\alpha=1}^{P}\,\phi_{\mu}(x_{\alpha})\,\phi_{\nu}(x_{\alpha}).
\end{align*}
The relation between the discrete modes $u_{\mu}$ and those on the
continuum is hence
\begin{align}
u_{\mu\alpha} & =\frac{1}{\sqrt{P}}\,\phi_{\mu}(x_{\alpha})\,.\label{eq:basis_relation}
\end{align}
Defining the expected target as
\begin{align*}
y(x) & :=\int dy\,y\,p(y|x)
\end{align*}
the first term in \eqref{eq:action_joint} therefore takes the form
\begin{align}
\sum_{\alpha=1}^{P}\big(f_{\alpha}-y_{\alpha}\big)^{2} & \simeq P\,\int\big(f(x)-y\big)^{2}\,p(y,x)\,dx\,dy\label{eq:data_term_in_EK}\\
 & =P\,\int\big(f(x)-y(x)-(y-y(x))\big)^{2}\,p(y,x)\,dx\,dy\nonumber \\
 & =P\,\int\big(f(x)-y(x)\big)^{2}\,p(x)\,dx\nonumber \\
 & +P\,\int\big(y-y(x)\big)^{2}\,p(y,x)\,dx\,dy\,,\nonumber 
\end{align}
where the latter term, the variance of the target, is independent
of $f$, hence does not influence its statistics and we used that
cross terms vanish. In the eigenspace of the kernel $K$ and expanding
$f$ and $y$ in these modes
\begin{align}
f(x) & =\sum_{\mu}\,f_{\mu}\,\phi_{\mu}(x)\,,\label{eq:expansion_f}\\
y(x) & =\sum_{\mu}\,y_{\mu}\,\phi_{\mu}(x)\,.\nonumber 
\end{align}
Evaluated at the data samples $x=x_{\alpha}$, the fields must assume
the same values as in the discrete. The coefficients $f_{\mu}$ and
$y_{\mu}$ are therefore related to those of the discrete system as
\begin{align}
F_{\mu} & =\sqrt{P}\,f_{\mu}\,,\label{eq:relation_eigen_EK}\\
Y_{\mu} & =\sqrt{P}\,y_{\mu}\,,\nonumber 
\end{align}
which is due to \eqref{eq:basis_relation}.

The $f$-dependent term in \eqref{eq:data_term_in_EK} of the action
reads
\begin{align}
P\,\int\big(f(x)-y(x)\big)^{2}\,p(x)\,dx & =P\,\int\big(\sum_{\mu}\,(f_{\mu}-y_{\mu})\,\phi_{\mu}(x)\big)^{2}\,p(x)\,dx\nonumber \\
 & =P\,\sum_{\mu}\,(f_{\mu}-y_{\mu})^{2}\,,\label{eq:first_term_action_equiv}
\end{align}
where we used the pairwise orthonormality \eqref{eq:orthonormality}
of the $\phi$.

The second term in the action \eqref{eq:action_joint} can be written
as what is known as the reproducing kernel Hilbert space (RKHS) norm
\begin{align*}
-\frac{1}{2}f^{\T}[C^{(xx)}]^{-1}f & \equiv-\frac{1}{2}\langle f,f\rangle\,,\\
\langle f,g\rangle & :=f^{\T}\,[C^{(xx)}]^{-1}g\,.
\end{align*}
The second term of the action in eigenspace becomes $-\frac{1}{2}\,\sum_{\mu}\,\frac{f_{\mu}^{2}}{\lambda_{\mu}}\,$,
so together with \eqref{eq:first_term_action_equiv}, we have the
approximate action
\begin{align}
S(f,y) & =-\frac{1}{2}\sum_{\mu}\,\frac{(f_{\mu}-y_{\mu})^{2}}{\kappa/P}+\frac{f_{\mu}^{2}}{\lambda_{\mu}}\,.\label{eq:action_EK}
\end{align}
Since the eigenvalues $\lambda$ are by definition independent of
$P$ (they only rely on the functional form of $K(x,x^{\prime})$
and the measure $p(x)$), this form has the advantage of exposing
the explicit $P$-dependence of the first term. The sum over $\mu$
here extends to infinity, as there are infinitely many modes of the
kernel. We also note that the mean of each mode $f_{\mu}$, by the
relation between $\Lambda_{\mu}=P\lambda_{\mu}$, is the same as in
the original action \eqref{eq:joint_action_eigen}, because both terms
are scaled with the same factor $P$. Also the action only depends
on $P/\kappa$ as an effective parameter. So we have
\begin{align}
\lambda_{\mu}= & \mu^{-1-\alpha}\,,\label{eq:power_law_rescaled}\\
y_{\mu}^{2}= & \mu^{-1-\beta}\,.\nonumber 
\end{align}

\section{Quartic theory from finite-width corrections}\label{sec:Quartic-theory-from-finite-size}

An alternative motivation to study a quartic theory similar to \prettyref{eq:action_quartic_final},
we may consider corrections due to finite width. In the simplest case,
one may consider a single hidden layer network architecture
\begin{align}
h_{\alpha}=Vx_{\alpha}\,,\label{eq:network_architecture}\\
f_{\alpha}=w^{\T}\phi(h_{\alpha})\,,\nonumber 
\end{align}
which is trained on $P$ tuples of training data $\mathcal{D}=\{(x_{\alpha},y_{\alpha})\}_{1\le\alpha\le P}$
with $x_{\alpha}\in\mathbb{\bR}^{d}$ and $y_{\alpha}\in\mathbb{\bR}$.
Here $\phi$ denotes a non-linear activation function and $f_{\alpha}\in\bR$
is the scalar network output. 

We study the Bayesian setting with Gaussian priors on the readin weights
$V\in\bR^{N\times d}$ as $V_{ij}\sim\mathcal{N}(0,g_{v}/d)$ and
the readout weights $w\in\bR^{N}$ as $w_{i}\sim\mathcal{N}(0,g_{w}/N)$.
To keep the notation concise, we use the shorthands $f_{\mathcal{D}}=(f_{\alpha})_{1\le\alpha\le P}$,
$X=(x_{\alpha})_{1\le\alpha\le P}$ and $y=(y_{\alpha})_{1\le\alpha\le P}$
in the following. Further, summations over repeated indices are implied
$V_{kl}x_{l}\equiv\sum_{l=1}^{N}V_{kl}x_{l}$.

Training is performed with a squared error loss function $\cL=\frac{1}{2}\|y-f\|^{2}$
and gradient descent with weight decay and Gaussian noise, so that
the weights follow the equilibrium distribution
\begin{align*}
V,w & \sim\frac{1}{\cZ}\exp\big(-\frac{1}{2\kappa}\,\cL\big[y,f(w,V|X)\big]-\frac{1}{2g_{v}}\|V\|^{2}-\frac{1}{2g_{w}}\|w\|^{2}\big)\,,
\end{align*}
where $\cZ$ is the partition function. After some standard manipulations,
the latter takes the form
\begin{align*}
\cZ(y) & =\int df\,\int dC\,\N(y|f,\kappa\bI)\,\N(f|0,C)\\
 & \times\int d\tC\,\exp\big(-\tr\,\tC^{\T}C+W(\tC|C^{(xx)})\big)\,,\\
\\W(\tC|C^{(xx)}) & =N\,\ln\,\big\langle\exp\big(\frac{g_{w}}{N}\phi(h)^{\T}\tC\phi(h)\big)\big\rangle_{h\sim\N(0,C^{(xx)})}\,,\\
C^{(xx)} & :=\frac{g_{v}}{D}XX^{\T}\,.
\end{align*}
The scaling form of the cumulant-generating function $W$ implies
that the mean $W^{(1)}\propto\order(1)$ and the variance $W^{(2)}\propto\order(N^{-1})$,
which shows that the inner kernel matrix $C$ concentrates. Keeping
fluctuation effects up to Gaussian order, one may expand $W$ into
the first two cumulants as
\begin{align}
W(\tC|C^{(xx)}) & =C_{\alpha\beta}^{(\phi\phi)}\,\tC_{\alpha\beta}+\frac{1}{2}S_{\alpha\beta,\gamma\delta}^{(\phi\phi,\phi\phi)}\,\tC_{\alpha\beta}\tC_{\gamma\delta}+\order(\tC^{3})\,,\label{eq:approx_W}\\
C_{\alpha\beta}^{(\phi\phi)} & :=g_{w}\,\big\langle\phi_{\alpha}\phi_{\beta}\big\rangle\,,\nonumber \\
S_{\alpha\beta,\gamma\delta}^{(\phi\phi,\phi\phi)} & :=\frac{g_{w}^{2}}{N}\,\Big[\big\langle\phi_{\alpha}\phi_{\beta}\phi_{\gamma}\phi_{\delta}\big\rangle-\big\langle\phi_{\alpha}\phi_{\beta}\big\rangle\,\big\langle\phi_{\gamma}\phi_{\delta}\big\rangle\Big]\,,\nonumber 
\end{align}
where all expectations are with regard to the Gaussian measure $\langle\ldots\rangle_{h\sim\N(0,C^{(xx)})}$.
The approximation \eqref{eq:approx_W} implies a Gaussian distribution
of the kernel $C_{\alpha\beta}$. Rewriting 
\begin{align}
\cZ(y)= & \int df\,\big\langle\N(y|0,C+\kappa\bI)\,\big\rangle_{C}\label{eq:Z_before_integration_C}\\
 & =\int d\tf\,\big\langle\exp\big(-\tf^{\T}y+\frac{1}{2}\tf^{\T}(C+\kappa\bI)\,\tf\big)\,\big\rangle_{C}\,,\nonumber 
\end{align}
we may perform the integral over $C$ with the cumulant-expansion
of $W$ \eqref{eq:approx_W}, which yields
\begin{align}
\cZ(y)= & \int d\tf\,\exp\big(-\tf^{\T}y+\frac{1}{2}\tf^{\T}(C^{(\phi\phi)}+\kappa\bI)\,\tf+\frac{1}{8}\,S_{\alpha\beta,\gamma\delta}^{(\phi\phi,\phi\phi)}\,\tf_{\alpha}\tf_{\beta}\tf_{\gamma}\tf_{\delta}\big)\,.\label{eq:Z_C_integrated_out}
\end{align}
For the special case of a linear activation function $\phi(h)=h$,
we obtain with Wick's theorem
\begin{align*}
S_{\alpha\beta,\gamma\delta} & =\frac{g_{w}^{2}}{N}\,\big[C_{\alpha\gamma}^{(xx)}C_{\beta\delta}^{(xx)}+C_{\alpha\delta}^{(xx)}C_{\beta\gamma}^{(xx)}\big]\,,
\end{align*}
which allows us to write 
\begin{align}
\cZ(y)= & \int d\tf\,\exp\big(-\tf^{\T}y+\frac{1}{2}\tf^{\T}(C^{(xx)}+\kappa\bI)\,\tf+\frac{g_{w}^{2}}{4N}\,\Big[C_{\alpha\beta}^{(xx)}\tf_{\alpha}\tf_{\beta}\Big]^{2}\big)\,.\label{eq:Z_3}
\end{align}
We may move into the eigenspace of $C^{(xx)}$ as $\tf=U\,\varphi$
so
\begin{align}
\cZ(y) & =\int d\varphi\,\exp\big(-y^{\T}U\,\varphi+\frac{1}{2}\,\sum_{\alpha}(\lambda_{\alpha}+\kappa)\,\varphi_{\alpha}^{2}+\frac{g_{w}^{2}}{4N}\,\Big[\sum_{\alpha}\lambda_{\alpha}\varphi_{\alpha}^{2}\Big]^{2}\big)\,.\label{eq:Z_4}
\end{align}
Now redefine the auxiliary variables as real variables $\Delta_{\alpha}:=i\,\sqrt{\kappa\,\lambda_{\alpha}}\,\varphi_{\alpha}\in\bR$,
to get the partition function
\begin{align}
\cZ(y) & =c\,\int d\Delta\,\exp\big(\,iy^{\T}U\,\sqrt{\kappa^{-1}\,\Lambda^{-1}}\,\Delta-\frac{1}{2}\,\sum_{\alpha}(\kappa^{-1}+\lambda_{\alpha}^{-1})\,\Delta_{\alpha}^{2}+\frac{g_{w}^{2}}{4N\kappa^{2}}\,\Big[\sum_{\alpha}\Delta_{\alpha}^{2}\Big]^{2}\big)\,,\label{eq:finite_size_net_action}
\end{align}
where we obtain an inconsequential factor $c$ due to the change of
measure and $\Lambda$ is a diagonal matrix with entries $\lambda_{\alpha}$.

The last form has an identical structure to our starting point \prettyref{eq:action_eigenspace},
with the only additional assumption that the transformed target $y^{\T}U$
in \eqref{eq:finite_size_net_action} has power law entries. Also
we get an imaginary unit in the mixed term $\propto y\Delta$ and
the sign of the interaction term is such that $U<0$.

\section{Continuum limit of the action}\label{sec:Continuum-limit}

To perform the renormalization group calculation it is useful to have
a continuum representation rather than having a sum over a discrete
set of modes. In this section we perform the step from the discrete
representation to a continuum representation by ensuring that all
observables maintain the same value in the discrete and in the continuum
theory.

\subsection{General setting}

We here consider the general setting of a theory that consists of
a Gaussian, solvable part and non-Gaussian perturbations. Let the
ground truth solvable theory represented as a sum over a discrete
set of modes be the centered Gaussian diagonal action
\begin{align*}
S_{0}(\phi) & =-\frac{1}{2}\sum_{\mu=1}^{N}G_{\mu}\phi_{\mu}^{2}\,.
\end{align*}
We assume a source term $J^{\T}\phi$ which allows us to also describe
the situation of a non-vanishing mean. The free energy then is
\begin{align}
F_{0}(J) & =\ln\int d^{N}\phi\,e^{S_{0}(\phi)+J^{\T}\phi}=\frac{1}{2}\sum_{k=1}^{N}J_{\mu}^{2}G_{\mu}^{-1}-\frac{1}{2}\,\sum_{\mu=1}^{N}\,\ln G_{\mu}\,.\label{eq:free_energy_free}
\end{align}
The theory implies connected correlation functions (cumulants)
\begin{align*}
\langle\phi_{\mu}\rangle & =G_{\mu}^{-1}J_{\mu}=:m_{\mu}\,,
\end{align*}
\begin{align*}
\langle\phi_{\mu}\phi_{\nu}\rangle_{c} & =\delta_{\mu\nu}\,G_{\mu}^{-1}=:\delta_{\mu\nu}\,c_{\mu},
\end{align*}
where the mean also follows from the stationary point of the action
including the source. In addition, we would like to compute the effect
of a perturbation
\begin{align}
S_{\mathrm{int}}(\phi) & =\sum_{\mu,\nu,\eta,\iota=1}^{N}V_{\mu,\nu,\eta,\iota}\,\phi_{\mu}\phi_{\nu}\phi_{\eta}\phi_{\iota}\,\label{eq:S_int}
\end{align}
and we are interested in expectation values of observables $O(\phi)$
\begin{align}
\langle O\rangle & =\frac{\int d^{N}\phi\,O(\phi)\,e^{S(\phi)}}{\int d^{N}\phi\,e^{S(\phi)}}\,.\label{eq:Observable}
\end{align}
We assume that observables can be decomposed into a series of field
monomials, so that it is sufficient to ensure that all cumulants of
the fields are treated faithfully to maintain the expectation value
of any such observable. Since correlation functions can be expressed
entirely in terms of the $J$-dependent part of the free energy in
\prettyref{eq:free_energy_free}, it is sufficient to ensure that
this part be treated faithfully when going from the discrete to the
continuum. 

The full free energy of the interacting system then is
\begin{align}
F(J) & =\ln\int d^{N}\phi\,e^{S_{0}+S_{\mathrm{int}}+J^{\T}\phi}\label{eq:free_energy_full}\\
 & =\ln\,\Big\langle e^{S_{\mathrm{int}}}\Big\rangle_{\phi\sim e^{S_{0}(J)-J^{\T}\phi}}\,,\nonumber 
\end{align}
where $\phi\sim e^{S_{0}+J^{\T}\phi}$ is meant as $\phi$ is distributed
as Gaussian implied by the quadratic action $S_{0}+J^{\T}\phi$, namely
$p(\phi)=e^{S_{0}(\phi)+J^{\T}\phi-F_{0}(J)}\,$, where $F_{0}$ \eqref{eq:free_energy_free}
is the normalization.

\subsection{Fine-grained free energy}\label{sec:Fine-grained-free-energy}

As an intermediate step of transitioning to the continuum, we insert
$n$ additional modes between any pair of original modes. Our aim
is to obtain a form of the free energy $F^{(n)}(J^{(n)})$ with the
following two properties: 
\begin{itemize}
\item First, the value of the source of the fine-grained system $J^{(n)}$
should be an interpolation of the source for the original system in
the sense 
\begin{align}
J_{i}^{(n)} & :=J_{\lfloor i/n\rfloor}\,,\label{eq:J_correspondence}
\end{align}
where $\lfloor\ldots\rfloor$ denotes down-rounding to the next lower
integer.
\item Second, the $J^{(n)}$-dependent part of the free energy of the fine-grained
system should be intensive in $n$ and its value should correspond
to the $J$-dependent part of the original discrete theory
\begin{align}
F(J) & \simeq F^{(n)}(J^{(n)})\,.\label{eq:F_correspondence}
\end{align}
Here $\simeq$ means up to $J$-independent terms.
\end{itemize}
The second property is required so that we may ultimately take the
limit $n\to\infty$ and obtain a finite result.

The properties \prettyref{eq:J_correspondence} and \prettyref{eq:F_correspondence}
imply by the chain rule that
\begin{align}
\frac{\partial F}{\partial J_{\mu}} & =\sum_{i:\,\lfloor i/n\rfloor=\mu}\,\frac{\partial F^{(n)}}{\partial J_{i}^{(n)}}\,.\label{eq:replacement_rule_deriv_source}
\end{align}
This shows that any derivative $\partial/\partial J_{\mu}$ needs
to be replaced by $\sum_{i::\,\lfloor i/n\rfloor=k}\partial/\partial J_{i}^{(n)}$,
because changing a single point $\mu$ of $J_{\mu}$ in the discrete
system, by \eqref{eq:J_correspondence}, implies a change for $n$
points $i$ of the fine-grained source $J_{i}^{(n)}$; we call this
set of modes the ``$n$-vicinity'' of mode $\mu$ in the following. 

An expectation value of any observable \eqref{eq:Observable} amounts
to computing the cumulants of the theory. The relation
\begin{align}
\langle\phi_{\mu}\ldots\phi_{\nu}\rangle_{c} & \equiv\frac{\partial}{\partial J_{\mu}}\cdots\frac{\partial}{\partial J_{\nu}}\,F\label{eq:replacement_rule_derivative}\\
 & =\sum_{i\,:\,\lfloor i/n\rfloor=\mu}\cdots\sum_{j\,:\,\lfloor j/n\rfloor=\nu}\,\frac{\partial}{\partial J_{\mu}}\cdots\frac{\partial}{\partial J_{\nu}}\,F^{(n)}\nonumber 
\end{align}
tells us how to compute cumulants of the original system in terms
of the fine-grained system: it incurs a summation over the fine-grained
modes in the $n$-vicinity of the original modes. In a sense this
means that the two theories have the same cumulant-density per degree
of freedom.

\subsection{Form of the Gaussian fine-grained action}\label{sec:fine_grained_Gaussian}

Next, we ask which form of the free part of the theory obeys properties
\eqref{eq:J_correspondence} and \eqref{eq:F_correspondence}. Consider
the choice
\begin{align*}
S_{0}^{(n)}(\phi) & =-\frac{1}{2n}\,\sum_{k=n}^{n\,N}G_{\lfloor k/n\rfloor}\phi_{k}^{2}\,,
\end{align*}
so the pairwise correlation is
\begin{align*}
c_{kl}^{(n)}=\langle\phi_{k}\phi_{l}\rangle & =G_{\lfloor k/n\rfloor}^{-1}\,n\,\delta_{kl}=n\,\delta_{kl}\,c_{\lfloor k/n\rfloor}
\end{align*}
and we set the source term as 
\begin{align}
\frac{1}{n}J^{(n)\T}\phi:= & \frac{1}{n}\,\sum_{k=n}^{n\,N}J_{k}^{(n)}\,\phi_{k}\,\label{eq:source_fine_grained}\\
\stackrel{(\ref{eq:J_correspondence})}{=} & \frac{1}{n}\,\sum_{k=n}^{n\,N}J_{\lfloor k/n\rfloor}\,\phi_{k}\,.\nonumber 
\end{align}
This source term assures that, assuming the same value for the source
by the interpolation property \eqref{eq:J_correspondence}, that the
mean of the field due to the Gaussian part stays the same, because
$S_{0}^{(n)}(\phi)+\frac{1}{n}J^{\T}\phi$ has the same stationary
point 
\begin{align*}
m_{k}^{(n)} & =\langle\phi_{k}\rangle=G_{\lfloor k/n\rfloor}^{-1}\,J_{\lfloor k/n\rfloor}=m_{\lfloor k/n\rfloor}
\end{align*}
as before. We obtain $F_{0}^{(n)}$ as
\begin{align*}
F_{0}^{(n)}(J): & =\ln\int d^{n\,N}\phi\,e^{S_{0}^{(n)}(\phi)+\frac{1}{n}J^{(n)\T}\phi}\\
 & =-\frac{1}{2}\,\sum_{k=1}^{n\,N}\,\ln(G_{\lfloor k/n\rfloor}/n)+\frac{1}{2n}\,\sum_{k=n}^{n\,N}\,J_{\lfloor k/n\rfloor}^{2}\,G_{\lfloor k/n\rfloor}^{-1}\,.
\end{align*}
The source-independent term $-\frac{1}{2}\,\sum_{k=1}^{n\,N}\,\ln(G_{\lfloor k/n\rfloor})$
has changed by approximately a factor $n$, because the function $G_{\mu}$
is sampled in the same range as before, but at $n$ additional intermediate
points. The source-dependent term of the free energy, however, is
the same as in the original system, thus it satisfies \eqref{eq:F_correspondence},
as desired.

\subsection{Non-Gaussian terms}\label{sec:continuum-limit-Non-Gaussian-terms}

Next, consider how non-Gaussian corrections in the original system
transfer to corresponding corrections in the fine-grained system.
We aim to show by induction in the number of interaction vertices
that the full free energy \eqref{eq:free_energy_full} maintains the
property \eqref{eq:F_correspondence}.

To this end, we follow the same steps as in the inductive proof of
the linked cluster theorem found in many text books (e.g., Zinn-Justin
or Kleinert), in the concrete form as presented in Kuehn \& Helias
2018, J Phys A, (Appendix A.3 https://iopscience.iop.org/article/10.1088/1751-8121/aad52e/pdf).

We here use this proof to show by induction in the number of interaction
vertices that all terms in $F^{(n)}$ share the property \eqref{eq:F_correspondence}.
The induction start is given by the result of the previous section,
the case of no interaction vertex. To prepare the induction step for
the fine-grained theory, we first derive what happens in the original,
discrete theory. So rewrite the full free energy of the interacting
system given by \eqref{eq:free_energy_full} as
\begin{align*}
e^{F(J)} & =\int d^{N}\phi\,e^{S_{\mathrm{int}}(\phi)+S_{0}(\phi)+J^{\T}\phi}\\
 & =e^{S_{\mathrm{int}}(\frac{\partial}{\partial J})}\,\int d^{N}\phi\,e^{S_{0}(\phi)+J^{\T}\phi}\\
 & =e^{S_{\mathrm{int}}(\frac{\partial}{\partial J})}\,e^{F_{0}(J)}\,.
\end{align*}
Next rewrite $e^{S_{\mathrm{int}}}=\lim_{K\to\infty}\big[1+\frac{1}{K}\,S_{\mathrm{int}}\big]^{K}$
and multiply the last expression from left with $e^{-F_{0}(J)}$

\begin{align}
e^{F(J)-F_{0}(J)} & =\lim_{K\to\infty}e^{-F_{0}(J)}\,\big[1+\frac{1}{K}\,S_{\mathrm{int}}(\frac{\partial}{\partial J})\big]^{K}\,e^{F_{0}(J)}.\label{eq:exp_as_limit}
\end{align}
The left hand side has in the exponent the difference of the free
energies of the full and the non-interacting system, the right hand
side produces all contributions to this difference as a sum of connected
graphs. The idea of the proof is to decompose
\begin{align}
\big[1+\frac{1}{K}\,S_{\mathrm{int}}\big]^{K} & =\big[1+\frac{1}{K}\,S_{\mathrm{int}}\big]\cdots\big[1+\frac{1}{K}\,S_{\mathrm{int}}\big]\,\label{eq:prod_operators}
\end{align}
into a product of $K$ identical factors (differential operators)
and track the additional diagrams produced by each factor. Each factor
contains $S_{\mathrm{int}}$ in linear power, so that each such term
has the potential to add one interaction vertex to the final result.
The induction step now considers one fixed value of $K$ and the step
is the application of the $k+1-$st of the $K$ factors in \eqref{eq:prod_operators}. 

For the induction step we assume we have applied $k$ factors of the
form $1+\frac{1}{K}\,S_{\mathrm{int}}$ and have already obtained
the free energy $F^{k}$ defined as
\begin{align}
e^{F^{k+1}(J)} & :=\big[1+\frac{1}{K}\,S_{\mathrm{int}}\big]^{k+1}\,e^{F_{0}(J)}\label{eq:exp_as_power}\\
 & =\big[1+\frac{1}{K}\,S_{\mathrm{int}}\big]\,e^{F^{k}(J)}\,.\nonumber 
\end{align}
Multiplying from left with $e^{-F^{k}(J)}$ and taking the logarithm
we have

\begin{align*}
F^{k+1}(J)-F^{k}(J) & =\ln\,\big[1+\frac{1}{K}\,e^{-F^{k}(J)}\,S_{\mathrm{int}}(\frac{\partial}{\partial J})\,e^{F^{k}(J)}\big]\,.
\end{align*}
Since we need to take the limit $K\to\infty$ ultimately in \eqref{eq:exp_as_limit},
we may expand the logarithm $\ln(1+K^{-1}\ldots)=K^{-1}\ldots$. This
is no additional approximation, because it is easy to show that the
omitted terms $\order(K^{-2})$ vanish in the limit. So each induction
step produces additional terms of the form
\begin{align}
F^{k+1}(J)-F^{k}(J) & =\frac{1}{K}\,e^{-F^{k}(J)}\,S_{\mathrm{int}}(\frac{\partial}{\partial J})\,e^{F^{k}(J)}\,.\label{eq:induction_step}
\end{align}
The right hand side, on the other hand, is the expectation value $\langle S_{\mathrm{int}}\rangle$
computed with the theory $F^{k}$. This is the result obtained by
the original discrete theory.

Now consider the same steps for the fine-grained theory. From \eqref{eq:replacement_rule_derivative}
we have the replacement $\frac{\partial}{\partial J_{\mu}}\rightarrow\sum_{i:\,\lfloor i/n\rfloor=\mu}\frac{\partial}{\partial J_{i}^{(n)}}$
which, applied to the free energy obeying \eqref{eq:F_correspondence}
yields identical results as the discrete theory. The induction assumption
is that $F^{(n),k}(J^{(n)})$ has this property (which is true at
induction start, as shown in \prettyref{sec:fine_grained_Gaussian}).
The step corresponding to \eqref{eq:induction_step} for the fine-grained
theory thus reads

\begin{align}
F^{(n),k+1}(J^{(n)})-F^{(n),k}(J^{(n)}) & =\frac{1}{K}\,e^{-F^{(n),k}(J^{(n)})}\,S_{\mathrm{int}}(\{\sum_{i:\,\lfloor i/n\rfloor=\mu}\frac{\partial}{\partial J_{i}^{(n)}}\})\,e^{F^{(n),k}(J^{(n)})}\,,\label{eq:induction_step-1}
\end{align}
which on the right hand side produces the same terms as in the original
theory \eqref{eq:induction_step} due to the property \prettyref{eq:replacement_rule_deriv_source}.
As a consequence, all additional terms corresponding to the difference
$F^{(n),k+1}-F^{(n),k}$ thus have the same value as in the original
theory. By induction the full free energy then has the properties
required in \prettyref{sec:fine_grained_Gaussian}.

As an example consider that the interaction be given by \eqref{eq:S_int}.
By the replacement rule \eqref{eq:replacement_rule_derivative} and
the source term $n^{-1}J^{(n)}\phi$, we have that each derivative
$\sum_{i:\,\lfloor i/n\rfloor=\mu}\frac{\partial}{\partial J_{i}^{(n)}}$
leads to $n^{-1}\,\sum_{i:\,\lfloor i/n\rfloor=\mu}\phi_{i}$, so
the interaction term takes the form 
\begin{align*}
S_{\mathrm{int}}(\phi) & =\frac{1}{n^{4}}\,\sum\sum_{i,j,k,l=1}^{n\,N}V_{\lfloor i/n\rfloor,\lfloor j/n\rfloor,\lfloor k/n\rfloor,\lfloor l/n\rfloor}\,\phi_{i}\phi_{j}\phi_{k}\phi_{l}\,,
\end{align*}
so that the additional perturbative corrections are to first order
in $V$
\begin{align*}
F_{\mathrm{int}}^{(n),1} & =\langle S_{\mathrm{int}}\rangle_{\phi\sim e^{S_{0}^{(n)}+\frac{1}{n}J^{\T}\phi}}\\
 & =\frac{1}{n^{4}}\,\sum_{i,j,k,l=1}^{N}V_{\lfloor i/n\rfloor,\lfloor j/n\rfloor,\lfloor k/n\rfloor,\lfloor l/n\rfloor}\,\langle\phi_{i}\phi_{j}\phi_{k}\phi_{l}\rangle_{\phi\sim e^{S_{0}^{(n)}+\frac{1}{n}J^{\T}\phi}}\,.
\end{align*}
For any function $V_{\lfloor i/n\rfloor,\lfloor j/n\rfloor,\lfloor k/n\rfloor,\lfloor l/n\rfloor}$,
a contraction with the mean of the field $\langle\phi_{k}\rangle$
together with the $n$-fold summation and the $1/n$ factor yields
the same contribution as in the original system. Each contraction
$\langle\phi_{k}\phi_{l}\rangle_{c}$ yields an $n\delta_{kl}$, which
eliminates one sum and one factor $1/n$, so that the contribution
is again the same as in the original system. Concretely, we obtain
\begin{align}
F_{\mathrm{int}}^{(n),1} & =\frac{1}{n^{4}}\,\sum_{i,j,k,l=n}^{n\,N}V_{\lfloor i/n\rfloor,\lfloor j/n\rfloor,\lfloor k/n\rfloor,\lfloor l/n\rfloor}\,\label{eq:expectation_V}\\
 & \big[m_{i}\,m_{j}\,m_{k}\,m_{l}\nonumber \\
 & +c_{ij}^{(n)}\,c_{kl}^{(n)}+2\,\text{permutations}\nonumber \\
 & +c_{ij}^{(n)}\,m_{k}m_{l}+5\,\text{permutations}\big]\,.\nonumber 
\end{align}
The result is thus the same as in the original system and in particular
all terms are intensive in $n$, so that the limit $n\to\infty$ can
be taken.

We may also consider the case that the $V_{ijkl}$ in the original
discrete system is partially diagonal, for example $V_{\mu\mu^{\prime}\,\nu\nu^{\prime}}=V^{(1)}\,\delta_{\mu\mu^{\prime}}\,\delta_{\nu\nu^{\prime}}\,$.
Such a constraint may meet a corresponding constraint from the covariance
in the Wick contraction, $\delta_{\mu\mu^{\prime}}^{2}=\delta_{\mu\mu^{\prime}}$.
In the fine-grained system, by the rule \eqref{eq:replacement_rule_derivative},
we get
\begin{align}
\sum_{\mu,\mu^{\prime}}\delta_{\mu\mu^{\prime}}\ldots & \to\sum_{\mu\mu^{\prime}}\,\delta_{\mu\mu^{\prime}}\,\sum_{i:\,\lfloor i/n\rfloor=\mu}\,\sum_{j:\,\lfloor j/n\rfloor=\mu^{\prime}}\ldots\nonumber \\
 & =\sum_{\mu}\,\sum_{i:\,\lfloor i/n\rfloor=\mu}\,\sum_{j:\,\lfloor j/n\rfloor=\mu}\ldots\,\nonumber \\
 & =\sum_{i}\,\sum_{j}\,\delta_{\lfloor i/n\rfloor\lfloor j/n\rfloor}\ldots\,.\label{eq:Kronecker_replacement}
\end{align}
We may interpret the last line as treating the Kronecker $\delta$
by point-splitting, namely decomposing each discrete interval into
$n$ sub-intervals and assigning a one if $\lfloor i/n\rfloor=\lfloor j/n\rfloor$;
so this corresponds to a particular form of smearing out the Kronecker
$\delta$. The contribution of the corresponding sums yields for a
contraction $\langle\phi_{i}\phi_{j}\rangle_{c}=n\,\delta_{ij}\,c_{i/n}$,
so 
\begin{align*}
 & \frac{1}{n^{2}}\sum_{i,j=n}^{n\,N}\,\delta_{\lfloor i/n\rfloor\lfloor j/n\rfloor}\,n\,\delta_{ij}\,c_{\lfloor i/n\rfloor}\\
= & \frac{1}{n}\,\sum_{i=n}^{n\,N}\,c_{\lfloor i/n\rfloor}=\sum_{\mu=1}^{N}\,c_{\mu}\,,
\end{align*}
which hence yields the value in the original system. A contraction
with two mean values, correspondingly, is
\begin{align*}
 & \frac{1}{n^{2}}\sum_{i,j=n}^{n\,N}\,\delta_{\lfloor i/n\rfloor\lfloor j/n\rfloor}\,m_{\lfloor i/n\rfloor}\,m_{\lfloor j/n\rfloor}\\
= & \sum_{\mu=1}^{N}\,m_{\mu}^{2}\,,
\end{align*}
The result is thus again the same as in the original system. This
example shows that also partially diagonal forms of interaction vertices
are treated correctly by the derived rules of discretization.

\subsection{Final form of action in the continuum}

In conclusion, we may take the limit $n\to\infty$ and replace $\frac{1}{n}\sum_{\mu=1}^{N}\,\sum_{i:\,\lfloor i/n\rfloor=\mu}=\frac{1}{n}\,\sum_{i=n}^{nN}\stackrel{n\to\infty}{\to}\int_{1}^{N}\,dk$
with $k=i/n$ to write the action as a functional
\begin{align}
S_{0}[\phi] & =-\frac{1}{2}\,\int_{1}^{N}dk\,G(k)\,\phi^{2}(k)\label{eq:continuum_final}\\
S_{\mathrm{int}}[\phi] & =\int_{1}^{N}dk_{1}\cdots\int_{1}^{N}dk_{4}\,V(k_{1},\ldots,k_{4})\,\phi(k_{1})\cdots\phi(k_{4})\,,\nonumber \\
J^{\T}\phi & =\int_{1}^{N}dk\,J(k)\,\phi(k)\,.\nonumber 
\end{align}
Here the function $G(k)$ is the interpolation of the original discrete
quadratic form $G_{k}$ and likewise for the function $V(k_{1},\ldots,k_{4})$
and the source term $J(k)$. The first two cumulants of the fields
due to $S_{0}$ are
\begin{align}
\langle\phi(k)\rangle & =G(k)^{-1}\,J(k)\,,\nonumber \\
\langle\phi(k)\phi(l)\rangle_{c} & =\delta(k-l)\,G(k)^{-1}\,.\label{eq:first_two_cum_S0}
\end{align}

The point-splitted Kronecker $\delta$ \eqref{eq:Kronecker_replacement}
$\delta_{\lfloor i/n\rfloor\,\lfloor i^{\prime}/n\rfloor}$ in the
limit $n\to\infty$ ensures that $i/n=:k$ and $i^{\prime}/n=k^{\prime}$
may only differ by their non-integer part, so we replace
\begin{align*}
\delta_{\lfloor i/n\rfloor\,\lfloor i^{\prime}/n\rfloor} & =\delta_{\lfloor k\rfloor,\lfloor k^{\prime}|}\,.
\end{align*}
We employ this replacement rule for interactions $V_{ii^{\prime}kk^{\prime}}\propto\delta_{ii^{\prime}}\delta_{kk^{\prime}}\cdots$

\begin{align}
\delta_{ii^{\prime}} & \to\delta_{\lfloor k\rfloor\lfloor k^{\prime}\rfloor}\,.\label{eq:point_splitting}
\end{align}
All contributions to the non-vacuum part (the one that depends on
sources) of $F$ are then identical to those of the discrete system.

Computing expectation values of observables $O(\phi)$ so that they
agree to their original value defined by \prettyref{eq:Observable},
by the same argument as used in the derivation of the replacement
rule \prettyref{eq:replacement_rule_derivative}, requires us to replace
any terms of the form $\phi_{k}^{2}$ by the point-splitted ones \prettyref{eq:point_splitting}.
For example with $k=l/n$
\begin{align}
\langle\phi_{l}^{2}\rangle & =\sum_{l^{\prime}}\,\delta_{ll^{\prime}}\,\langle\phi_{l}\phi_{l^{\prime}}\rangle\label{eq:Kronecker_replace}\\
 & \to\int\,dk^{\prime}\,\delta_{\lfloor k\rfloor\lfloor k^{\prime}\rfloor}\,\langle\phi(k)\phi(k^{\prime})\rangle\nonumber \\
 & =\int_{\lfloor k\rfloor}^{\lfloor k\rfloor+1}\,dk^{\prime}\,\langle\phi(k)\phi(k^{\prime})\rangle\,.\nonumber 
\end{align}
In the free theory this for example yields with $\langle\phi(k)\phi(k^{\prime})\rangle=\delta(k-k^{\prime})\,G(k)^{-1}$
\begin{align*}
\langle\phi_{l}^{2}\rangle & \to G^{-1}(k)
\end{align*}
a finite result, as it has to be.

Applied to the problem at hand \eqref{eq:action_EK} including the
interaction term \eqref{eq:action_eigenspace_F} the action is

\begin{align}
S(f,y) & =-\frac{1}{2}\int_{1}^{\Lambda}\,\frac{P}{\kappa}\,(y(k)-f(k))^{2}+\frac{f(k)^{2}}{\lambda(k)}\,dk\label{eq:action_cont_EK}\\
 & -UP\,\Big[\int_{1}^{\Lambda}\,\int_{\lfloor k\rfloor}^{\lfloor k\rfloor+1}\,\big(y(k)-f(k)\big)\,\big(y(k^{\prime})-f(k^{\prime})\big)\,dk\,dk^{\prime}\,\Big]^{2}\,.\nonumber 
\end{align}
Written in terms of the discrepancies $\Delta(k):=y(k)-f(k)$ this
is

\begin{align}
S(\Delta,y) & =-\frac{1}{2}\int\,\frac{P}{\kappa}\,\Delta^{2}(k)+\frac{\big(y(k)-\Delta(k)\big)^{2}}{\lambda(k)}\,dk\label{eq:action_cont_as_Delta}\\
 & -UP\,\Big[\int_{1}^{\Lambda}\,\int_{\lfloor k\rfloor}^{\lfloor k\rfloor+1}\,\Delta(k)\,\Delta(k^{\prime})\,dk\,dk^{\prime}\,\Big]^{2}\,.\nonumber 
\end{align}
For the perturbative computation of the RG equations we need the mean
and covariance of the Gaussian part which are
\begin{align}
m(k) & :=\langle f(k)\rangle=\frac{\lambda(k)}{\lambda(k)+\kappa/P}\,y(k)\,,\nonumber \\
\nonumber \\d(k):= & \langle\Delta(k)\rangle=y(k)-\langle f(k)\rangle\label{eq:mean_Delta_cont}\\
 & =\frac{\kappa/P}{\lambda(k)+\kappa/P}\,y(k)\,,\nonumber \\
\nonumber \\c(k,l) & =\langle f(k)f(l)\rangle^{c}=\langle\Delta(k)\Delta(l)\rangle^{c}\label{eq:cov_Delta_cont}\\
 & =\delta(k-l)\,\frac{\kappa/P\,\lambda(k)}{\lambda(k)+\kappa/P}\,.\nonumber 
\end{align}

\subsection{Non-diagonal correction terms to the quadratic part}\label{sec:Non-diagonal-correction-terms}

A final sublety arises when computing perturbative corrections to
the Gaussian part that stem from an interaction vertex that is partially
diagonal in the original discrete system and has been replaced by
\prettyref{eq:Kronecker_replacement}. If these fields remain non-contracted
they may constitute a Gaussian term to the action which, however,
is not perfectly diagonal, but only diagonal in a point-splitted manner,
namely of the form

\begin{align}
S_{2}(\phi) & =\frac{a}{2}\,\int_{1}^{\Lambda}dk\,\int_{\lfloor k\rfloor}^{\lfloor k\rfloor+1}dk^{\prime}\,\phi(k)\,\phi(k^{\prime})\,.\label{eq:non_diag}
\end{align}
We will here show that these may indeed be replaced by properly diagonal
terms. To see this, remember that all observables of interest originate
from the discrete system, so that they can be written in terms of
pairs of the fields that come with momenta closeby in the sense
\begin{align}
\int_{1}^{\Lambda}dk\,\int_{\lfloor k\rfloor}^{\lfloor k\rfloor+1}dk^{\prime}\,\phi(k)\,\phi(k^{\prime})\,,\label{eq:off_diag_obs}
\end{align}
where hence the momenta $k^{\prime}$ are summed over within range
$[\lfloor k\rfloor,\lfloor k\rfloor+1]$. We will now show that a
non-diagonal term such as \eqref{eq:non_diag} can be absorbed into
an effectively diagonal term without changing any expectation value
of either an observable or a perturbative correction term.

To see this, we again move to the discretized version of \eqref{eq:non_diag}
together with a diagonal Gaussian part $S_{0}^{(n)}$

\begin{align}
S & =S_{0}^{(n)}+S_{2}^{(n)}\label{eq:S_0_and_S_2}\\
S_{0}^{(n)}(\phi) & =-\frac{1}{2n}\,\sum_{k=n}^{n\,N}G_{\lfloor k/n\rfloor}\phi_{k}^{2}\,,\nonumber \\
S_{2}^{(n)}(\phi) & =\frac{a}{2n}\sum_{i=n}^{n\,N}\,\frac{1}{n}\,\sum_{j=n\,\lfloor i/n\rfloor}^{n\,\lfloor i/n\rfloor+n}\,\phi_{i}\,\phi_{j}\,.\nonumber 
\end{align}
The resulting quadratic part is therefore block-diagonal with blocks
that are homogeneous. Within one block of $n$ fine-grained modes
there are $n\times n$ matrices of the form
\begin{align*}
\bar{G}_{\lfloor k/n\rfloor} & =G_{\lfloor k/n\rfloor}\,\bI-\frac{a}{n}\,1\,,
\end{align*}
where $1$ is an $n\times n$-matrix of all ones. An expectation value
of a term of the form \eqref{eq:off_diag_obs} leads to expressions
\begin{align}
\langle I\rangle= & \sum_{i=n}^{n\,N}\,\sum_{j=n\,\lfloor i/n\rfloor}^{\lfloor i/n\rfloor+n}\,\langle\phi_{i}\,\phi_{j}\rangle=\sum_{i=n}^{n\,N}\,\sum_{j=n\,\lfloor i/n\rfloor}^{n\,\lfloor i/n\rfloor+n}\,[\bar{G}^{-1}]_{ij}\label{eq:exp_non_diag}\\
= & \sum_{i=n}^{n\,N}\,[\bar{G}^{-1}\,{\bf 1}^{(i)}]_{i}\,,\nonumber 
\end{align}
where
\begin{align*}
{\bf 1}^{(i)} & =(0,\ldots,\overbrace{\underbrace{1}_{n\,\lfloor i/n\rfloor-\text{th position}},\ldots,1}^{n\text{ ones}},0,\ldots0)\,.
\end{align*}
To obtain the inverse $\bar{G}^{-1}$ applied to ${\bf 1}^{(i)}$,
we need solutions ${\bf x}$ to equations of the form
\begin{align}
\bar{G}\,{\bf x}^{(i)} & ={\bf 1}^{(i)}\,.\label{eq:lineq}
\end{align}
Now observe that by
\begin{align*}
\bar{G}\,{\bf 1}^{(i)} & =\big[G_{\lfloor i/n\rfloor}-a\big]\,{\bf 1}^{(i)}\,,
\end{align*}
the ${\bf 1}^{(i)}$ are eigenvectors of $\bar{G}$, so that the solution
${\bf x}^{(i)}$ of \eqref{eq:lineq} is
\begin{align}
{\bf x}^{(i)} & =\big[G_{\lfloor i/n\rfloor}-a\big]^{-1}\,{\bf 1}^{(i)}\,\label{eq:x_sol}
\end{align}
and \eqref{eq:exp_non_diag} becomes
\begin{align*}
\langle I\rangle & =\sum_{i=n}^{n\,N}\,{\bf x}_{i}^{(i)}=n\,\big[G_{\lfloor i/n\rfloor}-a\big]^{-1}\,.
\end{align*}
The result for an expectation value of the form \eqref{eq:exp_non_diag}
in the presence of both terms $S_{0}$ and $S_{2}$ in \eqref{eq:S_0_and_S_2}
will hence be the same as produced by a diagonal action of the form
\begin{align}
\tilde{S}_{2}^{(n)} & =\frac{1}{2n}\,\sum_{i=n}^{n\,N}\,\big[G_{\lfloor k/n\rfloor}-a\big]\,\phi_{i}^{2}\,.\label{eq:S_tilde}
\end{align}
So far we neglected the presence of a source term \eqref{eq:source_fine_grained}
$\frac{1}{n}\,\sum_{k=n}^{n\,N}J_{\lfloor k/n\rfloor}\,\phi_{k}$,
which also resides in the same subspace spanned by the ${\bf 1}^{(i)}$.
The mean of the Gaussian part of the theory
\begin{align*}
S & =S_{0}+S_{2}+\frac{1}{n}\,\sum_{k=n}^{n\,N}J_{\lfloor k/n\rfloor}\,\phi_{k}\,
\end{align*}
is given by its stationary point
\begin{align*}
m & =\sum_{k=n}^{nN}J_{\lfloor k/n\rfloor}\,\bar{G}^{-1}{\bf 1}^{(k)}=\sum_{k=n}^{nN}J_{\lfloor k/n\rfloor}\,{\bf x}^{(k)}\,,
\end{align*}
which is a superposition of solutions ${\bf x}^{(k)}$ \eqref{eq:x_sol}.
Such solutions are identical whether we consider \eqref{eq:S_0_and_S_2}
or \eqref{eq:S_tilde}, so we may use the latter instead of the former,
as before. Since the mean is the same in both cases, all expectation
values involving this mean turn out to be identical.

In summary, this shows that we may replace terms that are diagonal
only in an $n$-vicinity, such as \prettyref{eq:non_diag}, by properly
diagonal term.

\section{Renormalization}\label{sec:Renormalization}

\subsection{Idea of the renormalization procedure}

This section recapitulates the main idea of renormalization in general
terms, before applying it to the problem of interest in the following
sections.

Quantities of interest follow from the free energy $F(\Theta)$, which
is a function of the parameters $\Theta$; in the example of non-linear
regression, these are $\Theta=\{r=P/\kappa,U,s\}$, where $s$ will
be a scale factor of the action to be introduced later. The free energy
can be defined as
\begin{align}
F(\Theta) & :=\ln\int\cD f\,\exp\big(S(f;\Theta)\big)\,,\label{eq:free_energy_orig}
\end{align}
where we denote the parameter dependence as arguments.

The idea is to split the degrees of freedom into two parts and to
perform the integration over one part only. To this end, we first
introduce an upper cutoff $\Lambda$ into the integrals
\begin{align*}
\int_{1}^{\infty}dk & \to\int_{1}^{\Lambda}dk\,.
\end{align*}
We will find that the theories considered here become close to Gaussian
for large $l$, so that the contributions from the finiteness of the
cutoff can be accounted for.

We here split the degrees of freedom in terms of the eigenmodes $f=(f_{<},f_{>})$,
where $f_{<}=f_{1\le k\le\Lambda/\ell}$ and $f_{>}=f_{\Lambda/\ell<k\le\Lambda}$.
It then holds that
\begin{align*}
F(\Theta) & \equiv\ln\int\cD f_{<}\,\int\cD f_{>}\,\exp\big(S(f_{<}+f_{>};\Theta)\big)\\
 & =\ln\int\cD f_{<}\,\exp\Big[\ln\,\int\cD f_{>}\,\exp\big(S(f_{<}+f_{>};\Theta)\big)\Big]\,,
\end{align*}
which motivates the definition of the action for the lower degrees
of freedom $f_{<}$ as the partial free energy of the higher degrees
of freedom
\begin{align*}
S_{<}(f_{<};\Theta) & :=\ln\,\int\cD f_{>}\,\exp\big(S(f_{<}+f_{>};\Theta)\big)\,.
\end{align*}
The partition function $e^{F}$ in both representations, by construction,
stays the same, namely
\begin{align}
\exp(F(\Theta)) & =\int\cD f\,\exp\big(S(f;\Theta)\big)\label{eq:equal_partition_function}\\
 & =\int\cD f_{<}\,\exp\big(S_{<}(f_{<};\Theta)\big)\,.\nonumber 
\end{align}
Let us now assume we had found a form of $S_{<}$ that is identical
to the form of the original $S$, but possibly with changed degrees
of freedom $f_{<}\to f^{\prime}$ and changed parameters $\Theta\to\Theta^{\prime}$.
The decimation step may also have produced terms $G(\Theta)$ that
are independent of the low degrees of freedom, but are still functions
of the parameters $\Theta$, respectively. We therefore have the action
\begin{align}
S[f^{\prime},\Theta^{\prime}] & +\ln G(\Theta)\,,\nonumber \\
\int\cD f_{<}\,\exp\big(S_{<}(f_{<};\Theta)\big) & =\int\cD f^{\prime}\,\exp\big(S(f^{\prime},\Theta^{\prime})+\ln G(\Theta^{\prime})\,\big)\,.\label{eq:decimation_rescaling_identity}
\end{align}
Because the functional form $S$ is the same as before and also the
integral boundaries are the same as before (after rescaling), the
integral on the right
\begin{align*}
F(\Theta^{\prime}) & =\ln\,\int\cD f^{\prime}\,\exp\big(S(f^{\prime},\Theta^{\prime})\big)\,
\end{align*}
is denoted by the same function $F(\Theta^{\prime})$ as the original
free energy \eqref{eq:free_energy_orig}. It follows from \eqref{eq:decimation_rescaling_identity}
that
\begin{align}
F(\Theta) & =F(\Theta^{\prime}(\ell))+G(\Theta^{\prime}(\ell))\,.\label{eq:free_energy_scaling}
\end{align}
Determining the dependencies of all parameters $\Theta^{\prime}(\ell)$
as functions of the coarse-graining variable $\ell$ allows us to
compute the free energy $F$ (and therefore the observables as its
derivatives) from any value of $\ell$ on the right hand side of \eqref{eq:free_energy_scaling}
that we like.

\subsection{Decimation step of RG: Gaussian part}

The following section will make the conceptual steps outlined in the
previous section concrete for the case of linear regression: We would
like to know how the parameters $\Theta=\{r:=P/\kappa,s\}$ change
as a function of the coarse-graining scale $\ell$. To make the analogy
to the usual RG procedure, the eigenvalue $\lambda(k)^{-1}$ here
plays the role of the kinetic term proportional to momentum squared;
so large $\lambda$ correspond to the nearly critical modes with momenta
close to zero. Likewise, $P/\kappa$ plays the role of a mass term,
limiting fluctuations of the field $f(k)$ when $\lambda(k)^{-1}$
becomes small. The linear term $\propto y(k)\,f(k)$ plays the role
of an external magnetic field. In this analogy, one would perform
the RG flow, starting at a high $k$ cutoff $\Lambda$ where one has
\begin{align}
\lambda(\Lambda)^{-1} & \gg r=P/\kappa\,,\label{eq:modes_above_cutoff}
\end{align}
so fluctuations are limited by the spectrum $\lambda^{-1}$ rather
then the regulator $\kappa$. As one progresses to smaller $k$, $\lambda^{-1}(k)$
declines and ultimately the mass term $P/\kappa$ limits fluctuations.
As long as one is sufficiently far in the UV regime, so that \eqref{eq:modes_above_cutoff}
holds, the presence of the regulator does not matter. To find the
point at which the regulator matters, we look for the $k_{\mathrm{min}}$
so that the two terms are of similar magnitude using the power-law
dependence \prettyref{eq:power_law_rescaled} of the eigenmodes

\begin{align}
r & \stackrel{!}{=}\lambda^{-1}(k_{\mathrm{min}})\,,\nonumber \\
P/\kappa=r & =k_{\mathrm{min}}^{1+\alpha}\,,\nonumber \\
k_{\mathrm{min}} & =\big[\frac{P}{\kappa}\big]^{\frac{1}{1+\alpha}}\,.\label{eq:l_min}
\end{align}
The factor $\frac{P}{\kappa}\gg1$, because we want to regularize
weak modes and the exponent $1+\alpha$ is in the vicinity of $1$,
so also $k_{\mathrm{min}}\gg1$. This is the effective low momentum
cutoff where the scaling region ends is thus typically above the strict
cutoff $k=1$, which is the lower bound of the integral in the action
\prettyref{eq:action_cont_EK} after rescaling.

We may wish to integrate out the high momentum modes $f_{>}$ for
$k>\Lambda/\ell$ with $\ell>1$ to obtain the action $S_{<}$ for
the remaining modes for $k<\Lambda/\ell$, splitting the functions
$f=f_{<}+f_{>}$ as well as $y=y_{<}+y_{>}$ into their low and high
momentum parts
\begin{align}
S_{<}(f_{<},y_{<}) & =\ln\int\cD f_{>}\,\exp\big(S(f_{<}+f_{>},y_{<}+y_{>})\big)\,.\label{eq:decimation}
\end{align}
The Gaussian part of the action \eqref{eq:action_cont_EK} can be
written in terms of linear and quadratic coefficients for the corresponding
mode $f(l)$, so we rewrite is as
\begin{align}
S[f,y]= & \int_{1}^{\Lambda}\,\big(-\frac{1}{2}\big)\,\big[r+\lambda(k)^{-1}\big]\,f(k)^{2}+r\,y(k)\,f(k)\label{eq:action_quad_lin}\\
- & \frac{1}{2}r\,y(k)^{2}\nonumber 
\end{align}
where the second line contains the term that is independent of the
field $f$ and hence does not influence its statistics.

Since the action is diagonal in $k$, it splits into high and low
modes (they are uncoupled)
\begin{align*}
S(f_{<}+f_{>},y_{<}+y_{>}) & =S(f_{<},y_{<})+S(f_{>},y_{>})\,.
\end{align*}
The decimation step \eqref{eq:decimation} therefore explicitly reads
\begin{align}
S_{<}(f_{<},y_{<}) & =S(f_{<},y_{<})+\ln\int df_{>}\,\exp\big(S(f_{>},y_{>})\big)\,.\label{eq:decimate_decomp}
\end{align}
The latter integral contains the term $-\frac{1}{2}\int_{\Lambda/\ell}^{\Lambda}\,r\,y(k)^{2}\,dk$
of \eqref{eq:action_quad_lin} that does not depend on $f_{>}$ as
well as the Gaussian integral
\begin{align*}
 & \ln\,\int\cD f_{>}\exp\big(\int_{\Lambda/\ell}^{\Lambda}\,\big(-\frac{1}{2}\big)\,\big[r+\lambda(k)^{-1}\big]\,f_{>}(k)^{2}+r\,y_{>}(k)\,f_{>}(k)\,dk\big)\\
 & =-\frac{1}{2}\int_{\Lambda/\ell}^{\Lambda}\,\ln\,\big[r+\lambda(k)^{-1}\big]-r^{2}\,y_{>}(k)^{2}\big[r+\lambda(k)^{-1}\big]^{-1}\,dk\,,
\end{align*}
so that we obtain from \eqref{eq:decimate_decomp}
\begin{align}
S_{<}(f_{<},y_{<}) & =S(f_{<},y_{<})+G(y_{>})\label{eq:decimated}\\
G(y_{>},r): & =-\frac{1}{2}\int_{\Lambda/\ell}^{\Lambda}\,\frac{y_{>}(k)^{2}}{\lambda(k)+r^{-1}}\,dk\,,\nonumber 
\end{align}
where we combined the terms in $G(y_{>},r)=-\frac{1}{2}\int_{\Lambda/\ell}^{\Lambda}\,\ln\,\big[r+\lambda(k)^{-1}\big]-r^{2}\,y_{>}(k)^{2}\big[r+\lambda(k)^{-1}\big]^{-1}+r\,y_{>}(k)^{2}\,dk$
so that the result is of similar form as \eqref{eq:free_energy_eigenspace};
in detail the coefficients in front of $y_{>}^{2}$ from the integral
and from the original part of the action are rewritten as $-r^{2}\,\big(\,\big[r+\lambda(l)^{-1}\big]^{-1}+r=\big[\lambda(l)+r^{-1}\big]^{-1}$.
In the next step we need to rescale the momentum range so that the
functional form of the action becomes identical to the beginning --
in particular, both actions needs to map functions $f(k\le\Lambda)$
of identical momentum ranges to the reals.

\subsection{Rescaling of Gaussian part: maintaining the target}\label{sec:Rescaling-maintaining-target}

We are choosing a rescaling that attempts to maintain the target field
$y$ in both systems. For momenta for which \eqref{eq:modes_above_cutoff}
holds, the action for the coarse-grained system \eqref{eq:decimated}
has the same form as for the original system, only with the cutoff
$\Lambda$ replaced by $\Lambda/\ell$. To make the actions comparable,
the range of modes must therefore be rescaled so that after rescaling
the cutoff is again at $\Lambda$. This can be achieved by defining
\begin{align}
k^{\prime} & =\ell\,k.\label{eq:rescaling_momentum}
\end{align}
Likewise we allow for a wavefunction renormalization factor
\begin{align}
z\,f^{\prime}(k^{\prime}) & :=f_{<}(k)\,\label{eq:rescaling_wavefunction}
\end{align}
demanding that the action in terms of the rescaled field be the same
as before
\begin{align*}
S^{\prime}(f^{\prime},y_{<}) & \stackrel{!}{=}S(f_{<},y_{<})\,.
\end{align*}
The left hand side is with \eqref{eq:action_quad_lin} $f_{<}(k)=f_{<}(\ell^{-1}k^{\prime})=z\,f^{\prime}(k^{\prime})$
and rewriting the action in its original form \eqref{eq:action_cont_EK}

\begin{align}
S^{\prime}(f^{\prime},y_{<}) & :=-\frac{s(\ell)}{2}\int_{\ell}^{\Lambda}\,\Big\{ r(\ell)\,(zf^{\prime}(k^{\prime})-y(\ell^{-1}k^{\prime}))^{2}+\frac{z^{2}f^{\prime}(k^{\prime})^{2}}{\lambda(\ell^{-1}k^{\prime})}\Big\}\,\frac{dk^{\prime}}{\ell}\,,\label{eq:rescaled_action-1}
\end{align}
where in addition we introduced an overall scale factor $s(\ell)$
with initial value $s\equiv s(1)\equiv1$ that controls the overall
scale of the fluctuations and $r(\ell)$ is given by $r(1)=P/\kappa$.
The low momentum cutoff has changed from $1$ to $\ell$; this only
means the we need to stop the flow earlier. We will come back to this
point after we know how the effective low momentum cutoff $r$ scales.
Otherwise, the integral over $k^{\prime}$ is now again of the same
form as the original integral over $k$, so we rename $k^{\prime}\to k$
in the following.

Demanding self-similarity in the momentum regime away from the lower
momentum cutoff implied by $r=P/\kappa$, we demand the term that
is quadratic in $y$ to be identical between the original action and
the rescaled one
\begin{align}
\underbrace{s(1)}_{=1}\,r(1)\,y(\ell^{-1}k)^{2}\,\frac{dk}{\ell} & \stackrel{!}{=}s(\ell)\,r(\ell)\,y(k)^{2}\,dk.\label{eq:rescale_y_term}
\end{align}
From the assumed power law \eqref{eq:power_law_y} one has $y(\ell^{-1}k)^{2}=\ell^{1+\beta}\,y(k)$,
so that $r(1)\,\ell^{\beta}\,y(k)^{2}\,dk\stackrel{!}{=}s(\ell)\,r(\ell)\,y(k)^{2}\,dk$
and hence 
\begin{align}
s(\ell)\,r(\ell) & =r(1)\,\ell^{\beta}\,.\label{eq:S_r_scaling}
\end{align}
The term $\propto f$ which is linearly combined with $y$ must scale
identically to $y$, so we need
\begin{align*}
z\,f^{\prime}(k)-y(\ell^{-1}k) & =\big(z\,f^{\prime}(k)-\ell^{\frac{1+\beta}{2}}\,y(k)\big)\,,
\end{align*}
so
\begin{align}
z(\ell) & =\ell^{\frac{1+\beta}{2}}\,.\label{eq:z_rescale}
\end{align}
This choice assures that all terms involving $f$ instead of $y$
scale identically to the terms involving $y$; the latter maintain
its form by the choice of $s(\ell)\,r(\ell$) above.

Considering the last the term $s\,\frac{z^{2}f^{\prime}(k^{\prime})^{2}}{\lambda(\ell^{-1}k^{\prime})}$

\begin{align*}
\underbrace{s(1)}_{1}\,\frac{\ell^{1+\beta}\,f^{\prime}(k)^{2}}{\lambda(\ell^{-1}k)}\,\frac{dk}{\ell} & =\frac{\ell^{\beta}f^{\prime}(k)^{2}}{\ell^{1+\alpha}\,\lambda(k)}\,dk\\
 & \stackrel{!}{=}s(\ell)\,\frac{f^{\prime}(k)^{2}}{\lambda(k)}\,dk\,,
\end{align*}
it follows that
\begin{align}
s(\ell) & =\ell^{\beta-\alpha-1}.\label{eq:s_ell}
\end{align}
Together with \eqref{eq:S_r_scaling} we have
\begin{align*}
r(\ell) & =r(1)\,\ell^{\beta}\,s(\ell)^{-1}\\
 & =r\,\ell^{\beta}\ell^{-\beta+\alpha+1}\\
 & =r\,\ell^{1+\alpha}\,,
\end{align*}
as expected from the mass term $r$ kicking in at some sufficiently
large $\ell$.

Taken together we get the rescaled action (renaming $f^{\prime}$
as $f$ again)
\begin{align}
S^{\prime}(f,y) & :=-\frac{s(\ell)}{2}\int_{\ell}^{\Lambda}\,\Big\{ r(\ell)\,\big[f(k)-y(k)\big]^{2}+\frac{f(k)^{2}}{\lambda(k)}\Big\}\,dk\,,\label{eq:renormed_action-2}\\
s(\ell) & =\ell^{\beta-\alpha-1}\,,\nonumber \\
r(\ell) & =r\,\ell^{1+\alpha}\,.\nonumber 
\end{align}
Splitting the action into the quadratic and the linear part, we have
the form

\begin{align}
S^{\prime}(f,y)= & \int_{\ell}^{\Lambda}\big(-\frac{s(\ell)}{2}\big)\,\big[r(\ell)+\lambda^{-1}(k)\big]\,f(k)^{2}+s(\ell)\,r(\ell)\,y(k)\,f(k)\,dk\label{eq:renormed_action}\\
 & +\const(f)\,,\nonumber 
\end{align}
which more clearly shows that $r(\ell)$ plays the role of a mass
term limiting fluctuations for small $k$, when $\lambda^{-1}$ is
small. The RG procedure leaves the scaling regime when the mass term
and the kinetic term are of similar magnitude, so when \prettyref{eq:l_min}
is fulfilled.

The propagators of the renormalized action \eqref{eq:renormed_action-2}
in the new variables $k^{\prime}=\ell\,k$ are

\begin{align}
d(k^{\prime};\ell) & :=\frac{1}{r(\ell)\,\lambda(k^{\prime})+1}\,y(k^{\prime})\,,\label{eq:mean_prop_rescaled}\\
c(k^{\prime},l^{\prime};\ell) & =\delta(k^{\prime}-l^{\prime})\,s(\ell)^{-1}\,\frac{\lambda(k^{\prime})}{r(\ell)\,\lambda(k^{\prime})+1}\,.\nonumber 
\end{align}

\subsection{Rescaling of the interaction term}\label{sec:Rescaling-of-interaction}

The interaction term in \eqref{eq:action_cont_as_Delta} rescales
under the above rescaling \eqref{eq:rescaling_momentum} $k^{\prime}=\ell\,k$
and \eqref{eq:rescaling_wavefunction} $z\,f^{\prime}(k^{\prime})=f_{<}(k)$
with \eqref{eq:z_rescale} $z(\ell)=\ell^{\frac{1+\beta}{2}}$ as
well as $y(\ell^{-1}k^{\prime})=\ell^{\frac{1+\beta}{2}}\,y^{\prime}(k^{\prime})$
as

\begin{align*}
S_{\mathrm{int}}(f^{\prime},y^{\prime}) & =-UP\,\int_{\ell}^{\Lambda}\frac{dk^{\prime}}{\ell}\,\int_{\ell\lfloor k^{\prime}/\ell\rfloor}^{\ell\lfloor k^{\prime}/\ell\rfloor+\ell}\frac{d\hat{k}^{\prime}}{\ell}\,\int_{\ell}^{\Lambda}\,\frac{dl^{\prime}}{\ell}\,\int_{\ell\lfloor l^{\prime}/\ell\rfloor}^{\ell\lfloor l^{\prime}/\ell\rfloor+\ell}\frac{d\hat{l}^{\prime}}{\ell}z(\ell)^{4}\\
 & \times\Delta^{\prime}(k^{\prime})\,\Delta^{\prime}(\hat{k}^{\prime})\,\Delta^{\prime}(l^{\prime})\,\Delta^{\prime}(\hat{l}^{\prime})\\
 & =-\ell^{2\beta-2}\,UP\,\int_{\ell}^{\Lambda}dk^{\prime}\,\int_{\ell\lfloor k^{\prime}/\ell\rfloor}^{\ell\lfloor k^{\prime}/\ell\rfloor+\ell}d\hat{k}^{\prime}\,\int_{\ell}^{\Lambda}\,dl^{\prime}\,\int_{\ell\lfloor l/\ell\rfloor}^{\ell\lfloor l/\ell\rfloor+\ell}d\hat{l}^{\prime}\\
 & \times\Delta^{\prime}(k^{\prime})\,\Delta^{\prime}(\hat{k}^{\prime})\,\Delta^{\prime}(l^{\prime})\,\Delta^{\prime}(\hat{l}^{\prime})\,.
\end{align*}
The change of the integral boundaries $\int_{\ell\lfloor k^{\prime}/\ell\rfloor}^{\ell\lfloor k^{\prime}/\ell\rfloor+\ell}\ldots$
increases the width of integration by a factor of $\ell$ compared
to before rescaling. 

We here need to distinguish two cases of the scaling, depending on
the smoothness of the integrand.
\begin{enumerate}
\item For a smooth integrand, we may replace $\ell\lfloor k^{\prime}/\ell\rfloor\simeq\lfloor k^{\prime}\rfloor$,
so we replace
\begin{align}
\int_{\ell\lfloor k^{\prime}/\ell\rfloor}^{\ell\lfloor k^{\prime}/\ell\rfloor+\ell}\ldots & \to\ell\,\int_{\lfloor k^{\prime}\rfloor}^{\lfloor k^{\prime}\rfloor+1}\,,\label{eq:replacement_bounds}
\end{align}
which again brings the interaction term to the same form as before
rescaling. So we obtain two additional factors $\ell$, one from each
of these integrals, so that the interaction term rescales as
\begin{align}
U(\ell) & =\ell^{2\beta}\,U\,.\label{eq:U_l}
\end{align}
Since $\beta>0$ this shows that the interaction term is IR relevant.
\item For a non-smooth integrand, for example when the pair of fields $\Delta^{\prime}(l^{\prime}),\Delta^{\prime}(\hat{l}^{\prime})$
or the pair of fields $\Delta^{\prime}(k^{\prime})\,\Delta^{\prime}(\hat{k}^{\prime})$
is contracted by a connected propagator $c$ \prettyref{eq:mean_prop_rescaled}
that contains a Dirac $\delta$ (as in \eqref{eq:contraction_self_energy}),
we cannot treat the extension of the integration interval by $\ell$
for a prefactor $\ell$, so we get the scaling $U(\ell)=\ell^{2\beta-1}\,U$
in this case, if the replacement \eqref{eq:replacement_bounds} can
be made only for a single momentum integral. This is giving rise to
what we call ``scaling intervals'' in the main text. Since this
case appears whenever a contraction with a connected propagator is
performed, in this appendix we take care of this factor by an additional
factor $\ell^{-1}$ in each contraction with $c$ and in return scale
the interaction by its native scaling dimension \prettyref{eq:U_l}.
\end{enumerate}

\subsection{Decimation step: Contribution of the interaction to the flow of the
quadratic part}\label{sec:Decimation-step-quandratic}

The decimation step for the interaction part considers the four-point
vertex \eqref{eq:quartic_interaction_main} written in terms of the
discrepancy $\Delta(k):=y(k)-f(k)$ as

\begin{align}
S_{\mathrm{int}}(\Delta,y)= & -U(\ell)\,P\,\int_{1}^{\Lambda}dk\,\int_{\lfloor k\rfloor}^{\lfloor k\rfloor+1}dk^{\prime}\,\int_{1}^{\Lambda}\,dl\,\int_{\lfloor l\rfloor}^{\lfloor l\rfloor+1}dl^{\prime}\label{eq:interaction_for_perturb}\\
 & \,\Delta(k)\,\Delta(k^{\prime})\,\Delta(l)\,\Delta(l^{\prime})\,.\nonumber 
\end{align}
Integrating out the highest mode at the cutoff $\Lambda$ in a thin
shell we choose $\ell=1+\epsilon$. To linear order in the interaction
vertex, we obtain contributions from the Wick-contractions with means
$\langle\Delta(k,\ell)\rangle$ and covariance $c(k,l,\ell)$ given
by \eqref{eq:mean_prop_rescaled}.

Writing the contracted (high momentum shell) momenta as $>$ and the
uncontracted (low momenta) as $<$, in terms of Wick's theorem, we
obtain the following contributions
\begin{align}
\mathrm{i)}\quad & \langle\Delta(>)\rangle\,\langle\Delta(>)\rangle\,\Delta(<)\Delta(<)\,,\label{eq:Wick_contractions}\\
\mathrm{ii)}\quad & c_{>>}\,\Delta(<)\Delta(<)\,,\nonumber 
\end{align}
where we dropped terms where all momenta are high and contracted,
because they only contribute a constant which does not affect the
statistics of the $f_{<}$ (this constant, however, typically affects
the $y_{>}$-dependence and hence the discrepancies at high momenta,
as in \eqref{eq:decimated}). Also we left out the contractions $\langle\Delta(>)\rangle\,\Delta(<)\,\Delta(<)\Delta(<)$
and $\langle\Delta(>)\rangle\,\langle\Delta(>)\rangle\,\langle\Delta(>)\rangle\,\Delta(<)$
as well as $c_{>>}\,\langle\Delta(>)\rangle\,\Delta(<)$ due to the
constraints on the momenta in the interaction vertex \eqref{eq:interaction_for_perturb},
which imply that there cannot be any contractions where only a single
momentum is low or high, because there must be at least one more momentum
in the low or high regime, respectively.

In addition to the two remaining Wick-contractions \eqref{eq:Wick_contractions},
we need to distinguish contractions in terms of the remaining low-momentum
integrals: Because of the constraint that $k^{\prime}\in[\lfloor k\rfloor,\lfloor k\rfloor+1]$
(and likewise for $l$ and $l^{\prime}$), if such a pair remains
uncontracted, the effective contribution is still diagonal in this
very sense. If, however, a $k$ and an $l$ (or likewise $k^{\prime}$
and $l^{\prime}$) remain uncontracted, their momenta are not necessarily
constrained to be diagonal.

Concretely, for contraction pattern $\mathrm{i)}$ we have only a
single variant
\begin{align}
\mathrm{i)}\quad & \langle\Delta(k_{>})\rangle\,\langle\Delta(k_{>}^{\prime})\rangle\,\Delta(l_{<})\Delta(l_{<}^{\prime})\,,\label{eq:contrib_i_pert}
\end{align}
because the other possibility $\langle\Delta(k_{>})\rangle\,\langle\Delta(l_{>})\rangle\,\Delta(k_{<}^{\prime})\Delta(l_{<}^{\prime})$
does not appear: if $k_{>}$ is high then also $k_{>}^{\prime}$ would
need to be high.

Considering contraction pattern $\mathrm{ii)}$, we have
\begin{align}
\mathrm{ii)}\quad & c(k_{>},k_{>}^{\prime})\,\Delta(l_{<})\Delta(l_{<}^{\prime})\,,\label{eq:contrib_ii_pert}
\end{align}
where again the other arrangement of momenta $c(k_{>},l_{>})\,\Delta(k_{<}^{\prime})\Delta(l_{<}^{\prime})$
cannot appear, because $k$ and $k^{\prime}$ both need to be high
or low at the same time. We are hence left with the contribution \eqref{eq:contrib_i_pert}
(with a factor $2$, because we may contract the pair $k,k^{\prime}$
or the pair $l,l^{\prime}$)
\begin{align}
\mathrm{i)}\quad & 2\cdot\,\Diagram{\vertexlabel^{l}\;hd & gu & \vertexlabel^{k=\Lambda}\\
\vertexlabel^{l^{\prime}}\;hu & gd & \vertexlabel^{k^{\prime}=\Lambda}
}
\label{eq:i_correction}\\
 & =-2\,U(\ell)\,P\,\langle\Delta(\Lambda,\ell)\rangle\,\langle\Delta(\Lambda,\ell)\rangle\,\int_{1}^{\Lambda/(1+\epsilon)}\,dl\,\int_{\lfloor l\rfloor}^{\lfloor l\rfloor+1}dl^{\prime}\,\Delta(l)\Delta(l^{\prime})\,,
\end{align}
where we denote the appearance of the mean as $\langle\Delta(\Lambda,\ell)\rangle=\feyn{g}$.

The other contribution \eqref{eq:contrib_ii_pert} yields
\begin{align}
\mathrm{ii)}\quad & 2\cdot\,\Diagram{\vertexlabel^{l}\;hd & f0flfluf0\\
\vertexlabel^{l^{\prime}}\;hu
}
\label{eq:ii_correction}\\
= & -2\,U(\ell)\,P\,\langle\Delta(\Lambda,\ell)\Delta(\Lambda,\ell)\rangle_{c}\,\int_{1}^{\Lambda/(1+\epsilon)}\,dl\,\int_{\lfloor l\rfloor}^{\lfloor l\rfloor+1}dl^{\prime}\,\Delta(l)\Delta(l^{\prime})\,.
\end{align}

So both terms contribute to the quadratic part of the action as they
are proportional to $\Delta^{2}$.

The difference of the new terms compared to those that preexisted
in the action is the non-diagonal quadratic interaction. It is, however,
only non-diagonal within an interval that belongs to the same discrete
index. When including such a new term in the action, the covariance
will hence be affected. As shown in \prettyref{sec:Non-diagonal-correction-terms}
such a term can effectively can be absorbed into a diagonal term if
all observables and all future perturbative corrections of interest
only depend on expectation values of the form $\int_{1}^{\Lambda/(1+\epsilon)}\,dk\,\int_{\lfloor k\rfloor}^{\lfloor k\rfloor+1}dk^{\prime}\,\langle\Delta(k)\Delta(k^{\prime})\rangle$,
that is to say that the non-diagonal terms only appear within such
sums. So we may replace the sum of \prettyref{eq:i_correction} and
\prettyref{eq:ii_correction} by
\begin{align}
\mathrm{}-2\,U(\ell)\,P\,[d(\Lambda,\ell)^{2}+c(\Lambda,\ell)/\ell]\,\int_{1}^{\Lambda/(1+\epsilon)}\,dk\,\Delta(k)^{2}\,,\label{eq:contraction_self_energy}
\end{align}
where an explicit factor $\ell^{-1}$ appears for the contraction
with the connected propagator, as explained above (close to \prettyref{eq:U_l}).
Treating the RG as an infinitesimal momentum shell integration, we
consider the difference of the actions before and after decimation
as
\begin{align}
\ell\frac{dS_{\ell}}{d\ell} & =\lim_{\epsilon\searrow0}\,\frac{1}{\epsilon}\,\big(S_{\ell(1+\epsilon)}-S_{\ell}\big).\label{eq:limit_diffeq_RG}
\end{align}
The decimation integrals are of the form
\begin{align*}
I_{\epsilon} & =\int_{\Lambda/(1+\epsilon)}^{\Lambda}\,dk\,\langle\Delta(k,\ell)\Delta(k,\ell)\rangle\,f(\Delta(<))\,,
\end{align*}
where $f$ is some function of $\Delta(<)$. Expanding to linear order
in $\epsilon$ to obtain the limit \eqref{eq:limit_diffeq_RG} we
have
\begin{align*}
\lim_{\epsilon\searrow0}\,\frac{I_{\epsilon}}{\epsilon} & =\Lambda\,\langle\Delta(\Lambda,\ell)\Delta(\Lambda,\ell)\rangle\,f(\Delta(<))\,.
\end{align*}
So the contributions to the flow equation for $r(\ell)$ in the renormalized
action \eqref{eq:renormed_action} written in terms of the fields
$\Delta$
\begin{align}
S_{0}(\Delta,y) & =-\frac{s(\ell)}{2}\int_{\ell}^{\Lambda}\,\Big\{ r(\ell)\,\Delta(l)^{2}+\frac{(\Delta(l)+y(l))^{2}}{\lambda(l)}\Big\}\,dl\label{eq:s_0_as_Delta}
\end{align}
are for $sr(\ell):=s(\ell)\,r(\ell)$
\begin{align}
\ell\,\frac{d\,sr(\ell)}{d\ell} & =\beta\,sr(\ell)\label{eq:d_sr}\\
 & +4\,\Lambda\,U(\ell)\,P\,\big[d(\Lambda,\ell)^{2}+c(\Lambda,\ell)/\ell\big]\,,\nonumber 
\end{align}
where the first line stems from the rescaling step \eqref{eq:S_r_scaling}.
We note that the angular bracket is proportional to the $\Lambda$-mode's
contribution to the expected loss.

\subsection{Decimation step: Contribution of the interaction to the flow of the
interaction}\label{sec:Decimation-interaction}

Likewise at one-loop order we expect a contribution to the interaction
$U$ which would be $\propto U^{2}$. For small $1\gg\beta>0$, the
interaction part is weakly relevant due to the rescaling term $U(\ell)=U(1)\,\ell^{2\beta}$.
The decimation will lead to a term $\propto-U^{2}$, which in an ordinary
$\phi^{4}$ theory is responsible for the appearance of a Wilson-Fischer
fixed point, where $U\propto\beta$, because we will get a flow equation
of the form
\begin{align*}
\ell\frac{dU}{d\ell} & =2\beta U-C\,U^{2}\\
 & =(2\beta-CU)\,U
\end{align*}
with some constant $C$ to be determined. Two differences, though,
will be that the mean value of the fields is non-vanishing here and
that the contraction with the connected propagator comes with an additional
factor $\ell^{-1}$ when the contraction collapses one weak non-diagonal integral, as explained above (close to \prettyref{eq:U_l}).
The latter leads to a suppression of the quadratic term, so that no
Wilson-Fisher fixed point exist here.

To compute the decimation contributions, one needs to contract two
pairs of fields $\Delta(>)$ from the interaction and leave four $\Delta(<)$
of them in the low momentum sector uncontracted. The interaction vertex
is

\begin{align*}
S_{\mathrm{int}}(\Delta,y) & =-UP\,\int_{1}^{\Lambda}dk\,\int_{\lfloor k\rfloor}^{\lfloor k\rfloor+1}dk^{\prime}\,\int_{1}^{\Lambda}\,dl\,\int_{\lfloor l\rfloor}^{\lfloor l\rfloor+1}dl^{\prime}\\
 & \Delta(k)\,\Delta(k^{\prime})\,\Delta(l)\,\Delta(l^{\prime})\,.
\end{align*}
Denote the fields of the two vertices in the diagram to be considered
as
\begin{align*}
\text{vertex 1)\ensuremath{\quad}} & \Delta_{1}(<)\quad\Delta_{1}(>)\,,\\
\text{vertex 2)\ensuremath{\quad}} & \Delta_{2}(<)\quad\Delta_{2}(>)\,,
\end{align*}
and likewise use subscripts $1,2$ for the momenta on their legs.
Only connected diagrams can appear due to the logarithm in the definition
of $S_{<}$.

So we have the following contributions
\begin{align}
\mathrm{i)}\quad & \Delta_{1}(<)\Delta_{1}(<)\Delta_{1}(<)\,\langle\Delta_{1}(>)\Delta_{2}(>)\rangle\,\Delta_{2}(<)\Delta_{2}(<)\Delta_{2}(<)-\text{unconnected part},\label{eq:six_point}\\
\nonumber \\\mathrm{ii)\quad} & \Delta_{1}(<)\Delta_{1}(<)\,\langle\Delta_{1}(>)\Delta_{2}(>)\rangle\,\langle\Delta_{1}(>)\Delta_{2}(>)\rangle\,\Delta_{1}(<)\Delta_{1}(<)-\text{unconnected part}.\label{eq:four_point}
\end{align}
Due to the momentum constraints of the interaction vertex there are
no contributions where only a single field $\Delta(>)$ of an interaction
vertex is in the high-momentum sector or a single field $\Delta(<)$
is in the low momentum sector and all others are in the respective
other sector.

The first diagram \eqref{eq:six_point} contributes a six-point vertex.
It rescales as $z(\ell)^{6}/\ell^{6}\,\ell^{3}=\ell^{6\frac{1+\beta}{2}-6+3}=\ell^{3\beta}$
but is driven only by $U^{2}$, so we will here neglect it first;
if we seek a theory where $U\ll1$, we may neglect this contribution.

The second diagram \eqref{eq:four_point} is a contribution to the
four-point vertex. Due to the momentum constraints implied by the
interaction vertices, the only possible momentum assignment is
\begin{align*}
 & \Delta_{1}(k_{1}<)\Delta_{1}(k_{1}^{\prime}<)\,\Delta_{2}(k_{2}<)\Delta_{2}(k_{2}^{\prime}<)\\
\times & \langle\Delta_{1}(l_{1}>)\Delta_{2}(l_{2}>)\rangle\,\langle\Delta_{1}(l_{1}^{\prime}>)\Delta_{2}(l_{2}^{\prime}>)\rangle-\text{unconnected part}.
\end{align*}
The diagram comes with a factor $2^{2}$, because at each vertex we
may choose either pair $(k,k^{\prime})$ or $(l,l^{\prime})$ to be
contracted to the respective other vertex. The value of this contribution
is
\begin{align*}
 & 2^{2}\cdot\frac{1}{2!}(UP)^{2}\\
 & \int_{1}^{\Lambda/(1+\epsilon)}dk_{1}\,\int_{\lfloor k_{1}\rfloor}^{\lfloor k_{1}\rfloor+1}dk_{1}^{\prime}\,\int_{1}^{\Lambda/(1+\epsilon)}dk_{2}\,\int_{\lfloor k_{2}\rfloor}^{\lfloor k_{2}\rfloor+1}dk_{2}^{\prime}\,\\
 & \times\Delta(k_{1})\,\Delta(k_{1}^{\prime})\,\Delta(k_{2})\,\Delta(k_{2}^{\prime})\\
\\ & \times\int_{\Lambda/(1+\epsilon)}^{\Lambda}dl_{1}\,\int_{\lfloor l_{1}\rfloor}^{\lfloor l_{1}\rfloor+1}dl_{1}^{\prime}\,\int_{\Lambda/(1+\epsilon)}^{\Lambda}dl_{2}\,\int_{\lfloor l_{2}\rfloor}^{\lfloor l_{2}\rfloor+1}dl_{2}^{\prime}\\
 & \times\big[\langle\Delta(l_{1})\,\Delta(l_{2})\rangle\,\langle\Delta(l_{1}^{\prime})\,\Delta(l_{2}^{\prime})\rangle+\langle\Delta(l_{1})\,\Delta(l_{2}^{\prime})\rangle\,\langle\Delta(l_{1})\,\Delta(l_{2}^{\prime})\rangle\\
 & -\langle\Delta(l_{1})\rangle\langle\Delta(l_{2}^{\prime})\rangle\,\langle\Delta(l_{1})\rangle\,\langle\Delta(l_{2}^{\prime})\rangle\big]\,,
\end{align*}
where the subtraction in the last line is the unconnected part and
the factor $1/2!$ comes from the diagram being second order (expansion
of $\exp(V)$ for two vertices $V$). 

Proceeding diagrammatically, we obtain the contributions (leaving
out the fields $\Delta(k)$ of the amputated $\Diagram{h}
$-legs as well as their corresponding low-momentum integrals for brevity
here)

\begin{align}
\Diagram{\vertexlabel^{k_{1}}\;hd & hu\;\vertexlabel^{l_{1}}\\
\vertexlabel^{k_{1}^{\prime}}\;hu & hd\;\vertexlabel^{l_{1}^{\prime}}
}
\quad\Diagram{\vertexlabel^{l_{2}}\;hd & hu\;\vertexlabel^{k_{2}}\\
\vertexlabel^{l_{2}^{\prime}}\;hu & hd\;\vertexlabel^{k_{2}^{\prime}}
}
 & \qquad\rightarrow\qquad2^{2}\cdot\,\Diagram{\vertexlabel^{k_{1}}\;hd & f0flfluf0 & hu\;\vertexlabel^{k_{2}}\\
\vertexlabel^{k_{1}^{\prime}}\;hu &  & hd\;\vertexlabel^{k_{2}^{\prime}}
}
\label{eq:contrib_flow_U_cov}\\
 & =2^{2}\cdot\frac{1}{2!}(UP)^{2}\,\int_{\Lambda/(1+\epsilon)}^{\Lambda}dl_{1}\,\int_{\lfloor l_{1}\rfloor}^{\lfloor l_{1}\rfloor+1}dl_{1}^{\prime}\,c(l_{1})\,c(l_{1}^{\prime})/\ell\nonumber \\
 & \simeq2\,(UP)^{2}\,\int_{\Lambda/(1+\epsilon)}^{\Lambda}dl_{1}\,c(l_{1})^{2}/\ell \nonumber \\
 & \propto\epsilon\,.\nonumber 
\end{align}
The latter contribution is $\propto\epsilon$, so it contributes to
the flow equation. Contracting the legs connecting the two vertices
cross-wise $(l_{1}\to l_{2}^{\prime};\,l_{1}^{\prime}\to l_{2})$
yields

\begin{align*}
\Diagram{\vertexlabel^{k_{1}}\;hd & hu\;\vertexlabel^{l_{1}}\\
\vertexlabel^{k_{1}^{\prime}}\;hu & hd\;\vertexlabel^{l_{1}^{\prime}}
}
\quad\Diagram{\vertexlabel^{l_{2}^{\prime}}\;hd & hu\;\vertexlabel^{k_{2}}\\
\vertexlabel^{l_{2}}\;hu & hd\;\vertexlabel^{k_{2}^{\prime}}
}
 & \qquad\rightarrow\qquad2^{2}\cdot\,\Diagram{\vertexlabel^{k_{1}}\;hd & f0flfluf0 & hu\;\vertexlabel^{k_{2}}\\
\vertexlabel^{k_{1}^{\prime}}\;hu &  & hd\;\vertexlabel^{k_{2}^{\prime}}
}
\\
 & =2^{2}\cdot\frac{1}{2!}(UP)^{2}\,\int_{\Lambda/(1+\epsilon)}^{\Lambda}dl_{1}\,\int_{\Lambda/(1+\epsilon)}^{\Lambda}dl_{2}\,c(l_{1})/\ell \,c(l_{2})/\ell\\
 & \propto\epsilon^{2}\,,
\end{align*}
the contribution is $\propto\epsilon^{2}$, because there are two
independent momentum integrals. This contribution thus vanishes in
the limit $\lim_{\epsilon\searrow0}\frac{1}{\epsilon}\ldots$, so
it does not contribute to the flow equation.

Likewise, we get contributions where two fields are contracted to
the mean (denoted as $\feyn{g}$)

\begin{align}
\Diagram{\vertexlabel^{k_{1}}\;hd & hu\;\vertexlabel^{l_{1}^{\prime}}\\
\vertexlabel^{k_{1}^{\prime}}\;hu & hd\;\vertexlabel^{l_{1}}
}
\quad\Diagram{\vertexlabel^{l_{2}^{\prime}}\;hd & hu\;\vertexlabel^{k_{2}}\\
\vertexlabel^{l_{2}}\;hu & hd\;\vertexlabel^{k_{2}^{\prime}}
}
 & \qquad\rightarrow\qquad2^{2}\cdot\,\Diagram{ & \vertexlabel^{l_{1}^{\prime}} &  & \vertexlabel^{l_{2}^{\prime}}\\
\vertexlabel^{k_{1}}\;hd & gv & f0fluf0 & gv & hu\;\vertexlabel^{k_{2}}\\
\vertexlabel^{k_{1}^{\prime}}\;hu &  &  &  & hd\;\vertexlabel^{k_{2}^{\prime}}
}
\label{eq:contrib_flow_U_mean}\\
 & =2^{2}\cdot\frac{1}{2!}(UP)^{2}\,\int_{\Lambda/(1+\epsilon)}^{\Lambda}dl_{1}\,\int_{\lfloor l_{1}\rfloor}^{\lfloor l_{1}\rfloor+1}dl_{1}^{\prime}\,\int_{\lfloor l_{1}\rfloor}^{\lfloor l_{1}\rfloor+1}dl_{2}^{\prime}\,c(l_{1})\,d(l_{1}^{\prime})\,d(l_{2}^{\prime})\nonumber \\
 & \simeq2\,(UP)^{2}\,\int_{\Lambda/(1+\epsilon)}^{\Lambda}dl_{1}\,c(l_{1})\,d(l_{1})^{2}\,,\nonumber 
\end{align}
which is a contribution $\propto\epsilon$, so it contributes to the
flow equation. Contracting the moments $l_{1}$ and $l_{2}$ to the
means, instead,

\begin{align*}
\Diagram{\vertexlabel^{k_{1}}\;hd & hu\;\vertexlabel^{l_{1}}\\
\vertexlabel^{k_{1}^{\prime}}\;hu & hd\;\vertexlabel^{l_{1}^{\prime}}
}
\quad\Diagram{\vertexlabel^{l_{2}}\;hd & hu\;\vertexlabel^{k_{2}}\\
\vertexlabel^{l_{2}^{\prime}}\;hu & hd\;\vertexlabel^{k_{2}^{\prime}}
}
 & \qquad\rightarrow\qquad2^{2}\cdot\,\Diagram{ & \vertexlabel^{l_{1}} &  & \vertexlabel^{l_{2}}\\
\vertexlabel^{k_{1}}\;hd & gv & f0fluf0 & gv & hu\;\vertexlabel^{k_{2}}\\
\vertexlabel^{k_{1}^{\prime}}\;hu &  &  &  & hd\;\vertexlabel^{k_{2}^{\prime}}
}
\\
 & =2^{2}\cdot\frac{1}{2!}(UP)^{2}\,\int_{\Lambda/(1+\epsilon)}^{\Lambda}dl_{1}\,\int_{\Lambda/(1+\epsilon)}^{\Lambda}dl_{2}\,\int_{\lfloor l_{1}\rfloor}^{\lfloor l_{1}\rfloor+1}dl_{1}^{\prime}\,c(l_{1}^{\prime})/\ell\,d(l_{1})\,d(l_{2})\\
 & \propto\epsilon^{2}\,,
\end{align*}
which is $\propto\epsilon^{2}$, because there are two independent
momentum shell integrations.

Lastly, we may contract the momenta cross-wise

\begin{align*}
\Diagram{\vertexlabel^{k_{1}}\;hd & hu\;\vertexlabel^{l_{1}}\\
\vertexlabel^{k_{1}^{\prime}}\;hu & hd\;\vertexlabel^{l_{1}^{\prime}}
}
\quad\Diagram{\vertexlabel^{l_{2}^{\prime}}\;hd & hu\;\vertexlabel^{k_{2}}\\
\vertexlabel^{l_{2}}\;hu & hd\;\vertexlabel^{k_{2}^{\prime}}
}
 & \qquad\rightarrow\qquad2^{2}\cdot\,\Diagram{ & \vertexlabel^{l_{1}} &  & \vertexlabel^{l_{2}^{\prime}}\\
\vertexlabel^{k_{1}}\;hd & gv & f0fluf0 & gv & hu\;\vertexlabel^{k_{2}}\\
\vertexlabel^{k_{1}^{\prime}}\;hu &  &  &  & hd\;\vertexlabel^{k_{2}^{\prime}}
}
\\
 & =2^{2}\cdot\frac{1}{2!}(UP)^{2}\,\int_{\Lambda/(1+\epsilon)}^{\Lambda}dl_{1}\,\int_{\Lambda/(1+\epsilon)}^{\Lambda}dl_{2}\,\int_{\lfloor l_{2}\rfloor}^{\lfloor l_{2}\rfloor+1}dl_{2}^{\prime}\,c(l_{2})/\ell\,d(l_{1})\,d(l_{2}^{\prime})\\
 & \propto\epsilon^{2}\,,
\end{align*}
which again contains two independent momentum integrations, so it
does not contribute to the flow.

Taken together, the non-vanishing contributions to the decimation
part of the flow equation of $U$ are \eqref{eq:contrib_flow_U_cov}
and \eqref{eq:contrib_flow_U_mean}
\begin{align}
\ell\frac{dU}{d\ell} & =2\beta\,U\label{eq:decimation_U}\\
 & -2U^{2}P\,\Lambda\,\big[c(\Lambda)^{2}/\ell+c(\Lambda)\,d(\Lambda)^{2}\big]\,,\nonumber 
\end{align}
where one factor $P$ is gone because the interaction term is $\propto P$
itself and the first line is the contribution from the rescaling \eqref{eq:U_l}.

\subsection{Complete set of RG equations}

We may now assemble the set of RG equations from \prettyref{eq:d_sr}
and \prettyref{eq:renormed_action-2} and \eqref{eq:decimation_U}
as 
\begin{align}
s(\ell)= & \ell^{\beta-\alpha-1}\,,\label{eq:RG_complete}\\
\ell\,\frac{d\,sr(\ell)}{d\ell}= & \beta\,sr(\ell)\nonumber \\
 & +4\,U(\ell)\,P\,\Lambda\,\big[d(\Lambda,\ell)^{2}+c(\Lambda,\ell)/\ell\big]\,,\nonumber \\
\ell\,\frac{dU(\ell)}{d\ell} & =2\beta\,U(\ell)\nonumber \\
 & -2U(\ell)^{2}\,P\,\Lambda\,\big[c(\Lambda,\ell)^{2}/\ell+c(\Lambda,\ell)\,d(\Lambda,\ell)^{2}\big]\,.\nonumber 
\end{align}
These flow equations depend explicitly on the cutoff $\Lambda$. The
equation for $U$ shows that as $\ell\to\infty$, the decimation contributions
decline due to the factors $\ell^{-1}$ and $\ell^{-2}$, respectively.
This shows that the flow equation is not homogeneous and in particular
it does not possess a fixed point, since for sufficiently large $\ell$,
always the rescaling term $2\beta\,U(\ell)$ will dominate.

\subsection{Analytical solution of the flow equations}

In order to make predictions about the scaling laws of the expected
loss we need an analytical expression of the solutions of \prettyref{eq:RG_complete}.
To simplify, we neglect second order corrections to the flow of $U(\ell)$
and here only keep the self-energy corrections implied by the flow
equation for $sr$.

Transforming the flow variable $\tau=\ln\,\ell$ we obtain from \prettyref{eq:RG_complete}
for $U$ the differential equation of an exponential function, solved
by
\begin{equation}
U(\tau)=U(0)\,\exp(2\beta\,\tau)\,.\label{eq:U_ell_rescaling}
\end{equation}
 Inserting this into the differential equation of $sr(\ell)$, yields\eqref{eq:sr_flow_U_rescaling}
\begin{equation}
\ell\,\dfrac{d\,sr(\ell)}{d\ell}=\beta\,sr(\ell)+4\,U(1)\,\ell^{2\beta}\,P\,\Lambda\,\left[\dfrac{y(\Lambda)^{2}}{[r(\ell)\lambda(\Lambda)+1]^{2}}+\ell^{-1}\,\dfrac{s(\ell)^{-1}\lambda(\Lambda)}{r(\ell)\lambda(\Lambda)+1}\right]\,,\label{eq:sr_flow_U_rescaling}
\end{equation}
where we used the explicit expressions for the mean and covariance
\prettyref{eq:mean_prop_rescaled}. This differential equation must
be solved together with $s(\ell)=\ell^{\beta-\alpha-1}$; it is a
first order non-linear differential equation with $\ell$-dependent
coefficients, which in general is not easy to solve.

\subsection{Perturbative approach to the flow equations}\label{sec:Perturbative-approach-to-flow-equations}

We are interested in the analytical solution of the flow equation
of $r(\ell)$ in \eqref{eq:sr_flow_U_rescaling}. To treat the non-linearities
in the equation, we will treat the flow of $r$ on the right hand
side in a perturbative manner. We perform the perturbative computation
around the Gaussian part (rescaling only) of the evolution of the
mass term, namely $sr(\ell)=sr(1)\,\ell^{\beta}$ and $s(\ell)=\ell^{\beta-\alpha-1}$
which together yields 
\begin{equation}
r^{\text{GP}}(\ell):=r(1)\,\ell^{1+\alpha}\,.\label{eq:r_GP}
\end{equation}
Formally, we are assuming that the main contribution comes from the
Gaussian process, such that we can write $r(\tau)=r^{\text{GP}}(\tau)+\epsilon(\tau)$,
where $r^{\text{GP}}(\tau)>\epsilon(\tau)$. The equation for the
perturbation $\epsilon(\ell)$ derived from \eqref{eq:sr_flow_U_rescaling}
reads
\begin{equation}
\frac{d\,\epsilon(\tau)}{d\tau}=(1+\alpha)\,\epsilon(\tau)+\underset{=:f(\tau)}{\underbrace{\left[A(P,\Lambda)\,\frac{s(\tau)^{-1}\,e^{2\beta\tau}}{\left[\frac{P}{\kappa}e^{(1+\alpha)\tau}\lambda(\Lambda)+1\right]^{2}}+B(P,\Lambda)\,\frac{s(\tau)^{-2}\,e^{(2\beta-1)\tau}}{\frac{P}{\kappa}e^{(1+\alpha)\tau}\,\lambda(\Lambda)+1}\right]}}\,,\label{eq:non_gauss_corr_r_flow}
\end{equation}
where $s(\tau)=e^{(\beta-\alpha-1)\tau}$ and 
\begin{align}
A(P,\Lambda) & :=4U(0)P\,\Lambda\,y(\Lambda)^{2}\,,\label{eq:def_A_B}\\
B(P,\Lambda) & :=4U(0)P\,\Lambda\,\lambda(\Lambda)\,.\nonumber 
\end{align}
The Gaussian contribution vanishes since it satisfies the homogeneous
part of the equation. As a result, we obtain a linear inhomogeneous
differential equation for $\epsilon$.

We know that the Green's function of the linear differential operator
$\big(\frac{d}{d\tau}-(1+\alpha)\big)$ is given by $H(\tau)\,e^{(1+\alpha)\,\tau}$,
so we find a particular solution by convolving this Green's function
with the inhomogeneity $f(\tau)$. As $r(0)=P/\kappa=r^{\text{GP}}(0)$
it follows that the initial value for $\epsilon(0)=0$, so the solution
obeying this initial condition is
\begin{align}
\epsilon(\tau) & =\int_{0}^{\tau}\,e^{(1+\alpha)\,(\tau-\tau^{\prime})}\,f(\tau^{\prime})\,d\tau^{\prime}\,.\label{eq:sol_epsilon_as_convolution}
\end{align}

Written explicitly
\begin{align*}
\epsilon(\tau) & =e^{(1+\alpha)\tau}\int_{0}^{\tau}e^{-(1+\alpha)\tau^{\prime}}\bigg[A(P,\Lambda)\frac{e^{(\beta+\alpha+1)\tau^{\prime}}}{\left[\frac{P}{\kappa}e^{(1+\alpha)\tau^{\prime}}\Lambda^{-(1+\alpha)}+1\right]^{2}}\\
 & \phantom{e^{(1+\alpha)\tau}\int_{0}^{\tau}e^{-(1+\alpha)\tau^{\prime}}\bigg[}+B(P,\Lambda)\frac{e^{(2(\alpha+1)-1)\tau^{\prime}}}{\frac{P}{\kappa}e^{(1+\alpha)\tau^{\prime}}\Lambda^{-(1+\alpha)}+1}\bigg]\,d\tau^{\prime}
\end{align*}
and undoing the transformation $\ell=e^{\tau}$ we have
\begin{align}
\epsilon(\ell) & =\ell^{(1+\alpha)}\int_{1}^{\ell}\,\bigg[A(P,\Lambda)\,\frac{\ell^{\prime\beta}}{\left[\frac{P}{\kappa}\ell^{\prime(1+\alpha)}\Lambda^{-(1+\alpha)}+1\right]^{2}}\nonumber \\
 & \phantom{\ell^{(1+\alpha)}\int_{1}^{\ell}\,\bigg[}+B(P,\Lambda)\,\frac{\ell^{\prime\alpha}}{\frac{P}{\kappa}\ell^{\prime(1+\alpha)}\Lambda^{-(1+\alpha)}+1}\bigg]\,\frac{d\ell^{\prime}}{\ell^{\prime}}\,.\label{eq:non_gauss_r_flow_solution}
\end{align}

\section{Use of RG for hyperparameter transfer}\label{sec:RG_parameter_transfer}

The RG flow provides a relation between the original theory given
in the form on an action $S(\Delta;r_{1},u_{1})$ for the degrees
of freedom $\Delta(1\le k\le P)$ and an action $S(\Delta';r_{\ell},u_{\ell})$
for the degrees of freedom $\Delta'$ which are obtained after integrating
out the upper $P-P^{\prime}$ degrees of freedom, controlled by $\ell=P/P^{\prime}$.
The rescaling part of the RG flow is made such that the functional
form of these two actions is the same, only the parameters change $(r_{1},u_{1})\mapsto(r_{\ell},u_{\ell})$.
To accomplish this, the degrees of freedom are rescaled versions of
the original ones, which undergo the transform \prettyref{eq:rescaling_momentum}
and \prettyref{eq:rescaling_wavefunction}
\begin{align}
\ell\,k^{\prime} & :=k\,,\label{eq:rescaling_momentum_transfer-1}\\
z_{\ell}\,\Delta^{\prime}(k^{\prime}) & :=\Delta(k)\,.\nonumber 
\end{align}
We would like to use the RG to map one system with $P$ degrees of
freedom to another system with $P^{\prime}<P$ degrees of freedom.
To do this, we need to set
\begin{align*}
\ell & =P/P^{\prime}\,.
\end{align*}
We start with the larger system with $P$ degrees of freedom. The
relation between the discrete system's the degrees of freedom and
the continuous ones is given by \eqref{eq:relation_eigen_EK}, so
\begin{align*}
D(k) & =\sqrt{P}\,\Delta(k)\,.
\end{align*}
Likewise, the target rescales as $Y=\sqrt{P}\,y$.

The action describing the $\Delta$ has the form \eqref{eq:RG_action_main_Delta}
with parameters $r(\ell=1)=P/\kappa$, $s(\ell=1)=1$, $U(\ell=1)=U$.
The RG transform integrates out the upper $P-P^{\prime}$ degrees
of freedom and writes the result as an action of the same functional
form as before. In particular, the momentum range again covers the
range $k\in[0,\Lambda=P]$. We would like to re-interpret this action
as one for the smaller number of $P^{\prime}$ of degrees of freedom.
We thus need to undo the rescaling, we thus need to re-express the
action \eqref{eq:RG_action_main_Delta} in terms of the original degrees
of freedom $\Delta$. This yields
\begin{align*}
\tilde{S}(\Delta;r_{\ell},u_{\ell},P^{\prime}):= & S(z_{\ell}^{-1}\,\Delta;r_{\ell},u_{\ell},P)\\
= & -\frac{s_{\ell}}{2}\,\int_{1}^{P^{\prime}}\,r_{\ell}\,z_{\ell}^{-2}\,\Delta^{2}(k)+\frac{\big(y(\ell\tilde{k})-z_{\ell}^{-1}\Delta(\tilde{k})\big)^{2}}{\lambda(\ell k)}\,\ell\,d\tilde{k}\,\\
 & -u_{\ell}\,P\,\Big[\int_{1}^{P^{\prime}}\,\int_{\lfloor\ell\tilde{k}\rfloor/\ell}^{\lfloor\ell\tilde{k}\rfloor/\ell+1/\ell}\,z_{\ell}^{-2}\,\Delta(\tilde{k})\,\Delta(\tilde{k}^{\prime})\,\ell^{2}\,d\tilde{k}\,d\tilde{k}^{\prime}\,\Big]^{2}\,.
\end{align*}
The goal is to interpret this system again as a system of $P^{\prime}$
degrees of freedom. We insert $z_{\ell}=\ell^{\frac{1+\beta}{2}}$
\eqref{eq:z_rescale} as well as $s_{\ell}=\ell^{\beta-\alpha-1}$
\eqref{eq:s_ell} and assume a smooth integrand to replace $\ell\,\int_{\lfloor\ell\tilde{k}\rfloor/\ell}^{\lfloor\ell\tilde{k}\rfloor/\ell+1/\ell}\to\int_{\lfloor\tilde{k}\rfloor}^{\lfloor\tilde{k}\rfloor+1}$,
thus obtaining

\begin{align*}
\tilde{S}(\Delta;r_{\ell},u_{\ell},P^{\prime})= & -\frac{1}{2}\,\int_{1}^{P^{\prime}}\,\underbrace{s_{\ell}r_{\ell}\,\ell^{-\beta}}_{=:\tilde{sr}_{\ell}}\,\Delta^{2}(k)+\ell^{\beta-\alpha}\,\frac{\big(\ell^{-\frac{1+\beta}{2}}\,y(\tilde{k})-\ell^{-\frac{1+\beta}{2}}\Delta(\tilde{k})\big)^{2}}{\ell^{-(1+\alpha)}\,\lambda(\tilde{k})}\,d\tilde{k}\,\\
 & -u_{\ell}\,P\,\Big[\int_{1}^{P^{\prime}}\,\int_{\lfloor\tilde{k}\rfloor}^{\lfloor\tilde{k}\rfloor+1}\,\ell^{-\beta}\,\Delta(\tilde{k})\,\Delta(\tilde{k}^{\prime})\,d\tilde{k}\,d\tilde{k}^{\prime}\,\Big]^{2}\\
= & -\frac{1}{2}\,\int_{1}^{P^{\prime}}\,\tilde{sr}_{\ell}\,\Delta^{2}(k)+\frac{\big(y(\tilde{k})-\Delta(\tilde{k})\big)^{2}}{\lambda(\tilde{k})}\,d\tilde{k}\\
 & -\underbrace{u_{\ell}\ell^{-2\beta}\,P}_{=:\tilde{u}_{\ell}\,P^{\prime}}\,\Big[\int_{1}^{P^{\prime}}\,\int_{\lfloor\tilde{k}\rfloor}^{\lfloor\tilde{k}\rfloor+1}\,\Delta(\tilde{k})\,\Delta(\tilde{k}^{\prime})\,d\tilde{k}\,d\tilde{k}^{\prime}\,\Big]^{2}\,.
\end{align*}
We have thus found the action
\begin{align}
\tilde{S}(\Delta;\tilde{sr}_{\ell},\tilde{u}_{\ell},P^{\prime}) & =-\frac{1}{2}\,\int_{1}^{P^{\prime}}\,\tilde{sr}_{\ell}\,\Delta^{2}(k)+\frac{\big(y(\tilde{k})-\Delta(\tilde{k})\big)^{2}}{\lambda(\tilde{k})}\,d\tilde{k}\label{eq:final_P_prime-2}\\
 & -\tilde{u}_{\ell}\,P^{\prime}\,\Big[\int_{1}^{P^{\prime}}\,\int_{\lfloor\tilde{k}\rfloor}^{\lfloor\tilde{k}\rfloor+1}\,\Delta(\tilde{k})\,\Delta(\tilde{k}^{\prime})\,d\tilde{k}\,d\tilde{k}^{\prime}\,\Big]^{2}\,,\nonumber \\
\tilde{sr}_{\ell} & :=s_{\ell}r_{\ell}\,\ell^{-\beta}=\ell^{-(1+\alpha)}\,r_{\ell}\,,\nonumber \\
\tilde{u}_{\ell} & :=\frac{P}{P^{\prime}}\,u_{\ell}\ell^{-2\beta}\,,\nonumber 
\end{align}
which is again of the same form as before, but the momentum range
extends only up to $P^{\prime}$. The rescaling of the $\tilde{sr}$
and $\tilde{u}$ corresponds the native rescaling due to dimensional
analysis, as it has to be, because we simply undid the rescaling performed
by the RG.

We may thus interpret the system \eqref{eq:final_P_prime-2} as a
small system with $P^{\prime}$ degrees of freedom. To map it to the
discrete numerics, we need to determine the parameters as
\begin{align*}
\tilde{sr}_{\ell} & =:P^{\prime}/\tilde{\kappa}\,,\\
\tilde{u}_{\ell} & =\frac{P}{P^{\prime}}\,u_{\ell}\ell^{-2\beta}\,,\\
\ell & =P/P^{\prime}.
\end{align*}
The discrepancies in the small system are
\begin{align*}
\langle D_{P}(k)\rangle/\sqrt{P} & =\langle\Delta(k)\rangle_{S(r_{1},u_{1},P)}\,,\\
\langle D_{P^{\prime}}(k)\rangle/\sqrt{P^{\prime}} & =\langle\Delta(k)\rangle_{\tilde{S}(\tilde{sr},\tilde{u},P^{\prime})}\,.
\end{align*}
These are the natural units to express the loss per sample $\langle\cL\rangle/P=f(\langle D_{P}(k)\rangle/\sqrt{P})$.

\section{Scaling of the loss with $P$}\label{sec:Scaling-of-the-loss}

In this section we derive how the loss scales with the number of training
patterns $P$ to obtain the neural scaling law. We begin with the
scaling for the Gaussian process and subsequently derive how the scaling
law changes due to non-Gaussian corrections. To this end we employ
the renormalization group to incorporate the effect of the non-Gaussian
corrections. We find that the loss is most sensitive to corrections
to the ridge parameter, while the direct effect of the interaction
term can be neglected. As a result, we obtain the loss from an effective
Gaussian process with a renormalized ridge parameter.

\subsection{Scaling for the Gaussian process}\label{sec:Scaling-for-the-loss-GP}

We would like to know how the loss per samples declines as a function
of $P$. To this end, we decompose the loss into its bias and variance
part. For the Gaussian process, we have 
\begin{align}
\cL_{\text{bias }}/P & =\frac{1}{2P}\sum_{k=1}^{P}\langle D(k)\rangle^{2}\,,\label{eq:L_bias-1}
\end{align}
where mean discrepancy $\langle D(k)\rangle$ measured in the discrete
system is related to the EK theory as
\begin{align}
D(k) & =\sqrt{P}\,\Delta(k)\label{eq:relation_D_Delta-1}
\end{align}
and $\langle\Delta(k)\rangle=\frac{y(k)}{\frac{P}{\kappa}\,\lambda(k)+1}\,$,
so
\begin{align}
\cL_{\text{bias }}/P & =\frac{1}{2}\sum_{k=1}^{P}\frac{y^{2}(k)}{\big[\frac{P}{\kappa}\,\lambda(k)+1\big]^{2}}\,.\label{eq:L_bias-1-1}
\end{align}
The asymptotics of the terms for small $k$ where $\frac{P}{\kappa}\,\lambda(k)\gg1$
and for large $k$, where $\frac{P}{\kappa}\,\lambda(k)\ll1$ are
\begin{align*}
\frac{y^{2}(k)}{\big[\frac{P}{\kappa}\,\lambda(k)+1\big]^{2}} & \simeq\begin{cases}
\big(\frac{\kappa}{P}\big)^{2}\,k^{1-\beta+2\alpha} & \frac{P}{\kappa}\,\lambda(k)\gg1\\
k^{-(1+\beta)} & \frac{P}{\kappa}\,\lambda(k)\ll1
\end{cases}\,,
\end{align*}
so the terms become small on either end of the summation interval.
We may therefore approximate the sum by an integral
\begin{align}
\cL_{\text{bias }}/P & \simeq\frac{1}{2}\int_{0}^{\infty}\,\frac{k^{-(1+\beta)}}{\big[\frac{P}{\kappa}\,k^{-(1+\alpha)}+1\big]^{2}}\,dk-\frac{1}{2}\int_{P}^{\infty}\,k^{-(1+\beta)}\,dk\label{eq:L_bias_as_int}\\
 & =\frac{1}{2}\int_{0}^{\infty}\,\frac{k^{-(1+\beta)}}{\big[\frac{P}{\kappa}\,k^{-(1+\alpha)}+1\big]^{2}}\,dk-\frac{P^{-\beta}}{2\beta}\,,\nonumber 
\end{align}
so that we could remove the $P$-dependence of the upper bound. The
correction $\frac{1}{2}\int_{0}^{1}\frac{k^{-(1+\beta)}}{\big[\frac{P}{\kappa}\,k^{-(1+\alpha)}+1\big]^{2}}\,dk\simeq\frac{1}{2}\big(\frac{\kappa}{P}\big)^{2}\,\int_{0}^{1}k^{(1-\beta+2\alpha)}\,dk=\frac{1}{2}\,\frac{1}{2-\beta+2\alpha}\,\big(\frac{\kappa}{P}\big)^{2}$
from replacing the lower bound $1\to0$ is negligible.

To extract the scaling of the remaining integral in \eqref{eq:L_bias_as_int}
with $P$ we perform a substitution, first combining $\frac{P}{\kappa}\,k^{-(1+\alpha)}=(a\,k)^{-(1+\alpha)}$
with $a=\big(\frac{P}{\kappa}\big)^{-\frac{1}{1+\alpha}}$ and then
substituting $ak\to\rho$ to get
\begin{align}
\cL_{\text{bias }}/P+\frac{P^{-\beta}}{2\beta} & \simeq\frac{1}{2}\int_{0}^{\infty}\,\frac{(\rho/a)^{-(1+\beta)}}{\big[\,\rho^{-(1+\alpha)}+1\big]^{2}}\,\frac{d\rho}{a}\label{eq:def_I_bias}\\
 & =a^{\beta}\,\frac{1}{2}\int_{0}^{\infty}\,\frac{\rho^{-(1+\beta)}}{\big[\,\rho^{-(1+\alpha)}+1\big]^{2}}\,d\rho\nonumber \\
 & =:\,\big(\frac{P}{\kappa}\big)^{-\frac{\beta}{1+\alpha}}\,I_{\mathrm{bias}}\,,\nonumber 
\end{align}
where the integral becomes a constant $I_{\text{bias }}$ as a function
of $P$. The bias term hence scales as
\begin{align}
\cL_{\text{bias }}/P & =I_{\text{bias }}\,\big(\frac{P}{\kappa}\big)^{-\frac{\beta}{1+\alpha}}-\frac{P^{-\beta}}{2\beta}\,.\label{eq:loss_bias_GP}
\end{align}
The variance contribution likewise can be written as
\begin{align*}
\cL_{\text{var}}/P & =\frac{1}{2}\,\sum_{k=1}^{P}\,\frac{\lambda(k)}{\frac{P}{\kappa}\,\lambda(k)+1}\,.
\end{align*}
The asymptotics of the summand is here different, namely
\begin{align*}
\frac{\lambda(k)}{\frac{P}{\kappa}\,\lambda(k)+1} & \simeq\begin{cases}
\frac{\kappa}{P} & \frac{P}{\kappa}\,\lambda(k)\gg1\\
k^{-(1+\alpha)} & \frac{P}{\kappa}\,\lambda(k)\ll1
\end{cases}\,,
\end{align*}
so that the integrand does not vanish at the lower bound. Performing
the analogous computation as for the bias term
\begin{align*}
\cL_{\text{var}}/P & \simeq\frac{1}{2}\,\int_{0}^{\infty}\,\frac{\lambda(k)}{\frac{P}{\kappa}\,\lambda(k)+1}\,dk-\frac{1}{2}\int_{P}^{\infty}\,\lambda(k)\,dk-\frac{\kappa}{2P}\int_{0}^{1}\,dk\\
 & =\frac{1}{2}\,\int_{a}^{\infty}\,\frac{(\rho/a)^{-(1+\alpha)}}{\rho^{-(1+\alpha)}+1}\,\frac{d\rho}{a}-\frac{1}{2}\int_{P}^{\infty}\,k^{-(1+\alpha)}\,dk-\frac{\kappa}{2P}\\
 & =a^{\alpha}\,\frac{1}{2}\,\int_{a}^{\infty}\,\frac{\rho{}^{-(1+\alpha)}}{\rho^{-(1+\alpha)}+1}\,d\rho-\frac{P^{-\alpha}}{2\alpha}-\frac{\kappa}{2P}\,,
\end{align*}
so that we may again approximate the integral by a constant $I_{\text{var}}$
to obtain the scaling of the variance term as
\begin{align}
\cL_{\text{var}}/P & \simeq I_{\text{var}}\,\big(\frac{P}{\kappa}\big)^{-\frac{\alpha}{1+\alpha}}-\frac{P^{-\alpha}}{2\alpha}-\frac{\kappa P^{-1}}{2}.\label{eq:loss_var_GP}
\end{align}

\subsection{Scaling for the non-Gaussian process}

To perform an analogous computation for the non-Gaussian process,
we employ the RG is to compute renormalized parameters of an effective
Gaussian process. To this end, we make use of the observation that
the mean discrepancies are effectively those of the Gaussian system,
albeit with a renormalized ridge parameter $r_{\ell}$ (see \prettyref{fig:saddle_mean}).
This means we may neglect the direct contribution of the interaction
term and only take into account that the interaction affects the flow
equation for the ridge parameter. To obtain the bias part of the loss,
we thus need to compute
\begin{align}
\cL_{\text{bias }} & =\frac{1}{2}\sum_{k=P}^{1}\langle D(k)\rangle^{2}\,.\label{eq:L_bias}
\end{align}
The mean discrepancy $\langle D(k)\rangle$ measured in the discrete
system is related to the EK as
\begin{align}
D(k) & =\sqrt{P}\,\Delta(k)\,.\label{eq:relation_D_Delta}
\end{align}
To compute $\Delta(k)$ the presence of the modes above $k$ needs
to be taken into account. This is done with help of the RG by integrating
out all modes above $k$. We assume a cutoff for the modes, $k\le\Lambda$.
This implies a flow parameter $\ell=\Lambda/k$ that depends on the
mode $k$ currently considered. The mean discrepancy is then given
by the highest mode $\Lambda$ in the renormalized system where all
modes beyond the mode $k$ of interest have been integrated out
\begin{align*}
\langle\Delta(k)\rangle & =z(\ell)\,\langle\Delta^{\prime}(\Lambda)\rangle\,\big|_{\,r(\ell),u(\ell),\ell=\frac{\Lambda}{k}}\,.
\end{align*}
We observe that the $\langle\Delta^{\prime}\rangle$ depends only
weakly on $u_{\ell}$ and is effectively given by the result for the
Gaussian theory \eqref{eq:mean_prop_Delta_main}, so that

\begin{align*}
\langle\Delta^{\prime}(\Lambda)\rangle & \simeq\frac{1}{r(\ell)\,\lambda(\Lambda)+1}\,y(\Lambda)\big|_{\,\ell=\frac{\Lambda}{k}}\,.
\end{align*}
Inserted into \eqref{eq:L_bias} one obtains with \eqref{eq:z_rescale}
$z(\ell)=\ell^{\frac{1+\beta}{2}}$ \label{eq:L_bias-2}
\begin{align}
\cL_{\text{bias }} & =\frac{P}{2}\,\sum_{k=P}^{1}\,z(\ell)^{2}\,\frac{y(\Lambda)^{2}}{[r(\ell)\,\lambda(\Lambda)+1]^{2}}\,\big|_{\ell=\frac{\Lambda}{k}},\nonumber \\
 & =\frac{P}{2}\,\sum_{k=P}^{1}\,\big(\frac{\Lambda}{k}\big)^{1+\beta}\,\frac{\Lambda^{-(1+\beta)}}{[r(\frac{\Lambda}{k})\,\Lambda^{-(1+\alpha)}+1]^{2}}\nonumber \\
 & =\frac{P}{2}\,\sum_{k=P}^{1}\,\frac{k^{-(1+\beta)}}{[r(\frac{\Lambda}{k})\,\Lambda^{-(1+\alpha)}+1]^{2}}\,,\label{eq:L_bias_discrete}
\end{align}
where $r(\ell=\frac{\Lambda}{k})$ is the solution of the RG equation
\eqref{eq:RG_U_main}.

As in the Gaussian case in \eqref{eq:L_bias_as_int}, we may replace
the sum by an integral and take the boundaries to $0$ and $\infty$
(correcting for the change of the upper bound) to get, analogous to
\prettyref{eq:L_bias_as_int} 
\begin{align}
\cL_{\text{bias }}/P & =\frac{1}{2}\,\int_{0}^{\infty}\,\frac{k^{-(1+\beta)}}{[r(\frac{\Lambda}{k})\,\Lambda^{-(1+\alpha)}+1]^{2}}\,dk-\frac{P^{-\beta}}{2\beta}\label{eq:L_bias_per_P}\\
 & =\frac{1}{2}\,\int_{0}^{\infty}\,\frac{k^{-(1+\beta)}}{[r(\frac{\Lambda}{k})\,\big(\frac{\Lambda}{k})^{-(1+\alpha)}\,k^{-(1+\alpha)}+1]^{2}}\,dk-\frac{P^{-\beta}}{2\beta}\,.\nonumber 
\end{align}
The $P$-dependence of the integral here appears in the form of $r(\ell)\,\ell{}^{-(1+\alpha)}$,
which for the GP reduces to $P/\kappa$, as it should.

For the variance contribution to the loss one obtains similarly
\begin{align}
\cL_{\text{var}} & =\frac{1}{2}\,\sum_{k=P}^{1}\,\langle D^{2}(k)\rangle^{c}=\frac{P}{2}\,\sum_{k=P}^{1}\,\langle\Delta^{2}(k)\rangle^{c}\,.\label{eq:L_var}
\end{align}

Expressing the variance of mode $k$ in terms of the variance of the
highest mode in the decimated system with $\ell=\Lambda/k$ we need
to take the factor $s(\ell)$ into account which scales the amplitude
of all fields. Also we need to take into account that the diagonal
part of the covariance comes with a Dirac-$\delta$ whose argument
gets rescaled by $\ell$ (see also the discussion close to \eqref{eq:U_ell_rescaling}
for the appearance of the factor $1/\ell$ for observables that contain
a Dirac $\delta$)
\begin{align*}
\langle\Delta^{2}(k)\rangle^{c}\,\delta(\circ) & =z(\ell)^{2}\,\langle\Delta^{\prime2}(\Lambda)\rangle^{c}\big|_{r(\ell),u(\ell),\ell=\frac{\Lambda}{k}}\,\delta(\ell\,\circ)\,.
\end{align*}
So together with $z^{2}(\ell)=\ell^{1+\beta}$ and $\delta(\ell\circ)=\ell^{-1}\,\delta(\circ)$
this is

\begin{align*}
\langle\Delta^{2}(k)\rangle^{c}\,\delta(\circ) & =\ell^{1+\beta}\,\ell^{-1}\,\langle\Delta^{\prime2}(\Lambda)\rangle^{c}\big|_{r(\ell),u(\ell),\ell=\frac{\Lambda}{k}}\,\delta(\circ)\\
 & =\ell^{\beta}\,\langle\Delta^{\prime2}(\Lambda)\rangle^{c}\big|_{r(\ell),u(\ell),\ell=\frac{\Lambda}{k}}\,\delta(\circ)\,.
\end{align*}
Assuming that we may neglect the explicit dependence of the variance
on $u_{\ell}$ we may use \eqref{eq:mean_prop_Delta_main} and \eqref{eq:RG_r_main} with $s(\ell)=\ell^{\beta-\alpha-1}$
to obtain

\begin{align}
\cL_{\text{var}} & =\frac{P}{2}\,\sum_{k=P}^{1}\,\ell^{\beta}\,\frac{s(\ell)^{-1}\,\lambda(\Lambda)}{r(\ell)\,\lambda(\Lambda)+1}\big|_{\ell=\frac{\Lambda}{k}}\label{eq:L_var_discrete}\\
 & =\frac{P}{2}\,\sum_{k=P}^{1}\,\ell^{1+\alpha}\,\frac{\Lambda^{-(1+\alpha)}}{r(\ell)\,\lambda(\Lambda)+1}\big|_{\ell=\frac{\Lambda}{k}}\nonumber \\
 & =\frac{P}{2}\,\sum_{k=P}^{1}\,\frac{k^{-(1+\alpha)}}{r(\frac{\Lambda}{k})\,\Lambda^{-(1+\alpha)}+1}\,.\nonumber 
\end{align}

Like in the bias contribution we can compute this as an integral by
correcting for the change in boundaries
\begin{align}
\cL_{\text{var }}/P & =\frac{1}{2}\,\int_{0}^{\infty}\,\frac{k^{-(1+\alpha)}}{r(\frac{\Lambda}{k})\,\Lambda^{-(1+\alpha)}+1}\,dk-\frac{P^{-\alpha}}{2\alpha}-\frac{P^{-1}}{2\kappa}\label{eq:L_var_per_P}\\
 & =\frac{1}{2}\,\int_{0}^{\infty}\,\frac{k^{-(1+\alpha)}}{r(\frac{\Lambda}{k})\,\big(\frac{\Lambda}{k})^{-(1+\alpha)}\,k^{-(1+\alpha)}+1}\,dk-\frac{P^{-\alpha}}{2\alpha}-\frac{P^{-1}}{2\kappa}\,.\nonumber 
\end{align}

\subsection{Extraction of the P-dependence of the training loss}\label{sec:Extraction-of-the-P-dependence}

To obtain the expected loss we need to determine the solution of $r(\ell)$
of the corresponding RG equation. Decomposing this solution into the
Gaussian part $r^{\text{GP}}$ and corrections $\epsilon$ due to
the interaction, leaves us with the problem to compute $\epsilon$
from \eqref{eq:sol_epsilon_as_convolution}, which cannot be solved
exactly. Our main goal here is to extract the scaling of the loss
with $P$. To this end, it turns out to suffice that we work with
the implicit analytical solution of \eqref{eq:sol_epsilon_as_convolution}.

\subsubsection{Bias part of the loss}

The goal is to extract the $P$-dependence of the loss. Starting with
\eqref{eq:L_bias_per_P}
\[
\langle\mathcal{L}_{\text{bias}}\rangle/P+\frac{P^{-\beta}}{2\beta}=\frac{1}{2}\int_{0}^{\infty}\frac{k^{-(1+\beta)}}{\left[r(\Lambda/k)\left(\frac{\Lambda}{k}\right)^{-(1+\alpha)}\,k^{-(1+\alpha)}+1\right]^{2}}\,dk\,,
\]
the idea is to decompose the ridge parameter into the Gaussian part
and corrections $r(\Lambda/k)=r^{\text{GP}}(\Lambda/k)+\epsilon(\Lambda/k)$,
where $r^{\text{GP}}(\ell)=P/\kappa\,\ell^{(1+\alpha)}$ is given
by \eqref{eq:r_GP}, such that we can expand for $r^{\text{GP}}(\Lambda/k)>\epsilon(\Lambda/k)$
the denominator
\begin{align}
\frac{1}{\left[r(\Lambda/k)\left(\frac{\Lambda}{k}\right)^{-(1+\alpha)}\,k^{-(1+\alpha)}+1\right]} & =\frac{1}{\left[\frac{P}{\kappa}\,k^{-(1+\alpha)}+1\right]}\label{eq:expansion_linear_order}\\
 & -\frac{k^{-(1+\alpha)}}{\left[\frac{P}{\kappa}\,k^{-(1+\alpha)}+1\right]^{2}}\left(\frac{\Lambda}{k}\right)^{-(1+\alpha)}\epsilon(\Lambda/k)+\mathcal{O}(\epsilon^{2})\,.\nonumber 
\end{align}
The first term corresponds to the Gaussian case of which we already
know the scaling as $I_{\text{bias}}^{\text{GP}}(P/\kappa)^{-\beta/(1+\alpha)}$
given by \eqref{eq:loss_bias_GP}.

We are thus left with finding the non-Gaussian corrections contained
in
\begin{equation}
\delta\langle\mathcal{L}_{\text{bias}}\rangle/P:=-\frac{2}{2}\int_{0}^{\infty}\frac{k^{-(1+\beta)}k^{-(1+\alpha)}}{\left[\frac{P}{\kappa}\,k^{-(1+\alpha)}+1\right]^{3}}\,\left(\frac{\Lambda}{k}\right)^{-(1+\alpha)}\epsilon(\Lambda/k)\,dk\,.\label{eq:non_gauss_corr_loss_bias_linear_term}
\end{equation}
Similarily to the Gaussian case, we want to do a substitution to extract
the $P$-dependence in $\epsilon(\Lambda/k)$ without necessarily
solving its convolution equation \eqref{eq:non_gauss_r_flow_solution}.
To this end, we perform a substitution in the integration variable,
namely $\tilde{k}^{-(1+\alpha)}:=\frac{P}{\kappa}\,k^{-(1+\alpha)}$
or
\begin{align}
\tilde{k}:= & \left(\frac{P}{\kappa}\right)^{-\frac{1}{1+\alpha}}k\,,\label{eq:subst_k_tilde-1}\\
d\tilde{k}= & \left(\frac{P}{\kappa}\right)^{-\frac{1}{1+\alpha}}\,dk\,,\nonumber 
\end{align}
so that we get
\begin{align}
\delta\langle\mathcal{L}_{\text{bias}}\rangle/P & =-\left(\frac{P}{\kappa}\right)^{-\frac{1+\alpha+\beta}{1+\alpha}}\int_{0}^{\infty}\frac{\tilde{k}^{-(2+\beta+\alpha)}}{\left[\tilde{k}^{-(1+\alpha)}+1\right]^{3}}\,\ell^{-(1+\alpha)}\,\epsilon(\ell)\big|_{\ell=\Lambda\left(\frac{P}{\kappa}\right)^{-\frac{1}{1+\alpha}}/\tilde{k}}\,d\tilde{k}\,.\label{eq:delta_loss_bias-2}
\end{align}
Next we need to determine $\ell^{-(1+\alpha)}\,\epsilon\big(\ell\big)$.
We can do this in two steps. First, we split the integral $\int_{1}^{\ell}$
in \eqref{eq:non_gauss_r_flow_solution} into $\int_{0}^{\ell}-\int_{0}^{1}$

\begin{align}
\ell^{-(1+\alpha)}\,\epsilon(\ell) & =\int_{0}^{\ell}\,\bigg[A(P,\Lambda)\,\frac{\ell^{\prime\beta-1}}{\left[\frac{P}{\kappa}\ell^{\prime(1+\alpha)}\Lambda^{-(1+\alpha)}+1\right]^{2}}\nonumber \\
 & \phantom{\ell^{(1+\alpha)}\int_{1}^{\ell}\,\bigg[}+B(P,\Lambda)\,\frac{\ell^{\prime\alpha-1}}{\frac{P}{\kappa}\ell^{\prime(1+\alpha)}\Lambda^{-(1+\alpha)}+1}\bigg]\,d\ell^{\prime}\label{eq:non_gauss_r_flow_solution-1}\\
- & \ell^{-(1+\alpha)}\,\delta\epsilon_{0}(\ell)\,,\nonumber 
\end{align}
We first compute
\begin{align*}
\ell^{-(1+\alpha)}\,\delta\epsilon_{0}(\ell) & :=\int_{0}^{1}\bigg[A(P,\Lambda)\frac{\ell^{\prime\beta-1}}{\left[\frac{P}{\kappa}\ell^{\prime(1+\alpha)}\Lambda^{-(1+\alpha)}+1\right]^{2}}\\
 & \phantom{=\int_{0}^{1}\bigg[}+B(P,\Lambda)\frac{\ell^{\prime\alpha-1}}{\frac{P}{\kappa}\ell^{\prime(1+\alpha)}\Lambda^{-(1+\alpha)}+1}\bigg]\,d\ell^{\prime}\,
\end{align*}
by exploiting that for small $\ell^{\prime}\in[0,1]$ we may approximate
the denominator by $1$ to obtain
\begin{align}
\ell^{-(1+\alpha)}\,\delta\epsilon_{0}(\ell) & \simeq\frac{A(P,\Lambda)}{\beta}+\frac{B(P,\Lambda)}{\alpha}\,.\label{eq:delta_epsilon_0}
\end{align}
For the remaining integral in \eqref{eq:non_gauss_r_flow_solution-1}
we substitute $\frac{P}{\kappa}\ell^{(1+\alpha)}\Lambda^{-(1+\alpha)}=:\tilde{\ell}^{(1+\alpha)}$
or likewise

\begin{align}
\tilde{\ell}:= & \left(\frac{P}{\kappa}\right)^{\frac{1}{1+\alpha}}\Lambda_{\text{ }}^{-1}\,\ell\,,\label{eq:l_tilde_subst}\\
d\tilde{\ell}= & \left(\frac{P}{\kappa}\right)^{\frac{1}{1+\alpha}}\Lambda^{-1}\,d\ell^{\prime}\,,\nonumber 
\end{align}
which yields

\begin{align}
\ell^{-(1+\alpha)}\,\epsilon(\ell) & =C_{I}(P)\,I_{\epsilon,I}(\tilde{\ell})\label{eq:eps_as_int_subst}\\
 & +C_{II}(P)\,I_{\epsilon,II}(\tilde{\ell})\\
 & -\ell^{-(1+\alpha)}\,\delta\epsilon_{0}(\ell)\,,\nonumber 
\end{align}
where we have defined constants $C_{I/II}$ that are still functions
of $P$ and $P$-independent integrals $I_{\epsilon,I/II}$ as

\begin{align}
I_{\epsilon,I}(\tilde{\ell}):= & \int_{0}^{\tilde{\ell}}\frac{\tilde{\ell}^{\prime\beta-1}}{\left[\tilde{\ell}^{\prime\alpha+1}+1\right]^{2}}\,d\tilde{\ell}^{\prime}\,,\label{eq:def_Is}\\
I_{\epsilon,II}(\tilde{\ell}):= & \int_{0}^{\tilde{\ell}}\,\frac{\tilde{\ell}^{\prime\alpha-1}}{\tilde{\ell}^{\prime\alpha+1}+1}\,d\tilde{\ell}^{\prime}\,,\nonumber \\
C_{I}(P):= & A(P,\Lambda)\,\Lambda^{\beta}\,\left(\frac{P}{\kappa}\right)^{-\frac{\beta}{1+\alpha}}=4U(1)\,\kappa\,\left(\frac{P}{\kappa}\right)^{-\frac{\beta-\alpha-1}{1+\alpha}}\,,\nonumber \\
C_{II}(P):= & B(P,\Lambda)\,\Lambda^{\alpha}\,\left(\frac{P}{\kappa}\right)^{-\frac{\alpha}{1+\alpha}}=4U(1)\,\kappa\,\left(\frac{P}{\kappa}\right)^{\frac{1}{1+\alpha}}\,,\nonumber 
\end{align}
where we used \eqref{eq:def_A_B}. In \eqref{eq:delta_loss_bias-2}
we need to evaluate $\ell^{-(1+\alpha)}\,\epsilon(\ell)\big|_{\ell=\Lambda\left(\frac{P}{\kappa}\right)^{-\frac{1}{1+\alpha}}/\tilde{k}}$,
so that the upper bound of the integral in \eqref{eq:eps_as_int_subst}
and hence in \eqref{eq:def_Is} becomes with \eqref{eq:l_tilde_subst}
\begin{align*}
\tilde{\ell}\Big(\ell=\Lambda\left(\frac{P}{\kappa}\right)^{-\frac{1}{1+\alpha}}/\tilde{k}\Big) & =1/\tilde{k}\,,
\end{align*}
so that the $P$-dependence is gone. The $P$-dependence has hence
been recast into $C_{I}$ and $C_{II}$ alone. In particular we see
that the integrand is also cutoff independent. 

The contribution from $\ell^{-(1+\alpha)}\,\delta\epsilon_{0}$ to
\eqref{eq:delta_loss_bias-2} with \eqref{eq:delta_epsilon_0} is
\begin{align*}
 & \left(\frac{A(P,\Lambda)}{\beta}+\frac{B(P,\Lambda)}{\alpha}\right)\,\left(\frac{P}{\kappa}\right)^{-\frac{1+\alpha+\beta}{1+\alpha}}\,I_{\mathrm{bias}}^{\epsilon_{0}}\,,
\end{align*}
where we defined the $P$-independent integral 
\begin{align*}
I_{\mathrm{bias}}^{\epsilon_{0}} & :=\int_{0}^{\infty}\frac{\tilde{k}^{-(2+\beta+\alpha)}}{\left[\tilde{k}^{-(1+\alpha)}+1\right]^{3}}\,d\tilde{k}\,.
\end{align*}
The contributions of $\epsilon$ proportional to $I_{\epsilon,I}(\tilde{\ell})$
and $I_{\epsilon,II}(\tilde{\ell})$ to \eqref{eq:delta_loss_bias-2}
likewise motivate the definition of the $P$-independent integrals
\begin{align*}
I_{\text{bias}}^{\epsilon,I} & :=\int_{0}^{\infty}\frac{\tilde{k}^{-(2+\alpha+\beta)}}{\left[\tilde{k}^{-(1+\alpha)}+1\right]^{3}}\,I_{\epsilon,I}(\tilde{k}^{-1})\,d\tilde{k}\,,\\
I_{\text{bias}}^{\epsilon,II} & :=\int_{0}^{\infty}\frac{\tilde{k}^{-(2+\alpha+\beta)}}{\left[\tilde{k}^{-(1+\alpha)}+1\right]^{3}}\,I_{\epsilon,II}(\tilde{k}^{-1})\,d\tilde{k}\,,
\end{align*}
which finally allow us to express the contribution $\delta\langle\mathcal{L}_{\text{bias}}\rangle/P$
of order $\order(\epsilon)$ to the bias part of the loss together
with the Gaussian part \eqref{eq:loss_bias_GP} as

\begin{align}
\langle\mathcal{L}_{\text{bias}}\rangle/P & =I_{\text{bias}}^{\text{GP}}\cdot\left(\frac{P}{\kappa}\right)^{-\frac{\beta}{1+\alpha}}-\frac{\kappa^{-\beta}}{2\beta}\cdot\left(\frac{P}{\kappa}\right)^{-\beta}+\delta\langle\mathcal{L}_{\text{bias}}\rangle/P\label{eq:loss_bias_final}\\
\delta\langle\mathcal{L}_{\text{bias}}\rangle/P= & -4U(1)\kappa\,I_{\text{bias}}^{\epsilon,I}\cdot\left(\frac{P}{\kappa}\right)^{-\frac{2\beta}{1+\alpha}}-4U(1)\kappa\,I_{\text{bias}}^{\epsilon,II}\cdot\left(\frac{P}{\kappa}\right)^{-\frac{\beta+\alpha}{1+\alpha}}\nonumber \\
 & +\left(\frac{4U(1)\Lambda^{-\beta}}{\beta}+\frac{4U(1)\Lambda^{-\alpha}}{\alpha}\right)I_{\mathrm{bias}}^{\epsilon_{0}}\,\kappa^{\frac{1+\alpha+\beta}{1+\alpha}}\,P^{-\frac{\beta}{1+\alpha}}\,,\nonumber 
\end{align}
where only the last line due to $\delta\epsilon_{0}$ depends on the
cutoff $\Lambda$ and vanishes for $\Lambda\to\infty$.

\subsubsection{Variance contribution to the loss}

For the variance contribution we can proceed analogously. We have
from \eqref{eq:L_var_per_P}
\[
\langle\mathcal{L}_{\text{var}}\rangle/P+\frac{P^{-\alpha}}{2\alpha}+\frac{P^{-1}}{2\kappa}=\frac{1}{2}\int_{0}^{\infty}\frac{k^{-(1+\alpha)}}{r(\Lambda/k)\left(\frac{\Lambda}{k}\right)^{-(1+\alpha)}k^{-(1+\alpha)}+1}\,dk\,.
\]
 We can use the same expansion up to linear order in $\epsilon(\ell)$
as in \eqref{eq:expansion_linear_order}. The leading order is again
the Gaussian contribution, which we determined in \eqref{eq:loss_var_GP}
to be $I_{\text{var}}^{\text{GP}}\,(P/\kappa)^{-\frac{\alpha}{1+\alpha}}$.
And we are interested in the non-Gaussian corrections $\propto\order(\epsilon)$
\begin{align}
\delta\langle\mathcal{L}_{\text{var}}\rangle/P & =-\frac{1}{2}\int_{0}^{\infty}\left[\frac{k^{-(1+\alpha)}}{\frac{P}{\kappa}k^{-(1+\alpha)}+1}\right]^{2}\,\left(\frac{\Lambda}{k}\right)^{-(1+\alpha)}\epsilon(\Lambda/k)\,dk\label{eq:non_gauss_corr_loss_var-1-1}\\
 & =-\frac{1}{2}\left(\frac{P}{\kappa}\right)^{-\frac{1+2\alpha}{1+\alpha}}\,\int_{0}^{\infty}\left[\frac{\tilde{k}^{-(1+\alpha)}}{\tilde{k}^{-(1+\alpha)}+1}\right]^{2}\,\ell^{-(1+\alpha)}\epsilon(\ell)\big|_{\ell=\Lambda\left(\frac{P}{\kappa}\right)^{-\frac{1}{1+\alpha}}/\tilde{k}}\,d\tilde{k},\nonumber 
\end{align}
where we used the same substitution \eqref{eq:subst_k_tilde-1} as
for the bias part. We proceed in the same manner as for the bias part,
using that $\epsilon$ is given by \eqref{eq:eps_as_int_subst}. In
summary we have
\begin{align*}
\delta\langle\mathcal{L}_{\text{var}}\rangle/P & =-\frac{1}{2}\left(\frac{P}{\kappa}\right)^{-\frac{1+2\alpha}{1+\alpha}}\,\int_{0}^{\infty}\left[\frac{\tilde{k}^{-(1+\alpha)}}{\tilde{k}^{-(1+\alpha)}+1}\right]^{2}\\
 & \times\,\bigg[C_{I}(P)\,I_{\epsilon,I}(\tilde{k}^{-1})+C_{II}(P)\,I_{\epsilon,II}(\tilde{k}^{-1})-\frac{A(P,\Lambda)}{\beta}+\frac{B(P,\Lambda)}{\alpha}\bigg]\,d\tilde{k}\,.
\end{align*}
Analogously, we define the following constants
\begin{align*}
I_{\text{var}}^{\epsilon,I} & :=\frac{1}{2}\int_{0}^{\infty}\left[\frac{\tilde{k}^{-(1+\alpha)}}{\tilde{k}^{-(1+\alpha)}+1}\right]^{2}\,I_{\epsilon,I}(\tilde{k}^{-1})\,d\tilde{k}\,,\\
I_{\text{var}}^{\epsilon,II} & :=\frac{1}{2}\int_{0}^{\infty}\left[\frac{\tilde{k}^{-(1+\alpha)}}{\tilde{k}^{-(1+\alpha)}+1}\right]^{2}\,I_{\epsilon,II}(\tilde{k}^{-1})\,d\tilde{k}\,,\\
I_{\text{var}}^{\epsilon_{0}} & :=\frac{1}{2}\int_{0}^{\infty}\left[\frac{\tilde{k}^{-(1+\alpha)}}{\tilde{k}^{-(1+\alpha)}+1}\right]^{2}\,d\tilde{k}\,.
\end{align*}
And putting everything together with the contribution \eqref{eq:loss_var_GP}
of the Gaussian process we finally find 

\begin{align}
\langle\mathcal{L}_{\text{var}}\rangle/P & =I_{\text{var}}^{\text{GP}}\,\left(\frac{P}{\kappa}\right)^{-\frac{\alpha}{1+\alpha}}-\frac{\kappa^{-\alpha}}{2\alpha}\,\left(\frac{P}{\kappa}\right)^{-\alpha}-\frac{1}{2}\,\left(\frac{P}{\kappa}\right)^{-1}+\delta\langle\mathcal{L}_{\text{var}}\rangle/P\,\label{eq:loss_var_final}\\
\delta\langle\mathcal{L}_{\text{var}}\rangle/P & \simeq-4U(1)\kappa\,I_{\text{var}}^{\epsilon,I}\cdot\left(\frac{P}{\kappa}\right)^{-\frac{\beta+\alpha}{1+\alpha}}-4U(1)\kappa\,I_{\text{var}}^{\epsilon,II}\cdot\left(\frac{P}{\kappa}\right)^{-\frac{2\alpha}{1+\alpha}}\nonumber \\
 & +I_{\text{var}}^{\epsilon_{0}}\cdot\left(\frac{4U(1)\Lambda^{-\beta}}{\beta}+\frac{4U(1)\Lambda^{-\alpha}}{\alpha}\right)\kappa^{\frac{1+2\alpha}{1+\alpha}}\,P^{-\frac{\alpha}{1+\alpha}}\,.\nonumber 
\end{align}
The first two lines are again independent of the cutoff $\Lambda$,
while the last line contains an explicit cutoff-dependence that, however,
vanishes for $\Lambda\to\infty$.

\textbf{}

\textbf{}

\textbf{}

\section{Saddle-point approximation of the renormalized action}\label{sec:Saddle-point-approximation-of-mean}

By decimating and re-scaling we got the renormalized action \eqref{eq:RG_action_main_Delta}.
Now we are interested decimating all modes beyond the mode of interest.
Decimation will contribute non-trivially until we get a sufficiently
massive theory, where fluctuations are suppressed by the mass-term.
At such a point, a saddle point approximation on the action is expected
to provide good results for the mean of the field. Based on the Gaussian
part of the action this condition translates to 
\begin{equation}
r(1)=\frac{P}{\kappa}\overset{!}{=}\lambda(k_{0})^{-1}=k_{0}^{1+\alpha}\,,\label{eq:cond_massive_th}
\end{equation}
 where $k_{0}$ is the mode up to where we decimate. Therefore, $\Lambda/\ell_{0}=k_{0}$
fixes the value of the flow parameter at which the non-trivial flow
is expected to stop.

By fixing $\ell$ and integrating the flow equations up to this point
to get renormalized parameters $r(\ell)$ and $U(\ell)$, we may use
the resulting action to determine an approximation of the mean of
the field by a saddle-point approximation, or, if $\ell>\ell_{0}$,
possibly take some fluctuation corrections into account. We will here
only perform the former, i.e. maximize the probability distribution
with respect to $\Delta$ as
\[
\max_{\Delta}\,p[\Delta]=\max_{\Delta}e^{S[\Delta]}\,.
\]
As the action is a negative quantity, this condition translates to
minimizing $S[\Delta]$ with respect to $\Delta$. The saddle-point
approximation yields an equation for the mean discrepancies $\Delta^{*}(m)$.
The minimization problem reads
\begin{equation}
\frac{\delta\,S[\Delta]}{\delta\Delta(m)}\overset{!}{=}0=\frac{\delta\,S_{0}[\Delta]}{\delta\Delta(m)}+\frac{\delta\,S_{\text{int}}[\Delta]}{\delta\Delta(m)}\,.\label{eq:saddle-point_cond}
\end{equation}
Such functional derivatives can be computed using the following identities
\begin{align*}
\frac{\delta\,\Delta(k)}{\delta\Delta(m)} & =\delta(k-m)\,,\\
\frac{\delta\,\Delta(k)^{2}}{\delta\Delta(m)} & =\int_{\ell}^{\Lambda}\frac{\partial\,\Delta^{2}}{\partial\Delta}|_{\Delta(s)}\,\frac{\delta\,\Delta(s)}{\delta\Delta(k)}\,ds\,\delta(k-m)=2\Delta(k)\,\delta(k-m)\,.
\end{align*}
 We can directly see by using these and the free part of \eqref{eq:RG_action_main_Delta}
that
\begin{equation}
\frac{\delta\,S_{0}[\Delta]}{\delta\Delta(m)}=-s(\ell)\,\big(r(\ell)+\lambda(m)^{-1}\Delta(m)+\lambda(m)^{-1}y(m)\big)\,.\label{eq:derivative_action_free}
\end{equation}

\begin{figure}
\begin{centering}
\includegraphics{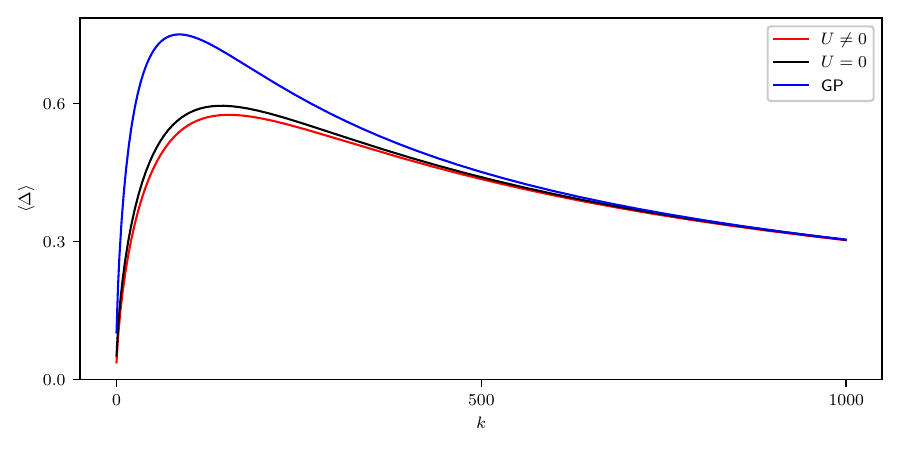}
\par\end{centering}
\caption{Mean of the interacting theory computed from the saddle point of the
action of the Gaussian process (blue GP), and of the non-Gaussian
theory (red: taking $U\protect\neq0$ into account with help of RG
and in solving the saddle point equation; black: taking $U\protect\neq0$
into account with help of RG but neglect $U$ when solving the saddle
point equation)}\label{fig:saddle_mean}
\end{figure}

In order to compute the derivative of the interacting part of the
action, we argue as in \prettyref{sec:Rescaling-of-interaction} that
we may effectively replace the non-diagonal terms described by the
integral boundaries $[\lfloor k\rfloor,\lfloor k\rfloor+1]$ with
diagonal terms, such that
\[
S_{\text{int}}\sim\int_{\ell}^{\Lambda}\Delta(k)^{2}\,dk\int_{\ell}^{\Lambda}\Delta(q)^{2}\,dq\,.
\]
 We can now derive with respect to the discrepancies, and by noting
that both integrals are symmetric we can simplify the expression.
The functional derivative gets rid of one integral and sets the index,
either $k$ or $q$, to $m$. The second integral remains, but both
terms of the product rule are equivalent, such that
\begin{equation}
\frac{\delta\,S_{\text{int}}[\Delta]}{\delta\Delta(m)}=-4U(\ell)P\,\Delta(m)\,\int_{\ell}^{\Lambda}\Delta(k)^{2}\,dk\:.\label{eq:derivative_action_interact}
\end{equation}
 Because of the interacting part, we have a non-local term which still
depends on $\Delta$. Therefore, we get a self-consistent equation
in terms of the $\Delta$-fields. We get the final expression by adding
\eqref{eq:derivative_action_free} and \eqref{eq:derivative_action_interact}
and setting the result to zero
\begin{equation}
\Delta^{*}(m)=\frac{y(m)}{r(\ell)\lambda(m)+1+4U(\ell)P\lambda(m)s(\ell)^{-1}\,R[\Delta^{*}]}\,,\label{eq:mean_discrepancies_saddle_point}
\end{equation}
 where we define the functional (non-locality) 
\[
R[\Delta^{*}]:=\int_{\ell}^{\Lambda}\Delta^{*}(k)^{2}\,dk\,.
\]
 We see that for $U\to0$ we get the Gaussian result \eqref{eq:mean_Delta_cont}.
A comparison of this saddle point to the one computed only from the
renormalized Gaussian part of the action is shown in \prettyref{fig:saddle_mean}.
The direct contribution of the non-Gaussian term

\subsection{Fixed parameters by artificial power law statistics}

In order to implement the theory, we generate artificially data whose
correlations show a power law decay in its eigenspectrum. In order
to find the parameters for the Langevin dynamics we compute the variance
of the output 
\begin{align*}
\langle f_{\alpha}f_{\beta}\rangle_{W} & =\langle\sum_{i}W_{i}\,x_{\alpha i}\,\sum_{j}W_{j}\,x_{\beta j}\rangle_{W}\\
 & =\sum_{i,j}x_{\alpha i}x_{\beta j}\,\langle W_{i}W_{j}\rangle_{W}\,,
\end{align*}
 which is fixed by the artificially generated power law in its spectrum.
Therefore, 
\[
\langle f_{\alpha}f_{\beta}\rangle_{W}\overset{!}{=}C^{(xx)}=\sum_{i,j}x_{\alpha i}x_{\beta j}\,,
\]
 yields a unit variance for the prior of the weights. This means $g_{w}/d=\mathcal{O}(1)$
and therefore, $\gamma$ must be equal to $\kappa$.

\section{Random data set with power law statistics}\label{sec:Random-data-set}

\begin{figure}
\begin{centering}
\includegraphics{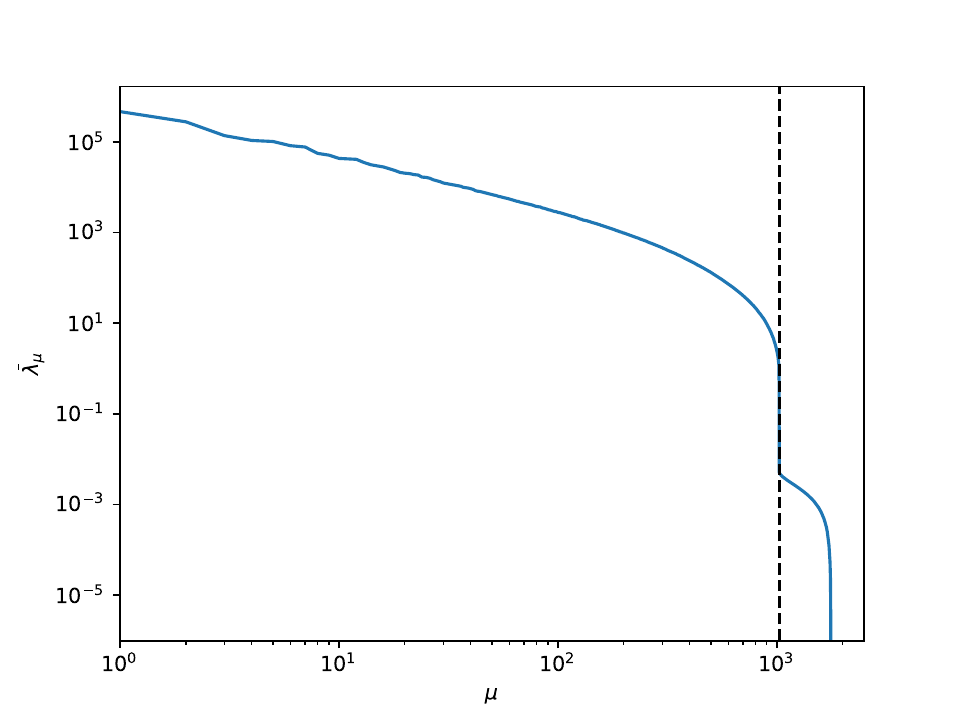}
\par\end{centering}
\caption{Power law falloff of eigenvalues constructed from the CIFAR-10 dataset.}\label{fig:Power-law-falloff-CIFAR}
\end{figure}

To test the theory, we use an artificial data set which allows us
to generate arbitrary amounts of training data that possesses power
law statistics. To this end, we first generate a matrix of preliminary
data vectors $\bar{X}_{\alpha}\in\bR^{d}$ for $1\le\alpha\le P$
\begin{align*}
\bar{X}_{\alpha i} & \stackrel{\text{i.i.d.}}{\sim}\N(0,1/d)\,.
\end{align*}
The covariance matrix $\bar{C}_{ij}:=\sum_{\alpha=1}^{P}\,\bar{X}_{\alpha i}\bar{X}_{\alpha j}\equiv\bar{X}^{\T}\bar{X}$
is then diagonalized
\begin{align*}
\bar{C}\,v_{\mu} & =\bar{\lambda}_{\mu}\,v_{\mu}\,,\\
v_{\mu}^{\T}v_{\nu} & =\delta_{\mu\nu}\,.
\end{align*}
Decomposing the data vectors into these modes one has
\begin{align*}
\bar{X}_{\alpha} & =\sum_{\mu=1}^{d}\,\bar{a}_{\alpha\mu}\,v_{\mu}\,.
\end{align*}
The coefficients obey
\begin{align}
\bar{\lambda}_{\mu}\,\delta_{\mu\nu} & =v_{\mu}^{\T}\bar{C}v_{\nu}\label{eq:eig_projection}\\
 & =v_{\mu}^{\T}\,\Big[\sum_{\mu^{\prime},\nu^{\prime}=1}^{d}\sum_{\alpha=1}^{P}\,u_{\mu^{\prime}}\bar{a}_{\alpha\mu^{\prime}}\bar{a}_{\alpha\nu^{\prime}}\,u_{\nu^{\prime}}^{\T}\Big]\,v_{\nu}\nonumber \\
 & =\sum_{\alpha=1}^{P}\bar{a}_{\alpha\mu}\bar{a}_{\alpha\nu}\,,\nonumber 
\end{align}
which shows that also the coefficient vectors are orthogonal
\begin{align*}
\bar{a}_{\mu}^{\T}\bar{a}_{\nu} & =\delta_{\mu\nu}\,\bar{\lambda}_{\mu}\,.
\end{align*}
Now assume we want to shape the spectrum to take on a desired form.
To this end, we may define new coefficients
\begin{align*}
a_{\circ\mu} & :=\sqrt{\frac{\lambda_{\mu}}{\bar{\lambda}_{\mu}}}\,\bar{a}_{\circ\mu}\,.
\end{align*}
Inserted into \eqref{eq:eig_projection} we obtain data vectors 
\begin{align}
X_{\alpha} & =\sum_{\mu=1}^{d}\,a_{\alpha\mu}v_{\mu}\label{eq:decomp_X}
\end{align}
with a covariance matrix $C:=X^{\T}X$ with the same eigenvectors
$v_{\mu}$ but different eigenvalues $\lambda_{\mu}$.

\subsection{Power law target}

Besides the data $x$ giving rise to a power law in the kernel eigenvalues,
we also want to assume that the coefficients $y_{\mu}$ of the target
follow a power law. To this end diagonalize the kernel $K=XX^{\T}$
as $K\,u_{\mu}=\lambda_{\mu}\,u_{\mu}$. We would like to construct
a test data set which realizes this condition. In particular, we would
like to have a linear teacher 
\begin{align*}
y_{\alpha} & =w^{\T}x_{\alpha}\,\quad\forall\alpha=1,\ldots,P.
\end{align*}
Now consider that the target $y$ is expanded into eigenmodes of the
kernel
\begin{align*}
y & =\sum_{\mu=1}^{P}\,y_{\mu}u_{\mu}\,,\\
y_{\mu} & =u_{\mu}^{\T}y\,.
\end{align*}
We would like these coefficients to decay with $\mu$ following some
prescribed form $y_{\mu}=f(\mu)$, for example a power law. We get
\begin{align*}
y_{\mu} & =\sum_{\alpha=1}^{P}u_{\mu\alpha}\sum_{i=1}^{d}\,X_{\alpha i}w_{i}\\
 & =u_{\mu}^{\T}Xw\,.
\end{align*}
Combining all eigenvectors $U=(u_{1},\ldots,u_{\mu})$ into a single
matrix $U\in\bR^{P\times\min(P,d)}$ and fixing the coefficients $y_{1\le\mu\le P}$
to prescribed values, the latter equation reads
\begin{align*}
y & =U^{\T}Xw\,.
\end{align*}
We may hence determine the coefficients $w$ as
\begin{align*}
w & =\big[U^{\T}X\big]^{-1}y\,.
\end{align*}
In case that $d>P$, we need to regularize this problem $U^{\T}X\to U^{\T}X+\epsilon\bI$.

\subsection{Overlaps}\label{sec:Overlaps}

\begin{figure}
\begin{centering}
\includegraphics{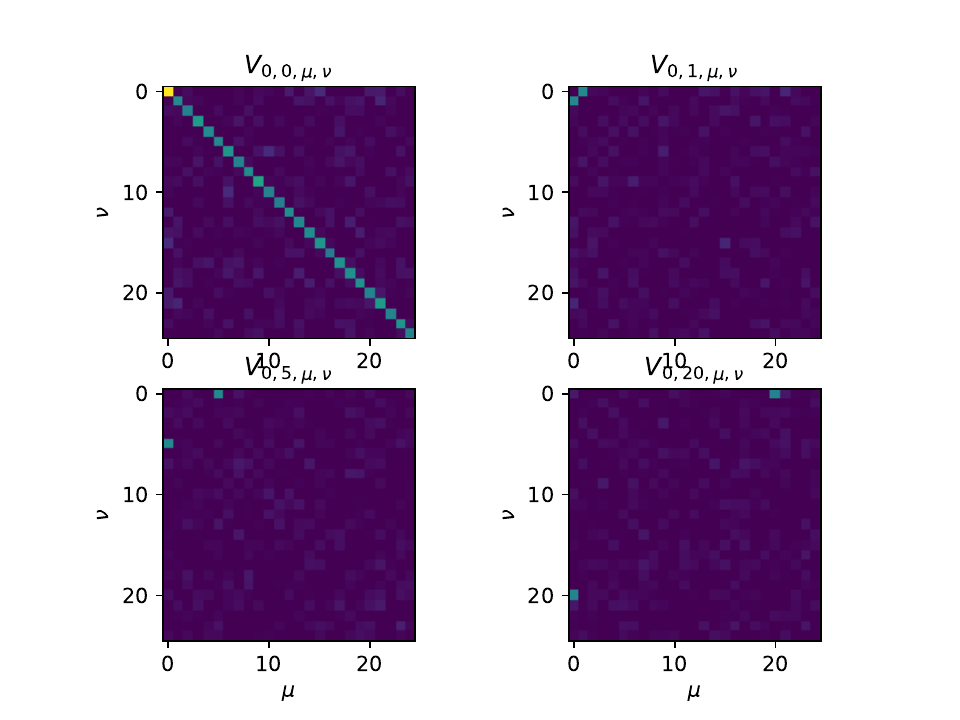}
\par\end{centering}
\caption{Overlap of four eigenvectors of Gaussian i.i.d. dataset given by \eqref{eq:overlap}.}\label{fig:Overlap-of-four}
\end{figure}

Another quantities we need for treating non-linear networks are the
feature-feature overlaps of higher order
\begin{align}
V_{\mu_{1}\mu_{2}\mu_{3}\mu_{4}} & =\sum_{\alpha=1}^{P}u_{\mu_{1}\alpha}u_{\mu_{2}\alpha}u_{\mu_{3}\alpha}u_{\mu_{4}\alpha}\,,\label{eq:overlap}
\end{align}
which is shown in \prettyref{fig:Overlap-of-four}. The overlaps are
partially diagonal
\begin{align}
V_{\mu_{1}\mu_{2}\mu_{3}\mu_{4}} & \simeq\begin{cases}
V^{(4)} & \text{all indices identical}\\
V^{(2)} & \text{two indices pairwise identical}\\
0 & \text{else}
\end{cases}\,,\label{eq:V_4_Gauss}
\end{align}
so they behave as overlaps of random i.i.d. vectors. Assuming the
effective model for the overlaps $v_{\mu\alpha}\stackrel{\text{i.i.d.}}{\sim}\N(0,1/P)$,
so that $\|v_{\mu}\|^{2}\simeq1$, one has for 
\begin{align}
V_{\mu_{1}\mu_{2}\mu_{3}\mu_{4}}^{\text{approx}} & =\sum_{\alpha=1}^{P}v{}_{\mu_{1}\alpha}v_{\mu_{2}\alpha}v_{\mu_{3}\alpha}v_{\mu_{4}\alpha}\label{eq:V4_appr}\\
 & \simeq\sum_{\alpha=1}^{P}\langle v{}_{\mu_{1}\alpha}v_{\mu_{2}\alpha}v_{\mu_{3}\alpha}v_{\mu_{4}\alpha}\rangle\nonumber \\
 & =3/P\,\delta_{\mu_{1}\mu_{2}\mu_{3}\mu_{4}}\nonumber \\
 & +1/P\,(1-\delta_{\mu_{1}\mu_{2}\mu_{3}\mu_{4}})\,\big(\delta_{\mu_{1}\mu_{2}}\,\delta_{\mu_{3}\mu_{4}}+\delta_{\mu_{1}\mu_{3}}\delta_{\mu_{2}\mu_{4}}+\delta_{\mu_{1}\mu_{4}}\delta_{\mu_{2}\mu_{3}}\big)\,.\nonumber 
\end{align}

\begin{figure}
\begin{centering}
\includegraphics{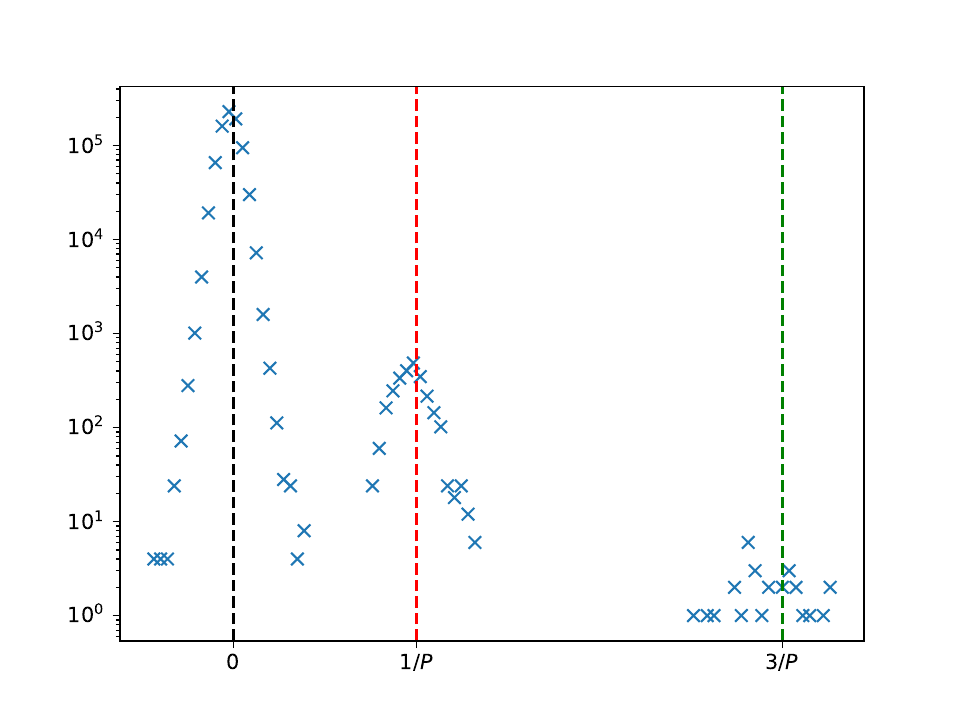}
\par\end{centering}
\caption{Histogram of entries in the overlap tensor $V_{\mu_{1}\mu_{2}\mu_{3}\mu_{4}}$
for Gaussian i.i.d. dataset together with prediction \prettyref{eq:V4_appr}
(dashed vertical lines) from the assumption of random vectors.}\label{fig:Overlap-of-four-1}
\end{figure}

\begin{figure}
\begin{centering}
\includegraphics{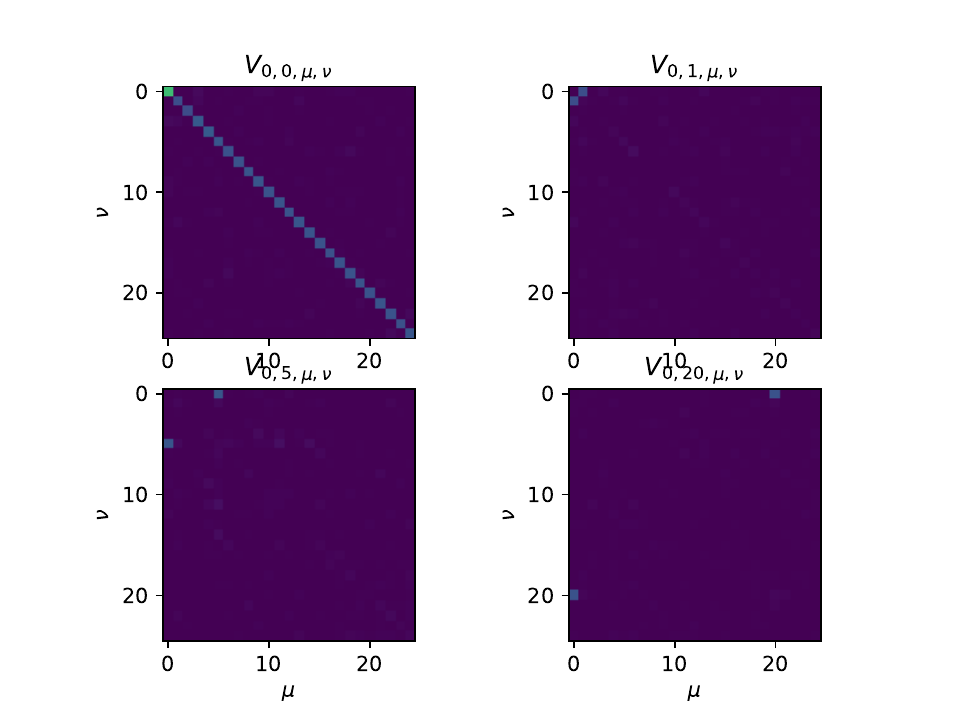}
\par\end{centering}
\caption{Overlap of four eigenvectors of CIFAR-10 given by \eqref{eq:overlap}.}\label{fig:Overlap-of-four-2}
\end{figure}

\begin{figure}
\begin{centering}
\includegraphics{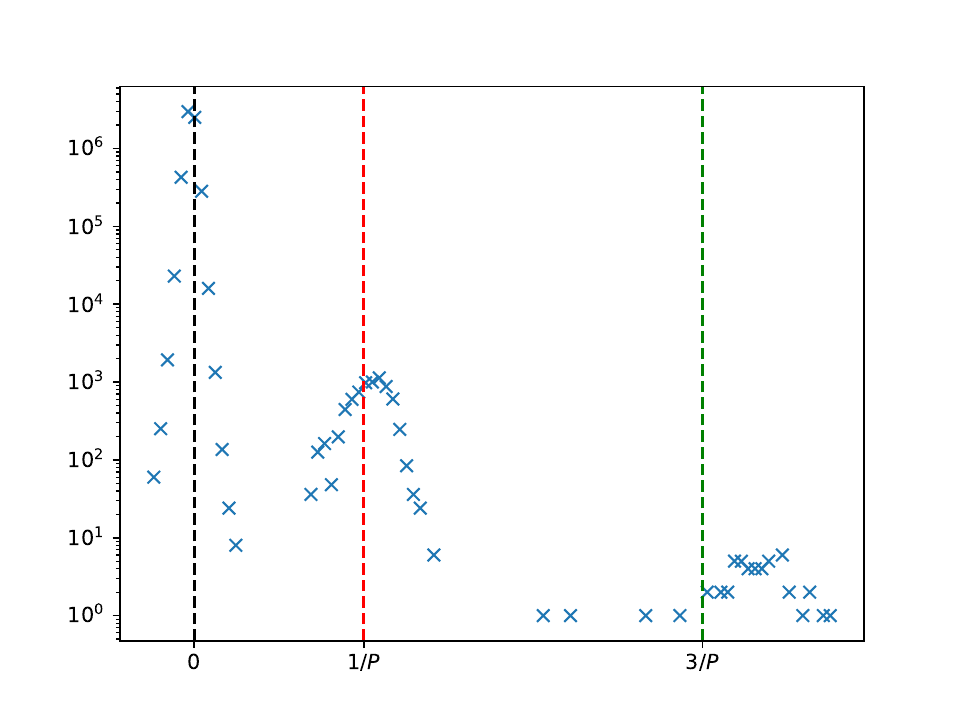}
\par\end{centering}
\caption{Histogram of entries in the overlap tensor $V_{\mu_{1}\mu_{2}\mu_{3}\mu_{4}}$
for CIFAR-10 dataset together with prediction \prettyref{eq:V4_appr}
(dashed vertical lines) from the assumption of random vectors.}\label{fig:Overlap-of-four-1-1}
\end{figure}
In the special case of a diagonal kernel $C^{(xx)}=\sum_{\mu=1}^{P}\Lambda_{\mu}\,e_{\mu}e_{\mu}^{\T}$,
where $e_{\mu}$ are canonical basis vectors, which are also eigenvectors
$e_{\mu\alpha}=\delta_{\mu\alpha}$, the overlaps obtain the exact
expression
\begin{align}
V_{\mu_{1}\mu_{2}\mu_{3}\mu_{4}} & =\sum_{\alpha=1}^{P}e_{\mu_{1}\alpha}e_{\mu_{2}\alpha}e_{\mu_{3}\alpha}e_{\mu_{4}\alpha}\,\label{eq:overlap_canonical}\\
 & =\sum_{\alpha=1}^{P}\delta_{\mu_{1}\alpha}\delta_{\mu_{2}\alpha}\delta_{\mu_{3}\alpha}\delta_{\mu_{4}\alpha}\,\nonumber \\
 & =\delta_{\mu_{1}\mu_{2}\mu_{3}\mu_{4}}\,.\nonumber 
\end{align}

\section{Numerical Workflow}

In this appendix, we provide a detailed description of the numerical
workflow used to ensure the correctness and reliability of the results
presented in this work. This includes the steps taken to validate
the theoretical predictions against numerical simulations and the
criteria used to assess the accuracy of the findings.

\begin{figure}
\begin{centering}
\includegraphics{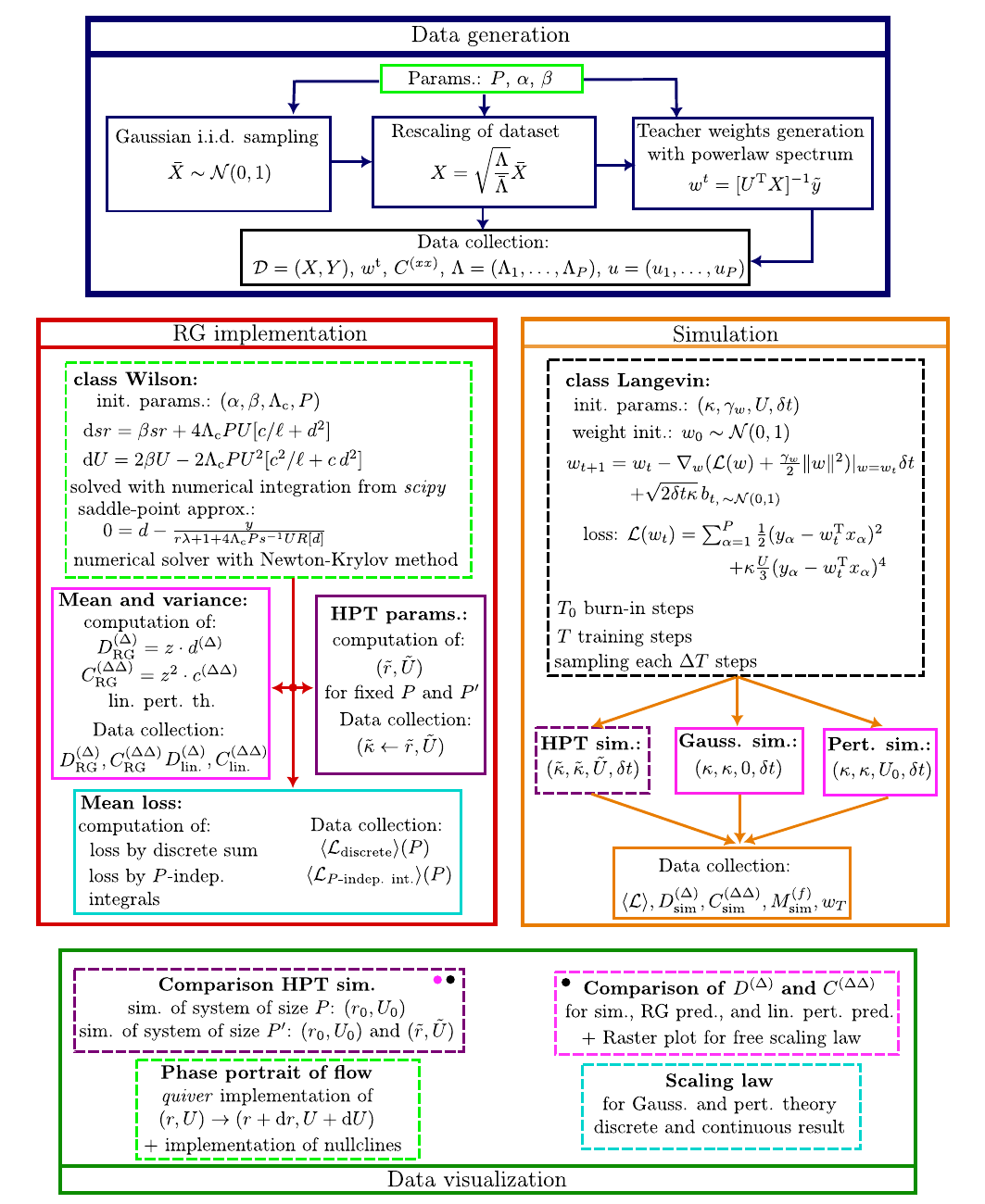}
\par\end{centering}
\caption{\textbf{Numerical workflow.} The validation procedure used to compare
theoretical predictions with numerical simulations comprises four
main stages: data generation (dark blue), network training (orange),
RG implementation (dark red), and data visualization (dark green).
The workflow within each stage is represented by arrows following
the same color scheme as the corresponding box. Conceptual relations
between different stages---which can run independently---are shown
by continuous boxes with matching colors. Hierarchical dependencies
are indicated by dashed boxes, meaning that the processes in dashed
boxes require the output of the corresponding solid-colored ones.
There are also colored dots, which denote intermediate outputs that
can be reused in different stages.}\label{fig:Numerical_workflow}
\end{figure}

In ·\prettyref{fig:Numerical_workflow}, we outline the main steps of the
verification process. It is separated into four main blocks: data
generation (navy blue), numerical simulations (orange), theoretical
implementation (dark red), and comparison of results or data visualization
(dark green).\\
Inside one process, the workflow is indicated by same colored arrows.
Conceptual relations between different processes are indicated by
solid boxes with the same color, meaning that the processes in these
boxes can run independently from each other. Dashed outlined boxes
indicate hierarchical dependencies, meaning that the processes in
such boxes require the output of the corresponding solid-colored ones.
Finally, colored dots indicate intermediate outputs that can be reused
in different processes. 

For instance, data generation (navy blue) is the first step in the
workflow. It involves the implementation of \prettyref{sec:Random-data-set}
to create artificial data sets with power law statistics. For this
we need to define the data dimension $d$, the number of data points
$P$, and the power law exponents $\alpha$ and $\beta$ of the covariance
matrix and the labels, respectively (light green box). The data is
necessary for the numerical simulations by Langevin Dynamics (dashed
black box), specifically to train the neural networks as described
in \prettyref{sec:Langevin-training}. I. e. the data generation's output
(black box) is a prerequisite for the numerical simulations. \\
Nevertheless, from the data generation we can directly extract the
eigenvector-matrix (in figure denoted as $u$) of the covariance matrix
($C^{(xx)}$), which is needed to project the output of the network
onto the eigenbasis and to compare it to the theoretical predictions.
Figures which require this projection, hence, have a black dot indicating
this connection. Hence, the \emph{data collection} box in the navy
blue block has a subleading dependency on the data visualization step.\\
The dashed black box contains the general algorithm to train the neural
networks. This was implemented in \emph{Python} as a class called
\emph{Langevin Dynamics}. One object is instantiated with the regularization
noise $\kappa$, the weight bias $\gamma_{w}$, the perturbation strength
$U$, and the learning rate $\delta t$. The main training loop has
$T_{0}$ burn-in steps, to ensure that the network reaches equilibrium,
followed by $T$ training steps where the weights are updated according
to the Langevin equation \prettyref{eq:Langevin_training}. The outputs
are sampled every $\Delta T$ steps to reduce correlations between
samples.\\
To simulate the free theory, we simply set $U=0$ (pink box left).
For the interacting theory, we set $U>0$ (pink box right). The outputs
of these cases will be plotted in the eigenbasis of the covariance
matrix (pink dashed box). See, for example, \prettyref{fig:Mean-discrepancies}.
There is a third simulation, which is implemented to show the validity
of the hyperparameter transfer in  \prettyref{sec:Hyperparameter-transfer}
(dashed violet box in simulation). To run this simulation, we first
need to determine the hyperparameters according \prettyref{eq:hyperparameter_transform}.\\
This requires the theoretical implementation (dark red box), where
we implemented the RG flow equations \eqref{eq:RG_r_main} and \eqref{eq:RG_U_main}
in a class called \emph{Wilson} (dashed light green box). An object
is instantiated with the assumed power law exponents $\alpha$ and
$\beta$, the cutoff $\Lambda$ and the dataset size $P$, that need
to coincide with the parameters in data generation. In addition to
the \emph{scipy} numerical integrator of the flow equations, we also
implemented the self-consistent saddle-point approximation in \prettyref{eq:mean_discrepancies_saddle_point}.
This allows us to predict the mean and variance of the discrepancies
(pink box) for either the free or interacting theory. These predictions
are then compared to the numerical simulations (pink dashed box).
\\
With the algorithm outlined in \prettyref{sec:Hyperparameter-transfer},
we can compute $(\tilde{r},\tilde{U})$ to compare two systems of
sizes $P$ and $P^{\prime}$ (violet box). The mass term $\tilde{r}$
implies an effective regularization noise $\tilde{\kappa}$ according
to the definition of the mass term, which is then used together with
$\tilde{U}$ to run the third simulation (violet dashed box in simulation).
The outputs are again projected onto the eigenbasis (black dot) and
compared to the theoretical predictions (pink dot) and plotted together
(violet dashed box in data visualization, see \prettyref{fig:Hyperparameter-transfer.}).
The last implementation is for the scaling laws of, both, the free
and interacting theory (light blue box). For this, we implemented,
both, the discrete summation in \prettyref{eq:loss_discrete} and the superposition
of power laws in \eqref{eq:loss_bias_final} and \eqref{eq:loss_var_final},
whose coefficients are $P$-independent integrals which can be integrated
numerically (see \prettyref{sec:Extraction-of-the-P-dependence}). These
results are plotted independently (light blue dashed box, see \prettyref{fig:Neural-scaling-law.}).

Finally, the data visualization block (dark green) contains the plotting
routines to visualize the results of the different simulations and
theoretical predictions. The only plot, we have not discussed yet,
is the phase portrait of the flow (light green dashed box), which
does not require explicit numerical integration. There we also plot
the separatrix, which is given by \eqref{eq:nullcline_limit_r} and
\eqref{eq:nullcline_limit_u} (see \prettyref{fig:Phase-diagram-flow}).

\end{document}